\documentclass[default,iicol]{sn-jnl}

\usepackage{graphicx}%
\usepackage{multirow}%
\usepackage{amsmath,amssymb,amsfonts}%
\usepackage{amsthm}%
\usepackage{mathrsfs}%
\usepackage[title]{appendix}%
\usepackage{xcolor}%
\usepackage{textcomp}%
\usepackage{manyfoot}%
\usepackage{booktabs}%
\usepackage{algorithm}%
\usepackage{algorithmicx}%
\usepackage{algpseudocode}%
\usepackage{listings}%

\def\Npart{$N_{\rm part}$}

\def\RAA{$R_{\rm AA}$}
\def\RCP{$R_{\rm CP}$}
\def\RpAu{$R_{\rm pAu}$}
\def\RdAu{$R_{\rm dAu}$}
\def\RAuAu{$R_{\rm AuAu}$}
\def\RpPb{$R_{\rm pPb}$}
\def\RPbPb{$R_{\rm PbPb}$}
\def\Jpsi{J/$\psi$}
\def\Upsi{$\Upsilon$}
\def\snn{$\sqrt{s_{\mathrm{NN}}}$}
\def\Npart{$N_\mathrm{part}$}

\newcommand{\be}{\begin{equation}}
\newcommand{\ee}{\end{equation}}
\newcommand{\ba}{\begin{eqnarray}}
\newcommand{\ea}{\end{eqnarray}}
\newcommand{\ban}{\begin{eqnarray*}}
\newcommand{\ean}{\end{eqnarray*}}

\newcommand{\req}[1]{({\ref{#1}})}
\newcommand{\jh}[1]{{\color{black}#1}}
\newcommand{\bm}[1]{{\color{black}#1}}

\begin{document}

\title[``QGP Signatures'' Revisited]{``QGP Signatures'' Revisited}

\author*[1]{\fnm{John W.} \sur{Harris}}\email{john.harris@yale.edu}

\author*[2]{\fnm{Berndt} \sur{M\"uller}}\email{berndt.mueller@duke.edu}


\affil*[1]{
\orgdiv{Wright Laboratory, Physics Department}, 
\orgname{Yale University}, 
\orgaddress{ 
\city{New Haven}, 
\postcode{06511}, 
\state{Connecticut}, 
\country{USA}}}

\affil*[2]{
\orgdiv{Department of Physics}, 
\orgname{Duke University}, 
\orgaddress{
\city{Durham}, 
\postcode{27705}, 
\state{North Carolina}, 
\country{USA}}}


\date{25 May 2023}

\abstract{
We revisit the graphic table of QCD signatures in our 1996 {\em Annual Reviews} article ``The Search for the Quark-Gluon Plasma''and assess the progress that has been made since its publication towards providing quantitative evidence for the formation of a quark-gluon plasma in relativistic heavy-ion collisions and its characteristic properties.
}

\keywords{Quark Gluon Plasma, Relativistic Heavy Ion collisions, Quantum Chromodynamics}

\maketitle

\section{Introduction}
\label{sec:Intro}

\subsection{Motivation}

\begin{figure}[ht]
	\centering
	\includegraphics[width=0.95\linewidth]{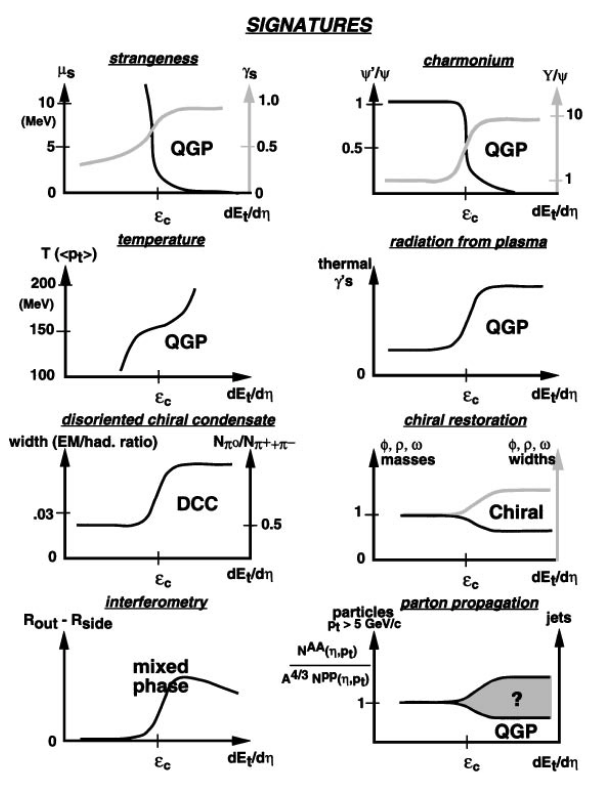}
	\caption{Schematic representations of the possible telltale signs (``signatures'') for the formation of a QGP in relativistic heavy-ion collisions.}
	\label{figures:Signatures1996}
\end{figure}

In our 1996 review article entitled ``The Search for the Quark-Gluon Plasma'' \cite{Harris:1996zx} we described the strategy \cite{Aprahamian:2015qub} adopted by the scientific community to produce, identify, and characterize the quark-gluon plasma (QGP), the predicted state of nuclear matter at temperatures resembling those that were prevalent in the early universe during the first 10 $\mu$s. 
Figure~\ref{figures:Signatures1996} of this review shows a list of observables that promised to be tell-tale signs or signatures for the formation of a QGP in relativistic heavy-ion collisions. 
At the time of our review fixed-target experiments at the CERN-SPS and BNL-AGS were being conducted, which provided first evidence for the prospects of several of these observables. The qualitative sketches in Fig.~\ref{figures:Signatures1996} represented the aspirations of the community of nuclear scientists at the time that were eager to begin the experimental search for the QGP at much higher energies at the Relativistic Heavy Ion Collider (RHIC) and, a decade later, at the Large Hadron Collider (LHC). These experiments began taking data in 2000 (RHIC) and 2010 (LHC), respectively, and have continued since then with regular upgrades of the accelerators and detectors, collecting a wealth of data over a large range of collision energies and for various collision systems. Early summaries of the experimental finding at RHIC were published in four collaboration ``white papers'' \cite{BRAHMS:2004adc,PHENIX:2004vcz,PHOBOS:2004zne,STAR:2005gfr}; a summary of results from the LHC was recently presented by ALICE \cite{ALICE:2022wpn}. It is thus worthwhile to assess the extent to which the expectations expressed in the 1996 review have been confirmed.

Even a casual look at the diagrams in Fig.~\ref{figures:Signatures1996} reveals several common features:
\begin{itemize}
\item The signature observables are shown to exhibit sudden drastic changes in magnitude or slope at a common threshold labeled as $\varepsilon_c$.
\item The abscissa axes are without quantitative numbers.
\item Some of the vertical axes do not give  quantitative information; in others detailed information is sparse.
\end{itemize}
To a certain extent, the absence of quantitative predictions was unavoidable because not enough was known at the time about the physical properties of the QGP, its threshold conditions, and the way in which its intrinsic properties would reveal themselves in experimental observables. It was therefore impossible to make reliable quantitative predictions, and even qualitative predictions required uncertain assumptions.

One of the assumptions that was commonly made at the time was that the transition from a hadron gas to a quark-gluon plasma is a discontinuous, possibly first-order, phase transition. This assumption was motivated by simplified models (c.f.~\cite{Karsch:1980mg}, Fig.~8) and by lattice simulations of SU(3) gauge theory without dynamical quarks (c.f.~\cite{Satz:1985vb}, Figs.~2 and 10). If this were the case in real QCD, rather abrupt changes of certain observables with changing external conditions might be expected, although they would be somewhat smoothened by the transverse nuclear density profile. We now know that the hadron-QGP transition in nature is a smooth, albeit rapid, crossover \cite{Borsanyi:2010cj}. Any characteristic changes in observables must therefore be much more gradual than originally anticipated, which is borne out by the data accumulated at SPS, RHIC, and LHC.

A common feature of all diagrams in Fig.~\ref{figures:Signatures1996} is that the abscissa axis is labeled by the transverse energy per unit pseudorapidity, $dE_t/d\eta$, with a symbol $\varepsilon_c$ that denotes the critical energy density at which hadronic matter transforms into a QGP.\footnote{The attentive reader will notice that the two quantities, $dE_t/d\eta$ and $\varepsilon_c$, have different units and thus should not be compared directly on the same axis. The resolution of this inconsistency is that what was meant to be shown as the axis label is $(dE_t/(A_\perp \tau_{\rm ini} d\eta)$, where $A_\perp$ is the transverse collision cross section and $\tau_{\rm ini}$ denotes the thermalization time.} The precise value of $\varepsilon_c$ was unknown at the time, but was anticipated to lie somewhat below 1 GeV/fm$^3$. Today it is known from lattice-QCD calculations \cite{Borsanyi:2010cj,HotQCD:2018pds} that $\varepsilon_c \approx 0.3-0.4~\textrm{GeV/fm}^3$ depending on the precise definition of the pseudocritical temperature $T_c$ where hadronic matter transitions into QGP. Because the transition is a continuous crossover, not a sharp discontinuity in the thermodynamic sense, an unambiguous and more precise definition of $\varepsilon_c$ is impossible.\footnote{\bm{It should be noted that these studies are focused on the thermal change of the quark condensate (the order parameter for chiral symmetry breaking). The lattice calculations also find a (more gradual) change in the expectation value of the Polyakov loop, which serves as an order parameter for quark confinement in the pure gauge theory and occurs around the same temperature. For real-world QCD with light quarks, however, rigorous probes of color deconfiement are not known.}}

In order to connect the energy density $\varepsilon$ reached in a heavy-ion collision with the measured transverse energy per unit pseudorapidity, $dE_t/d\eta$, one needs to make certain model assumptions. It is most common to invoke the Bjorken model of boost invariant longitudinal hydrodynamics \cite{Bjorken:1982qr} to make this connection. In the Bjorken model the energy density varies with proper time $\tau$ as 
\be
\varepsilon(\tau) = \varepsilon_\mathrm{ini} (\tau_\mathrm{ini}/\tau)^{1+c_s^2} ,
\label{eq:epsBJ}
\ee
where $\tau_\mathrm{ini}$ is the formation time of the QGP, and $c_s^2 = \partial P/\partial\varepsilon$ denotes the speed of sound in the QGP. \bm{In the weak coupling, high temperature limit,} the sound velocity approaches $c_s^2 = 1/3$ (we denote all quantities in natural units $\hbar = c = 1$.) We will use this value here for the sake of simplicity, effectively assuming the initial temperature to be much higher than $T_c$. This implies that the product $\tau\varepsilon(\tau)$ is not constant, but gradually drops as the plasma expands. \bm{This fall-off occurs because the plasma does mechanical work $dW = -pdV$ on itself during the expansion process, causing the decrease of its internal energy as it develops collective flow, primarily in the longitudinal direction.} At late times and at lower collision energies the expansion in transverse directions also becomes important, leading to an even faster drop in $\tau\varepsilon(\tau)$. In order not to complicate things too much, we ignore this effect here. 

The full evolution of the energy density during the nuclear collision can be realistically modeled with relativistic viscous hydrodynamics. Not only does $\varepsilon$ vary with (proper) time $\tau$, it also depends on the position within the fireball. A collision event thus cannot be characterized by a single value of the energy density. Furthermore, the energy density distribution varies from event to event, because both the nuclear density distribution at the moment of collision and the energy deposition are subject to quantum fluctuations.

Over the past two decades, we have learned much about how to model these processes effectively, and how to use detailed comparisons between model predictions and experimental data to constrain the initial conditions and other parameters that govern the dynamical evolution of the QGP. The application of these techniques, which apply Bayesian inference to extract the underlying physics from the data, is a main line of inquiry today. Here we will base our assessment on a more qualitative interpretation of the existing data, which is better suited for a ``big picture'' view that compares our current insight with the expectations in 1996.

This article is intended as an assessment of the progress that has been made since 1996 in the use of various observables to determine the physical properties of the QGP, ascertain its fleeting existence, and map the boundary between normal hadronic matter and the QGP. Our focus will be on the signatures shown in Fig.~\ref{figures:Signatures1996}, however we also point out additional observables, such as elliptic flow, that have become recognized as significant to the field and future investigations. We recognize that a large fraction of research with relativistic heavy ions, especially at the highest energies, has increasingly shifted in the intervening two-and-a-half decades away from the study of equilibrium properties of the QGP to the quest for an understanding of the dynamical processes involved in its formation and evolution. 
We will only touch on this aspect, which has sometimes been described as a ``paradigm shift'',  in the concluding section and refer readers interested in the current perspective of the questions to be addressed by future research in this field to the recent review article \cite{Busza:2018rrf}. Tremendous progress has been made to understand the equilibrium properties of the QGP through theoretical modeling.  Future progress will depend even more so on dynamical modeling and on continuing the close intellectual exchange between experiment and theory.

\subsection{Initial conditions}

The single-particle entropy per unit of pseudorapidity at midrapidity can be related to the charged-particle multiplicity $dN_\mathrm{ch}/d\eta$ as follows \cite{Muller:2005en}:
\be
\frac{dS}{dy} \approx 7\, J\, \frac{dN_\mathrm{ch}}{d\eta} .
\label{eq:dSdy}
\ee
where $J$ is the Jacobian relating a central pseudorapidity interval $d\eta$ to the corresponding rapidity interval $dy$. For energies of interest here, $1 < J < 1.35$ \cite{PHENIX:2015tbb}. For example, for Pb+Pb at a $\sqrt{s_{\rm NN}}$ with $dN_\mathrm{ch}/d\eta \approx 1600$ this yields $dS/dy \approx 12,500$. Alternatively, one can use the volume obtained in the thermal hadron model fit \cite{Andronic:2021dkw} $dV/dy = 4175\pm 380~\mathrm{fm}^3$ and the chemical freeze-out temperature $T_c = 156.6$ MeV to get an independent, consistent estimate:
\be
\frac{dS}{dy}~\approx ~5.5\, T_c^3\, \frac{dV}{dy} ~\approx ~11,500.
\ee

Encouraged by this result we use the relation (\ref{eq:dSdy}) to derive estimates for the entropy density $s_f$ at freeze-out from the measured charged-particle multiplicities $dN_\mathrm{ch}/d\eta$. Assuming approximate entropy conservation expressed by the relation $\tau s(\tau) = {\rm constant}$ for a boost-invariant expansion, we can then estimate the entropy density and temperature at the time of initial thermalization.

The entropy density $s_f$ at the freeze-out time $\tau_f$ can be related to the final entropy per unit rapidity \cite{PHENIX:2015tbb,ALICE:2016igk}
\be
s_f \approx  \frac{dS/dy}{A_\perp\tau_f},
\label{eq:sf-dSdy}
\ee
where $A_\perp$ is the transverse area of the QGP, and $\tau_f$ is the freeze-out proper time. The transverse area $A_\perp$ can be estimated within the Glauber model. For central collisions of identical nuclei $A_\perp \approx \pi R^2$, where $R\approx 7$ fm is the nuclear radius (for $^{197}$Au or $^{208}$Pb). 

The choice of the proper time of initial thermalization $\tau_\mathrm{ini}$ is somewhat more ambiguous. A common choice for the QGP formation time is $\tau_\mathrm{ini} \approx 0.6$ fm/c \cite{Song:2007ux}. This choice is appropriate at energies where the colliding Au or Pb nuclei are Lorentz contracted to less than 0.6 fm in the longitudinal direction, which is the case for collision energies $\sqrt{s_\mathrm{NN}} \ge 45$ GeV. At lower energies, the colliding nuclei are less strongly contracted. We therefore choose the formation time to be at least the transit time of the two nuclei,
\be
\tau_\mathrm{ini} =\mathrm{max}[\mathrm{0.6~fm/c},2R/\gamma],
\ee
where $\gamma$ is the Lorentz factor for a given collision energy in the center-of-mass frame. 

\bm{We then use the thermal expression for the entropy density $s = bT^3$ with $b=(4p+I)/T^4$ derived from lattice-QCD results (see Table 5 in \cite{Borsanyi:2010cj} for values of $p/T^4$ and $I/T^4$)} to be $b_c \approx 5.5$ at $T_c$ and $b_\mathrm{ini} \approx 15.5$ at $T_\mathrm{ini}$.  Since total entropy can only increase, the entropy at $\tau_\mathrm{ini}$ cannot be larger than that at chemical freeze-out. In fact, both values should be approximately equal since the QGP has a low specific viscosity, which implies that the expansion is approximately isentropic. Combining everything and assuming isentropic expansion, we obtain the initial temperature as
\be
T_\mathrm{ini}^3 \approx \frac{dS/dy}{A_\perp b_\mathrm{ini} \tau_\mathrm{ini}} \, .
\label{eq:Tini}
\ee

\begin{figure}[ht]
	\centering
	\includegraphics[width=0.95\linewidth]{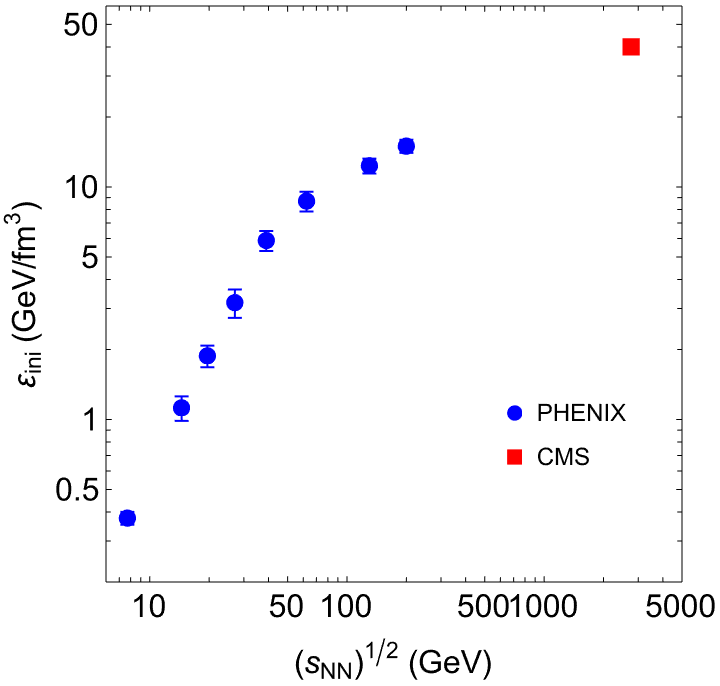}
	\caption{Average initial energy density reached in the 5\% most central Au+Au (Pb+Pb) collisions in the collision energy range $7.7~\mathrm{GeV} \le \sqrt{s_\mathrm{NN}} \le 2.76~\mathrm{TeV}$. The data are from \cite{PHENIX:2015tbb} for RHIC energies and \cite{CMS:2012krf} for the LHC energy. The steeper fall-off for \snn\ $<$ 10 GeV is caused by the incomplete Lorentz contraction of the colliding nuclei at lower energies.}
	\label{figures:eps-ini}
\end{figure}

Many heavy-ion experiments at SPS, RHIC, and LHC have reported measurements of the charged-particle multiplicity $dN_\mathrm{ch}/d\eta$. Here, we only consider data for the heaviest collision systems, Au+Au at RHIC \cite{STAR:2004moz,PHENIX:2004vdg,PHENIX:2013ehw,PHENIX:2015tbb} and Pb+Pb at SPS \cite{NA49:1994hfj,WA98:2000mvt} and LHC \cite{CMS:2012krf,ALICE:2016igk}. We use these data, together with (\ref{eq:sf-dSdy}) and entropy conservation, to convert the measured values of charged-particle multiplicity per unit pseudorapidity into estimates for the average initial energy density using the thermodynamic relation $sT=\varepsilon+p \approx (4/3)\varepsilon$:
\be
\varepsilon_\mathrm{ini} \approx (3/4)s_{\rm ini}T_{\rm ini} .
\ee
The resulting estimates of $\varepsilon_\mathrm{ini}$ covering the range $7.7~\mathrm{GeV} \le \sqrt{s_\mathrm{NN}} \le 2.76~\mathrm{TeV}$ are shown in Fig.~\ref{figures:eps-ini}.\footnote{\bm{The estimate of $T_{\rm ini}$ in \cite{CMS:2012krf} neglects the effect of the mechanical work done by the QGP during the longitudinal expansion, which results in the higher value of $T_{\rm ini}$ obtained here.}} The initial energy density for the lowest RHIC collision energy in the collider mode, $\sqrt{s_\mathrm{NN}} = 7.7~\mathrm{GeV}$, approximately coincides with the threshold for production of a QGP. It is worth mentioning that even if a QGP is formed at this energy, its lifetime must be extremely short, and most of the evolution of the fireball will occur in the hadronic phase.

\section{Strangeness}

A large increase in the production of strange antibaryons was predicted early on as a signature of quark deconfinement in baryon-rich quark matter \cite{Rafelski:1982ii}. More generally, the chemical saturation of strangeness in QGP is understood as a consequence of the presence of abundant thermal gluons \cite{Rafelski:1982pu}. As a result, hadrons containing strange quarks are expected to be produced with chemical equilibrium yields during the hadronization of a sufficiently long-lived QGP \cite{Koch:1986ud}.

The widely prevailing view today is that, once achieved during the QGP phase, the equilibrium would be maintained through hadronization. This implies that the measured hadron yields reflect {\it hadronic} equilibrium, not weakly interacting {\it partonic} equilibrium. If bulk hadronization would proceed very fast as a sudden disintegration process, vestiges of the earlier partonic equilibrium might survive in the measured hadron yield \cite{Petran:2013lja}. The general consensus today is that chemical equilibrium is maintained through hadronization. This view is amply supported by the data.\footnote{Here we are concerned with the hadronization of the bulk QGP, which is dominated by low-$p_T$ hadrons, not with the formation of higher-$p_T$ hadrons, which can be described by a sudden hadronization mechanism (see Section \ref{sec:elliptic}.} In this scenario then, the attainment of flavor equilibrium is seen as a QGP signature, even though the observed hadron yield ratios reflect the thermodynamics of the hadron gas at $T_c$.

\begin{figure}[ht]
	\centering
	\includegraphics[width=0.95\linewidth]{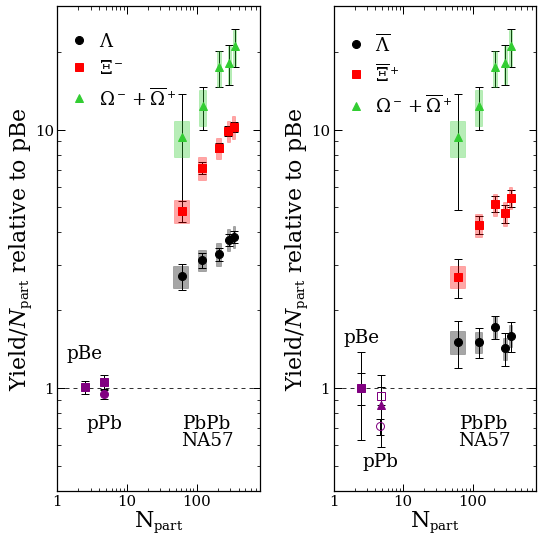}
	\caption{Multistrange baryon enhancement measured by NA57 in $\sqrt{s_\mathrm{NN}} = 17.3$ GeV Pb+Pb collisions as a function of the number of participant nucleons $N_{\rm part}$  (taken from the number of wounded nucleons $N_{w}$ in \cite{NA57:2006aux}).}
	\label{figures:NA57-multistrange}
\end{figure}

\begin{figure}[ht]
	\centering
	\includegraphics[width=0.95\linewidth]{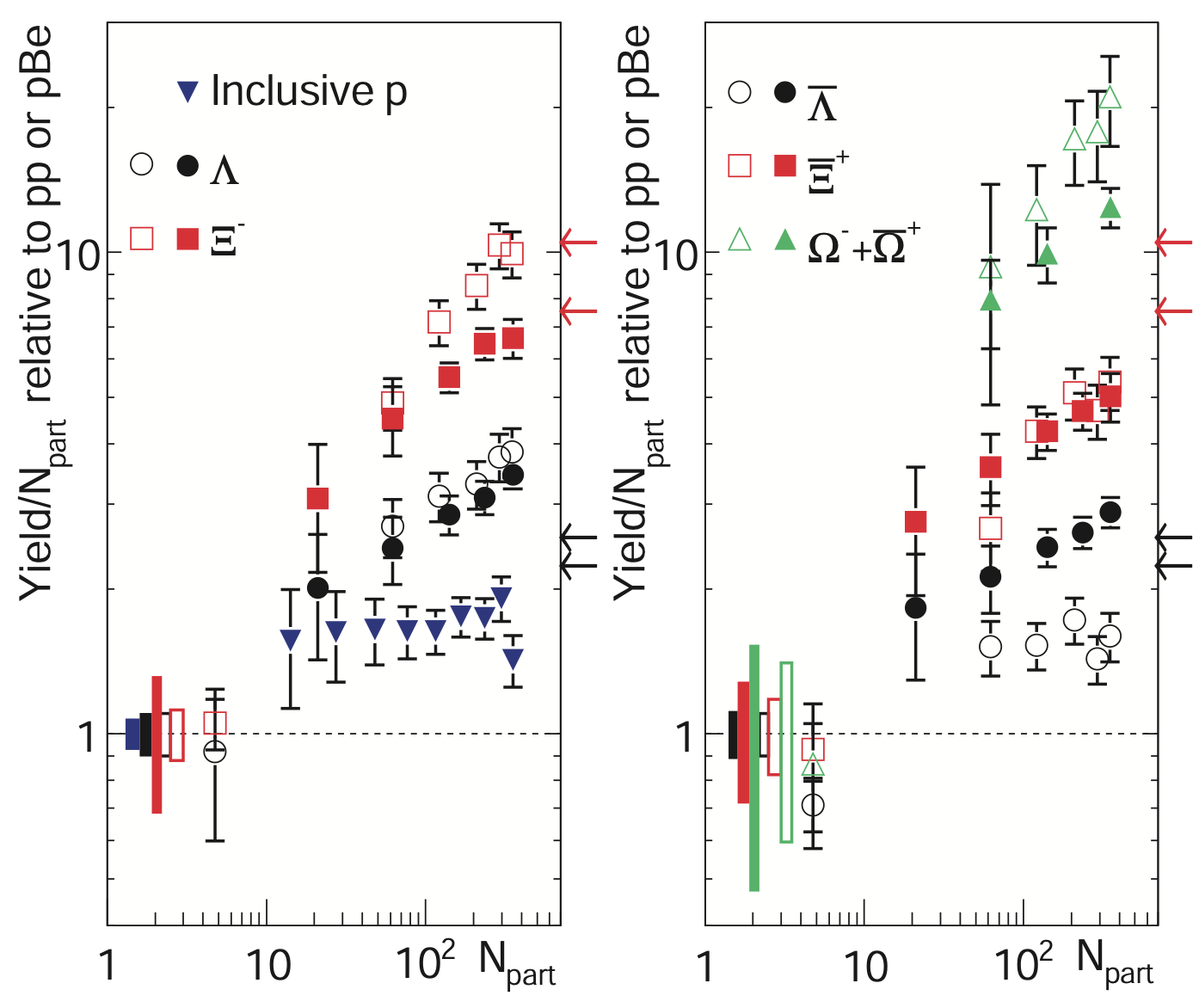}
	\caption{Multistrange baryon enhancement measured by STAR in $\sqrt{s_\mathrm{NN}} = 200$ GeV Au+Au collisions as a function of the number of participant nucleons $N_{\rm part}$ \cite{STAR:2007cqw}.}
	\label{figures:STAR-multistrange}
\end{figure}

The first confirmation of these expectations came from the WA97 experiment \cite{WA97:1999uwz}, which found a 20-fold enhancement of the production of $\Omega$ and $\overline\Omega$ hyperons in central fixed-target Pb+Pb collisions at $\sqrt{s_\mathrm{NN}} = 17.3$ GeV compared with extrapolations from p+Pb collisions. These results were subsequently confirmed by the NA57 experiment \cite{NA57:2006aux} (see Fig.~\ref{figures:NA57-multistrange}). A similar pattern was observed at the higher RHIC energy of $\sqrt{s_\mathrm{NN}} = 200$ GeV by the STAR experiment \cite{STAR:2007cqw}, as shown in Fig.~\ref{figures:STAR-multistrange}.

\begin{figure}[ht]
	\centering
	\includegraphics[width=0.95\linewidth]{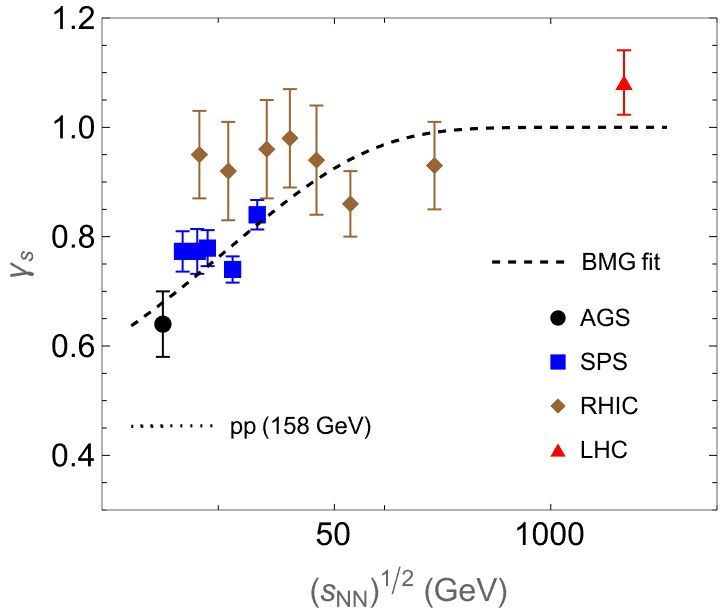}
	\caption{Evolution of the strangeness fugacity $\gamma_s$ as a function of $\sqrt{s_\mathrm{NN}}$ in central Au+Au or Pb+Pb collisions based on chemical fits using the grand canonical ensemble \cite{Becattini:2005xt,STAR:2017sal,Becattini:2014hla}. The dashed curve shows the analytic fit \req{eq:BMG} to $\gamma_s(\sqrt{s_\mathrm{NN}})$.}
	\label{figures:gammas}
\end{figure}

The overall degree of saturation of the strangeness flavor in the abundances of emitted hadrons can be assessed by a thermal fit to all particle yields with temperature $T_c$, chemical potentials $\mu_B$ and $\mu_s$ for baryon number and strangeness, and a strangeness fugacity $\gamma_s$ as adjustable parameters.\footnote{\bm{Here strangeness fugacity $\gamma_s$ is defined with respect to the macrocanonical equilibrium. It is well known that strangeness neutrality provides for a natural mechanism leading to $\gamma_s < 1$ in small thermal systems, which likely contributes to the apparent undersaturation of strangeness in p+p collisions.}} The evolution of $\gamma_s$ as a function of collision energy from AGS energies to LHC energies is shown in Fig.~\ref{figures:gammas}.  
The dashed curve in Fig.~\ref{figures:gammas} represents the analytic fit
\be
\gamma_s(A,\sqrt{s_\mathrm{NN}}) 
= 1 - \zeta\exp\left(-\xi\sqrt{A\sqrt{s_\mathrm{NN}}}\right) 
\label{eq:BMG}
\ee
provided in \cite{Becattini:2005xt}, where $A$ is the mass number of the colliding nuclei and $\zeta = 0.606$ and $\xi = 0.0209$ are fit parameters. The data presented in Fig.~\ref{figures:gammas} show an increase of $\gamma_s$ toward unity with increasing collision energy, implying full saturation of the strange quark density at hadronization in the top RHIC and LHC energy range and thus confirming the expectation depicted schematically for strangeness in Fig.~\ref{figures:Signatures1996}. \bm{The drop of $\gamma_s$ below unity for collision energies \snn $<$ 50 GeV rests on the data from SPS and AGS, but is not evident in the data from the first RHIC beam energy scan. It will be interesting to see the results of chemical analyses of data from the second beam energy scan, which covered the full range down to AGS energies.}

\begin{figure}[ht]
	\centering
	\includegraphics[width=0.95\linewidth]{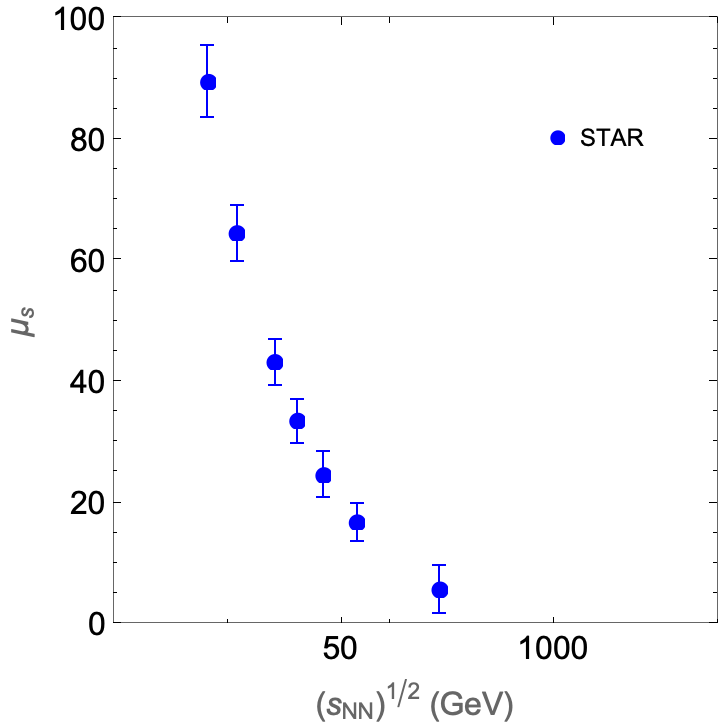}
	\caption{Evolution of the strangeness chemical potential $\mu_s$ as function of $\sqrt{s_\mathrm{NN}}$ in central Au+Au collisions based on chemical fits using the grand canonical ensemble \cite{STAR:2017sal}.}
	\label{figures:mus}
\end{figure}

As the collision energy increases and the net baryon density in the QGP falls, the chemical potential $\mu_s$ associated with strangeness drops rapidly as anticipated in Fig.~\ref{figures:Signatures1996}. The results from chemical fits to the RHIC data from central Au+Au collisions are shown in Fig.~\ref{figures:mus} and again confirm the original expectations.

One expects strangeness saturation to increase with the size and longevity of the QGP fireball. An apparent suppression for small volumes can be attributed to the conservation of net strangeness within the fireball volume, which requires that strange particles are produced in pairs, and is known as {\em canonical suppression} \cite{Cleymans:1990mn,Hamieh:2000tk}. One also expects the abundance of strange quarks to relax to the equilibrium value on a time-scale of order $1-2$ fm/c \cite{Rafelski:1982pu}. As the size, as well as the life-time, of the QGP depends in similar ways on the size $A$ of the colliding nuclei and on the collision energy $\sqrt{s_\mathrm{NN}}$, both effects contribute inextricably to the analytical formula \req{eq:BMG}.

\begin{figure}[ht]
	\centering
	\includegraphics[width=0.95\linewidth]{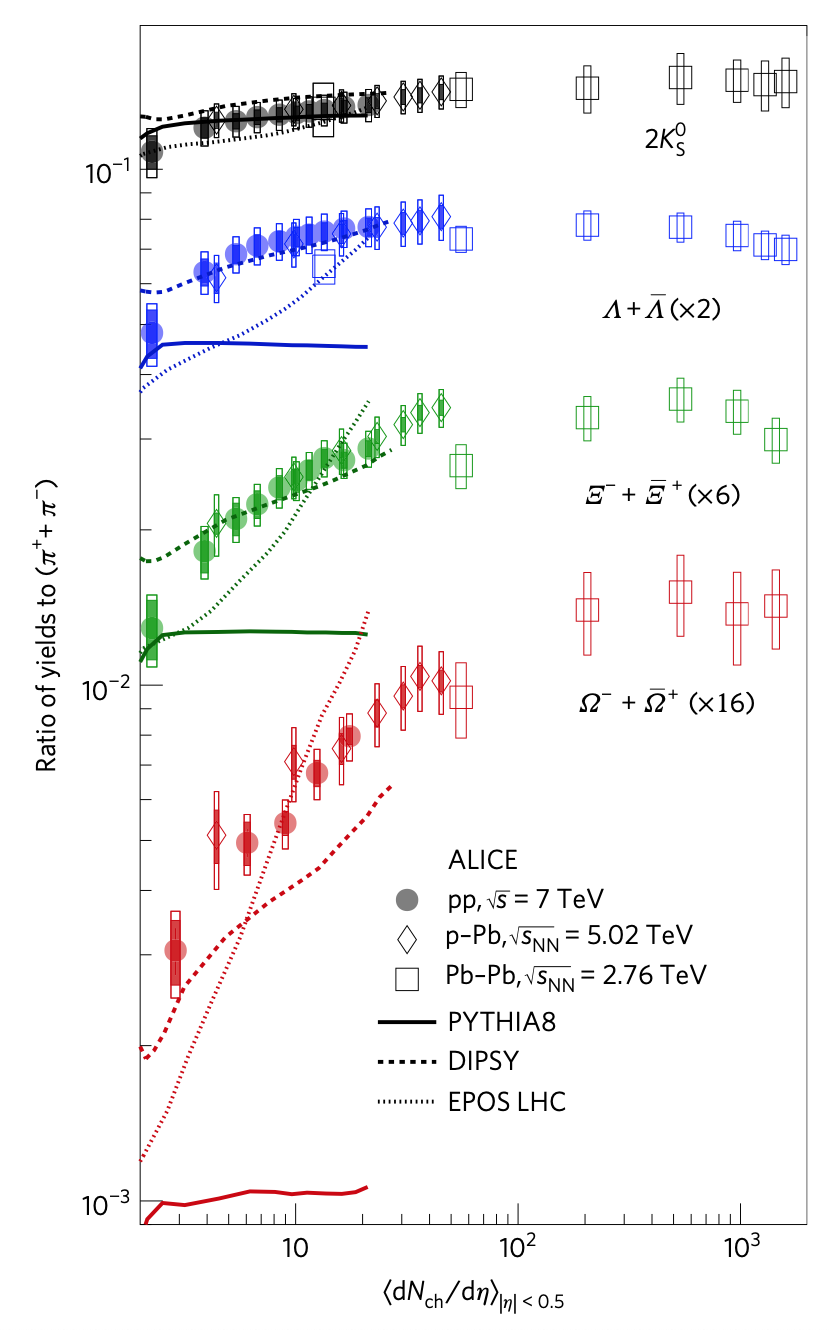}
	\caption{Multi-strange baryon enhancement measured by ALICE in p+p, p+Pb, and Pb+Pb collisions versus charged particle pseudorapidity density $dN_\mathrm{ch}/d\eta$. See \cite{ALICE:2016fzo} for details.}
	\label{figures:ALICE-multistrange}
\end{figure}

As it has become commonly accepted that hadron yields in Pb+Pb collisions at LHC energies are well described by setting $\gamma_s = 1$, the focus has more recently shifted to the system size dependence of strangeness saturation. Data on the dependence of strange baryon enhancement on system size has been reported by ALICE for p+p, p+Pb, and Pb+Pb collisions\footnote{The notation A+B, AB, and A--B for collision systems, where A, B denote the nuclei in the colliding beams, varies between experiments. Here we adopt the uniform notation A+B for consistency. We also omit the nuclear mass number in most instances, unless it is important to distinguish between different isotopic beams.} \cite{ALICE:2016fzo} and is shown in Fig.~\ref{figures:ALICE-multistrange}. The data exhibit a systematic increase with system size of multi-strange baryon yields relative to the charged pion yield, which becomes more pronounced as the number of strange valence quarks in the baryon grows. \bm{An analysis of the ALICE data in the framework of the canonical suppression model including a heuristic multiplicity dependent correlation volume and various interaction corrections has been published in \cite{Cleymans:2020fsc}. The correlation volume is not independently known and must currently be viewed as a fit parameter.}

\begin{figure}[ht]
	\centering
	\includegraphics[width=0.95\linewidth]{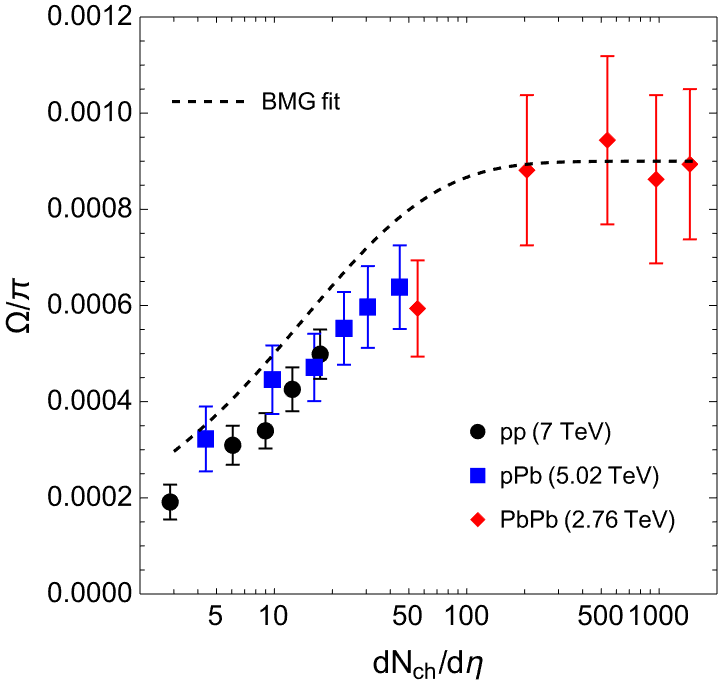}
	\caption{$\Omega$ hyperon-to-charged-pion ratio measured by ALICE in p+p, p+Pb, and Pb+Pb collisions at LHC \cite{ALICE:2016fzo} as a function of charged-particle pseudorapidity density in comparison with the analytical fit \req{eq:BMG} (see text for details).}
	\label{figures:ALICE-Omega}
\end{figure}

In order to explore whether the trend seen in Fig.~\ref{figures:ALICE-multistrange} is consistent with the results obtained at lower energies, we compare the $\Omega$ hyperon to charged pion ratio $(\Omega+\overline{\Omega})/(\pi^{+}+\pi^{-})$ measured by ALICE as a function of charged-particle pseudorapidity density $dN_\mathrm{ch}/d\eta$ with the analytical fit \req{eq:BMG}. In order to make contact with the data, we replace the nuclear mass $A$ in \req{eq:BMG} with the scaled charged-particle density ${\frac{1}{8}}dN_\mathrm{ch}/d\eta$. The scaling factor ${\frac{1}{8}}$ relates the charged-particle multiplicity $dN_\mathrm{ch}/d\eta$ to the number of participant nucleons \cite{ALICE:2010khr}. The comparison is shown in Fig.~\ref{figures:ALICE-Omega}, where the dashed curve is given by $0.0009\,\gamma_s^3$ accounting for the strangeness $|S|=3$ of the $\Omega$ hyperon. Given the vast extrapolation in energy and the heuristic substitution for $A$ in the analytical formula, the system size dependence is remarkably well represented.

The overall conclusion is that the prediction of enhanced production of baryons containing multiple strange quarks in nuclear collisions at high energy, which results in chemical equilibrium yields for large collision systems, has been consistently confirmed by the data from SPS, RHIC, and LHC.

\section{Quarkonium}

\subsection{Conceptual overview}

The suppression of \Jpsi\ ($c\bar{c}$) production due to color screening has long been recognized as a promising signature of quark deconfinement in heavy-ion collisions \cite{Matsui:1986dk}. Its excited states are predicted to dissociate more easily in a QGP and thus to be more strongly suppressed because of their lower binding energies and larger radii. The same applies for the \Upsi\ states ($b\bar{b}$) with the proviso that  $b\bar{b}$ states have  different binding energies and will therefore dissociate at different temperatures in the QGP than $c\bar{c}$ bound states. The concept of {\em sequential melting} of excited states of the \Jpsi\ and \Upsi\ states has been substantiated in lattice calculations \cite{Digal:2001ue}.

The original expectation was that the suppression would be strongest for low quarkonium momenta, where the $Q\bar{Q}$ pair is quasi-statically imbedded in the QGP and feels the full effect of color screening. At high $p_T$, the suppression of \Jpsi\ was expected to weaken and eventually disappear, because the $c\bar{c}$ bound state is then formed outside the QGP due to relativistic time delay, and the small-sized color-singlet $c\bar{c}$ precursor does not feel the effect of color screening.

Additional theoretical insights and experimental observations have led to a significant revision of this picture. It was recognized that there exist additional mechanisms  for quarkonium melting than just the static $Q\bar{Q}$ potential. Instead, the relevant quantity is the in-medium spectral function that includes non-static effects such as thermal ionization. The spectral function, which needs to be deduced from static lattice simulations by analytic continuation, has been widely studied (see \cite{Jakovac:2006sf} for charmonium and \cite{Petreczky:2021zmz} for bottomonium). These studies confirmed that the principle of sequential melting transcends the simplified color screening picture.

Furthermore, feed-down from higher-lying, less strongly bound states will influence the degree of suppression of any lower-lying states. Other effects that can affect the degree of suppression have been labeled “cold nuclear matter” (CNM) effects. These include nuclear shadowing of the initial gluon distributions, momentum broadening of the initial-state partons, and final-state absorption by spectator nucleons. These effects are also present in hadron-nucleus interactions, where they may be studied to determine their contributions to the suppression observed in heavy-ion collisions.

The idea that high-$p_T$ quarkonia should be less suppressed has also been revised on account of the insight that quarkonium production at high $p_T$ proceeds mostly through the color-octet $c\bar{c}$ channel via gluon fragmentation \cite{Braaten:1996pv}. This still implies a growing formation time at high $p_T$, but the color-octet nature of the precursor state means that it will suffer strong energy loss on its passage through the QGP. High-$p_T$ charmonium is thus expected to be similarly suppressed as open-charm mesons, contrary to the original expectations.

Finally, at very high collision energies, the number of produced $c\bar{c}$ pairs is large enough to engender substantial {\em regeneration} of charmonium states at hadronization \cite{Thews:2000rj,Braun-Munzinger:2000csl}. At sufficiently high collision energy, charmonium yields are then expected to obey the same thermal equilibrium law as other hadron yields, except that their overall yield is governed by the production cross section for $c\bar{c}$ pairs in the nuclear collision, providing further proof of decofinement. This mechanism is most effective at low $p_T$ where the density of $c\bar{c}$ pairs is largest.

\subsection{Sequential suppression}

Processes involving the production of heavy quarks are characterized by an energy scale $2m_Qc^2 \gg \Lambda_{\rm QCD}$ far above the QCD scale, and thus should na{\"i}vely scale as the number of binary nucleon-nucleon collisions. One therefore characterizes their yield in heavy-ion collisions by a {\em nuclear modification factor} \RAA\ defined as the ratio of the inclusive yields per unit rapidity in $A+A$ collisions and in proton-proton collisions scaled by the number of binary nucleon-nucleon interactions in a nuclear collision: 

\be
R_{\rm AA}(p_T) = \frac{dN_{AA}/dp_Tdy}{\langle T_{\rm AA}\rangle\cdot d^2\sigma_{pp}/dp_Tdy} ,
\label{equation:RAA}
\ee
where $\langle T_{\rm AA} \rangle$ is the longitudinally integrated nuclear density averaged over the experimentally selected events in a certain collision centrality window. A value $R_{\rm AA}<1$ implies suppression in the nuclear collision relative to the extrapolation from independent proton-proton collisions.

\Jpsi\ suppression in heavy-ion collisions was initially studied and observed at the CERN SPS in experiments NA38 \cite{NA38:1998udo}, NA50 \cite{NA50:2000brc, NA50:2004sgj}, and NA60 \cite{NA60:2007ewx}. Di-muon spectra were measured for invariant masses above 2.9 GeV/c$^2$, encompassing \Jpsi, $\psi'$, Drell-Yan pairs, and open charm decays in Pb+Pb (In+In) fixed-target collisions at \snn\, = 17.3 GeV at the CERN SPS. 

\begin{figure}[ht]
	\centering
	\includegraphics[width=0.95\linewidth]{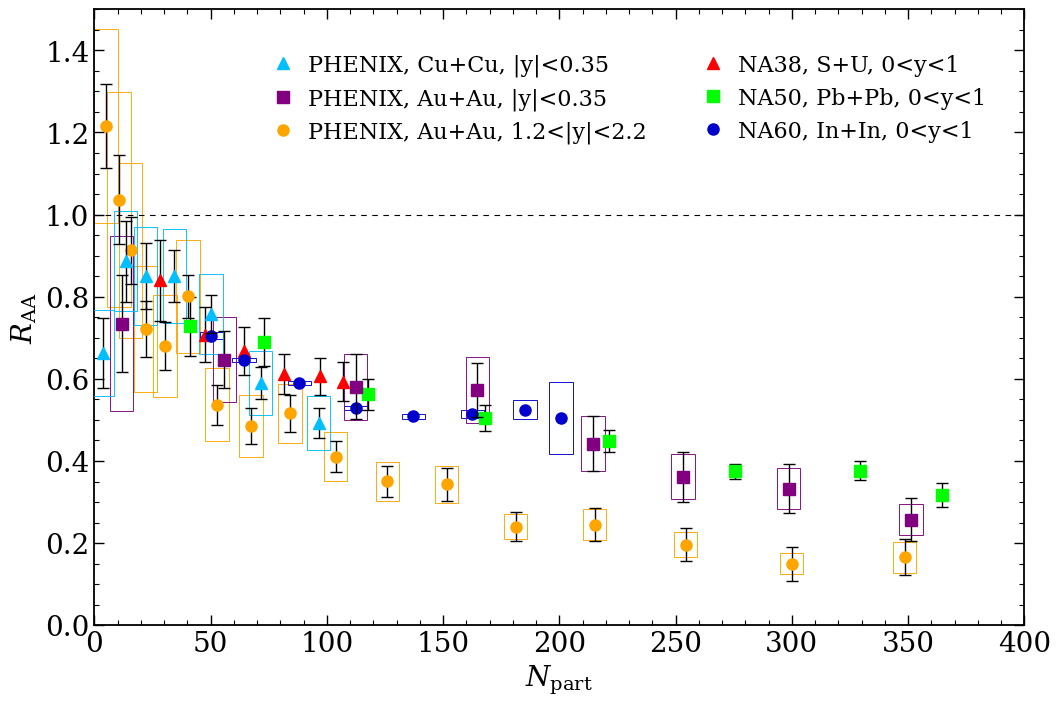}
	\caption{The nuclear modification factor \RAA(J/$\psi$) as a function of the number of participating nucleons \Npart\ at mid-rapidity from NA38, NA50, NA60 and PHENIX \cite{Kluberg:2009wc}. Collision system, rapidity window and centrality are given for each in the legend. See text for more details.}
	\label{figures:JPsi_NA50_NA60_PHENIX}
\end{figure}

The dependence of the nuclear modification factor $R_{\rm AA}$ for \Jpsi\ production as a function of centrality, expressed in terms of the number of participant nucleons \Npart, is displayed in Fig.~\ref{figures:JPsi_NA50_NA60_PHENIX} for data from NA38 \cite{NA38:1998udo}, NA50 \cite{ NA50:2004sgj}, and NA60 \cite{NA60:2007ewx} at collision energy \snn\, = 17.3 GeV. A clear pattern of suppression of the \Jpsi\ is seen above \Npart\, $\approx  100$, increasing steadily up to the most central collisions of \Npart\, $> 350$. This was the first experimental verification of melting of the \Jpsi\ in the presence of nuclear matter at high densities in heavy-ion collisions, although many questions about competing effects remained.  The cross-section ratios measured for \Npart$<100$ are in good agreement with the pattern of normal nuclear absorption extrapolated from proton-nucleus collisions \cite{NA50:2004sgj} (see next subsection).

With the advent of high-energy heavy-ion colliders, experimental studies of the quarkonium states and the degree of their melting have flourished. Figure \ref{figures:JPsi_NA50_NA60_PHENIX} also shows the centrality dependence of \RAA(\Jpsi) at the collision energy \snn\, = 200 GeV at RHIC from PHENIX \cite{PHENIX:2008jgc}, with almost identical suppression at RHIC as at the SPS. It is difficult to draw unambiguous conclusions from this observation, because the QGP conditions at the two energies are quite different (see Fig.~\ref{figures:eps-ini}). 

For Au-Au collisions at \snn = 200 GeV, \Jpsi\ is found to be more suppressed at forward rapidity than at mid-rapidity as can be seen in Fig.~\ref{figures:JPsi_NA50_NA60_PHENIX}, which shows PHENIX data for \RAA(\Jpsi) measured at midrapidity (purple squares) and measurements at forward rapidity \cite{PHENIX:2011img} (orange dots). One reason for the enhanced suppression at forward rapidity may be stronger gluon shadowing in the nucleus that moves in the backward direction. Production of a $c\bar{c}$ pair in the forward-rapidity window probes the nuclear gluon distribution in the backward-going nucleus in the range $x \sim (1.5-5)\times 10^{-3}$, where the nuclear gluon distribution is strongly suppressed.

\begin{figure}[ht]
	\centering
	\includegraphics[width=0.95\linewidth]{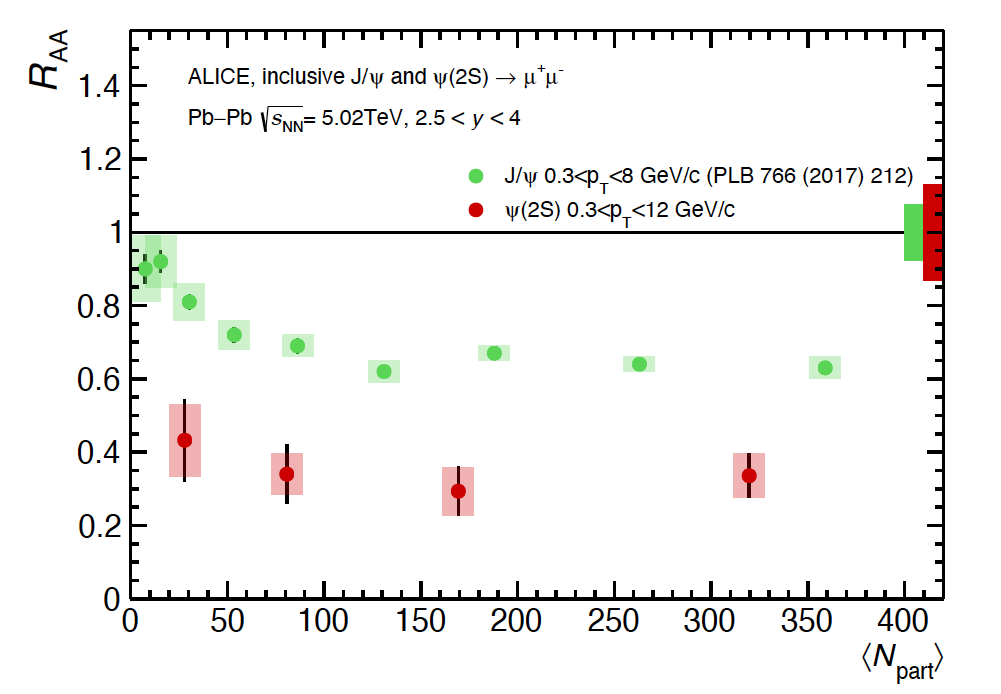}
	\caption{\Jpsi\ and $\psi$' (denoted as $\Psi$(2s) here) suppression results at forward-rapidity from ALICE in Pb+Pb collisions at \snn = 5.02 TeV \cite{ALICE:2022_Psi-prime}.}
	\label{figures:ALICE_Jpsi_Psi-prime.png}
\end{figure}

\begin{figure}[ht]
	\centering
	\includegraphics[width=0.95\linewidth]{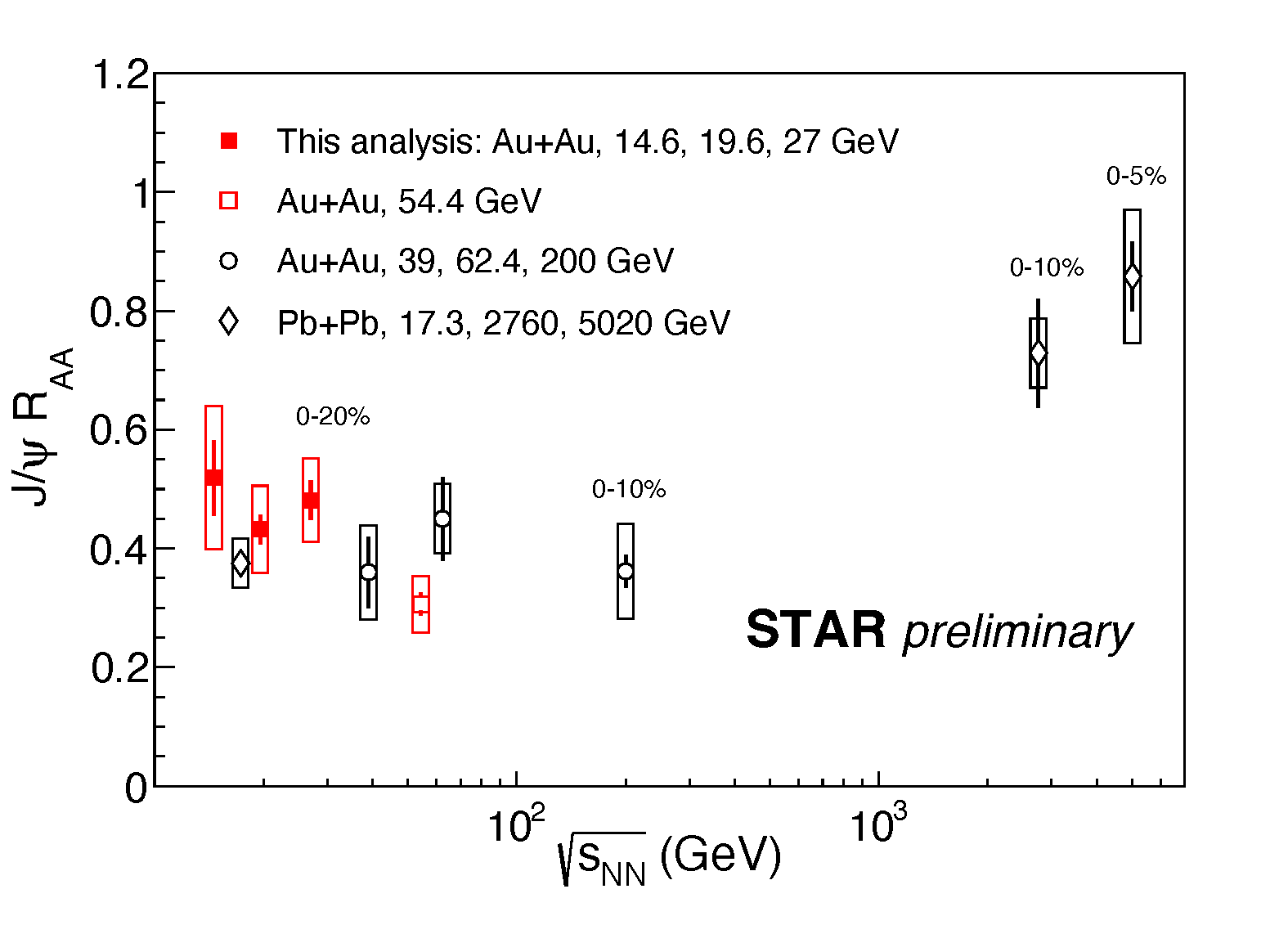}
	\caption{Compilation from STAR of \RAA(\Jpsi) results at mid-rapidity as a function of \snn ~ \cite{NA50:2004sgj,STAR:2019fge,ALICE:2019nrq}.}
	\label{figures:STAR_J-psi_vs_root_s.png}
\end{figure}

A new suppression pattern is observed in Pb+Pb collisions at the LHC. The \RAA(\Jpsi) and \RAA($\psi'$) measured at forward rapidity by ALICE at \snn = 5.02 TeV \cite{ALICE:2019lga} exhibits suppression and is rather flat for $N_{\rm part} > 100$ as seen in Fig.~\ref{figures:ALICE_Jpsi_Psi-prime.png}. The data clearly reveal a sequential suppression pattern showing stronger suppression of the excited charmonium state with $R_{\rm AA}(\psi')/R_{\rm AA}({\rm J}/\psi) \approx 0.5$. However, when comparing the \RAA(\Jpsi) to that measured at RHIC in Fig.~\ref{figures:JPsi_NA50_NA60_PHENIX}, it is clear that the suppression of the \Jpsi\ is less pronounced at the LHC energy than at RHIC. A detailed comparison of the $p_T$- and $N_{\rm part}$-dependence of \Jpsi\ production at RHIC and LHC can be found in \cite{STAR:2019fge} (STAR data and discussion of their Figs.~4 and 5). The most striking difference is observed for \Jpsi\ production at mid-rapidity integrated over all $p_T$, which is dominated by low-$p_T$ production and found to be much less suppressed at LHC \cite{ALICE:2019nrq} than at RHIC. This is seen clearly in Fig.~\ref{figures:STAR_J-psi_vs_root_s.png}, where the \RAA(\Jpsi) is suppressed (\RAA\ $\sim$ 0.4) and fairly flat at energies from the SPS through RHIC and becomes less suppressed (\RAA\ $\sim$ 0.8--0.9)
at LHC energies.

However, at higher $p_T$ (up to 40 and 50 GeV/c, respectively) ATLAS and CMS data on prompt \Jpsi\ suppression exhibit a strong increase of suppression with \Npart\ consistent with the path-length dependent energy loss of the precursor color-octet $c\bar{c}$ state \cite{ATLAS:2018hqe,CMS:2017uuv}.

The difference between the $p_T$-integrated \Jpsi\ suppression at RHIC and LHC cannot be explained by gluon shadowing as the charm production in the forward-rapidity window selected by ALICE probes the nuclear gluon distribution at $x < 10^{-4}$, where shadowing should be even stronger than at RHIC. The widely accepted explanation for this effect is that it reveals a new production mechanism for \Jpsi\ at the LHC collision energies. During hadronization of the QGP, {\em regeneration} by coalescence of $c\overline{c}$ pairs copiously produced by hard QCD processes in the initial phase of the collision increases the yield of \Jpsi\ at higher energies \cite{Braun-Munzinger:2000csl,Thews:2000rj}. It is possible that this mechanism already contributes to the observation that $R_{\rm AA}^{\rm mid} > R_{\rm AA}^{\rm forward}$ at the top RHIC energy (see Fig.~\ref{figures:JPsi_NA50_NA60_PHENIX}). 

\begin{figure}[htb]
	\centering
	\includegraphics[width=0.95\linewidth]{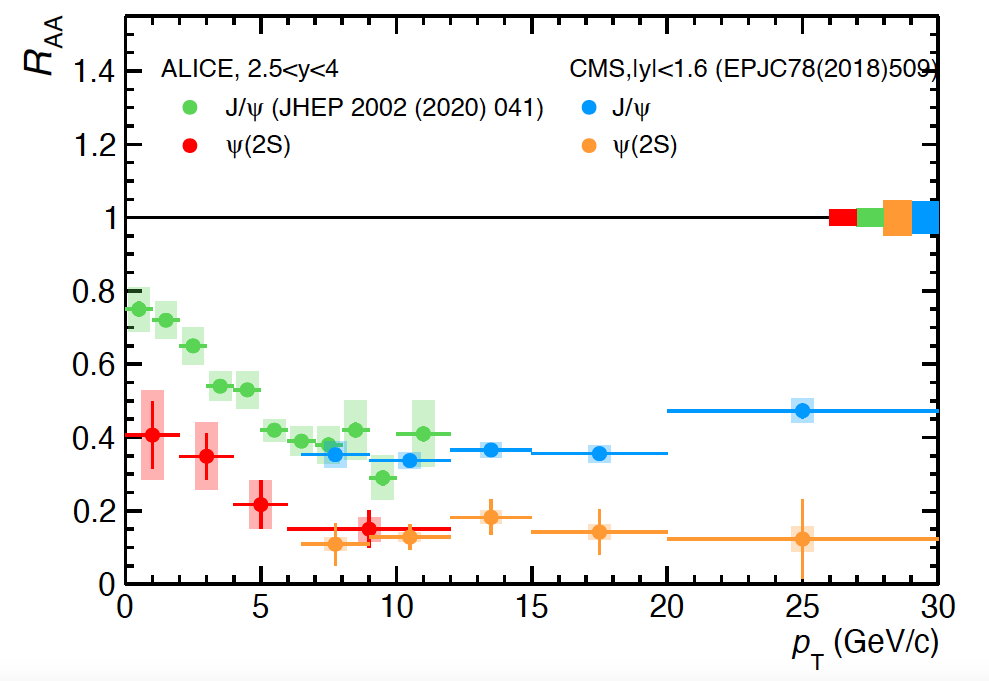}
	\caption{Inclusive \Jpsi\ and $\psi$' (denoted as $\Psi$(2s) here) suppression results at forward-rapidity from ALICE \cite{ALICE:2022_Psi-prime} and mid-rapidity from CMS \cite{CMS:2017uuv} in Pb+Pb collisions at \snn = 5.02 TeV. It is interesting to note that the measurements at forward- and mid-rapidity are consistent with each other in their range of overlap ($p_T \approx 7-11$ GeV/c).}
	\label{figures:ALICE-CMS_Jpsi_Psi-prime.png}
\end{figure}

Since most charm quark pairs are produced at low transverse momenta, regeneration should be most effective at low $p_T$ and cease to be a significant contribution at momenta above a few GeV/c. This expectation is confirmed by data for $p_T$-differential \RAA(\Jpsi) and \RAA($\psi'$) at \snn\, = 5.02 TeV from ALICE \cite{ALICE:2022_Psi-prime} and CMS \cite{CMS:2017uuv} presented in Fig.~\ref{figures:ALICE-CMS_Jpsi_Psi-prime.png}.
Both the \Jpsi\ and $\psi'$ initially exhibit a steep drop of their \RAA\ as p$_T$ increases but then level off at their lowest values for $p_T > 6$ GeV/c. The hierarchy of suppression is again evident with the $\psi'$ being suppressed by an additional factor $2-3$ relative to the \Jpsi\ over the entire p$_T$ range. 

\begin{figure}[!htb]
    \centering
	\includegraphics[width=0.95\linewidth]{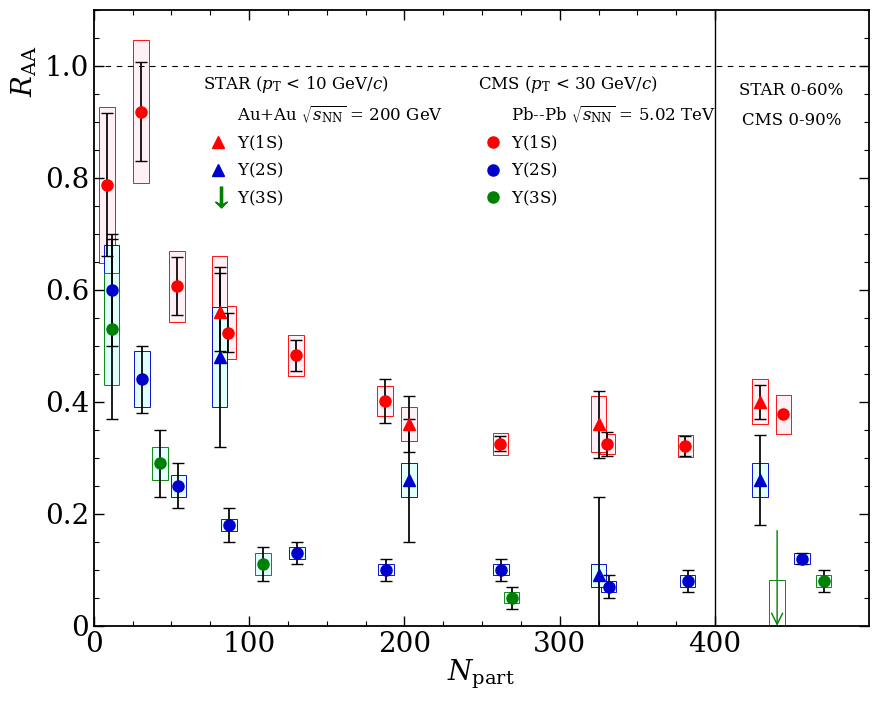}
	\caption{\RAA($\Upsilon$) results for \Upsi\ states at mid-rapidity in Pb+Pb collisions at \snn = 5.02 TeV for p$_T(\Upsilon) <$ 30 GeV/c from CMS \cite{CMS:2018zza} and in Au-Au collisions at \snn = 200 GeV for p$_T(\Upsilon) <$ 10 GeV/c from STAR \cite{STAR_Upsilon_paper}. The left segment of the figure shows the centrality differential data for \Upsi\ suppression; the integrated data are shown in the right segment. Both the CMS and the STAR data confirm the theoretical expectation of sequential suppression in the order of the binding energy and size of the bound state.}
	\label{figures:Upsilon_CMS_STAR}
\end{figure}

A clear pattern of sequential suppression of the \Upsi\ and its excited states is observed in Pb+Pb collisions at \snn\, = 5.02 TeV. The results from CMS \cite{CMS:2018zza} for Pb+Pb collisions at \snn\, = 5.02 TeV at LHC, presented as a function of \Npart\ in  Fig.~\ref{figures:Upsilon_CMS_STAR}, indicate that \RAA($\Upsilon$(2s)) at LHC is lower by a factor of 2 or more than \RAA($\Upsilon$(1s)) at mid-rapidity. As Fig.~\ref{figures:Upsilon_CMS_STAR} shows, the $\Upsilon$(1s) suppression factors at the top LHC energy (CMS data) and the top RHIC energy (STAR) data are identical within the experimental uncertainties. This is true for both the centrality differential data shown in the left segment of the figure and the integrated data shown in the right segment. For the $\Upsilon$(2s) the suppression at LHC appears to be stronger than at RHIC, although the large error bars of the STAR data do not permit a definite conclusion.

The observed pattern exhibited in Fig.~\ref{figures:Upsilon_CMS_STAR}  is consistent with a hierarchy of sequential melting of the \Upsi\ states. The $\Upsilon$(3s) is more suppressed than the $\Upsilon$(2s) which, in turn, is more suppressed than the $\Upsilon$(1s). Data from the ATLAS experiment on $\Upsilon$(1s) and $\Upsilon$(2s) \cite{ATLAS:2022xso} agree with those from CMS as a function of \Npart. Likewise, all experiments find the suppression to be independent of rapidity over the rapidity range $0<y<4$. The dependence on $p_T$ is found to be rather flat in minimum bias data with a slight rise observed by ATLAS in the range $p_T = 2-10$ GeV/c \cite{ATLAS:2022xso}. This is consistent with the expected absence of a contribution from regeneration for \Upsi\ states.

\subsection{Cold nuclear matter effects}

\begin{figure}[ht]
    \centering
	\includegraphics[width=1.\linewidth]{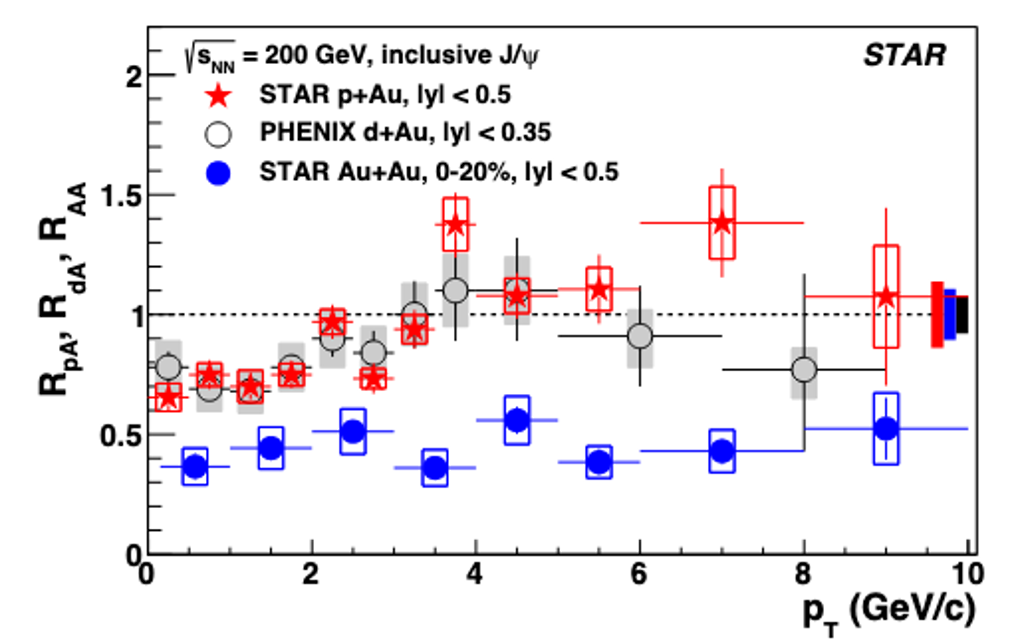}
	\caption{Inclusive J/$\psi$ data at mid-rapidity from STAR \cite{STAR:2021zvb} and PHENIX \cite{PHENIX:2012czk} at \snn = 200 GeV.
	\RpAu, \RdAu, and \RAuAu~data are shown versus p$_T$ \cite{STAR:2021zvb}.}
	\label{figures:STAR_PHENIX_CNM}
\end{figure}

One way to investigate the extent to which cold nuclear matter (CNM) effects play a role in the measured \RAA~suppression patterns of quarkonia in A+A collisions is to compare those with p+A and other light-particle induced reactions. STAR \RpAu\ and PHENIX \RdAu\ data \cite{PHENIX:2019brm} for \Jpsi, together with central \RAuAu\ data from STAR, are displayed in Fig.~\ref{figures:STAR_PHENIX_CNM} as a function of $p_T$ \cite{STAR:2021zvb}. While the Au+Au data show nearly constant suppression at $R_{\rm AuAu} \approx 0.4-0.5$ over the entire range of $p_T < 10$ GeV/c, the p+Au (d+Au) data are consistent with unity, $R_{\rm pAu} \approx R_{\rm dAu} \approx 1$,  for $p_T > 2$ GeV/c within the measurement uncertainties. However, a modest suppression with values $R_{\rm pAu} \approx R_{\rm dAu} \approx 0.6-0.8$ is observed for $p_T < 2$ GeV/c. These results leave little room for CNM effects in the range $p_T > 2$ GeV/c and help establish the strong suppression of \Jpsi\ seen in Au+Au collisions that is a final-state effect caused by \Jpsi\ melting in the QGP. The most likely reason for the \Jpsi\ suppression found in p+Au and d+Au collisions at low $p_T$ is gluon shadowing in the Au nucleus for $x \le 0.03$ \cite{Eskola:2021nhw}. Higher values of $p_T$ correspond to larger values of $x$ where gluons are not shadowed in nuclei.

\begin{figure}[ht]
	\centering
	\includegraphics[width=0.95\linewidth]{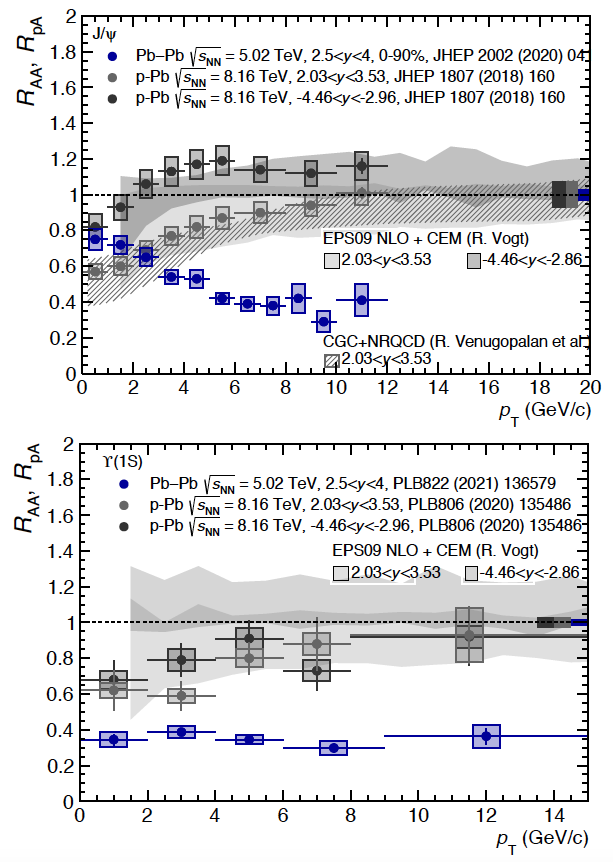}
	\caption{\RAA\ and $R_{\rm pA}$ as a function of $p_T$ for \Jpsi\ (top) and \Upsi(1s) (bottom) production in \snn\ = 5.02 TeV Pb+Pb collisions integrated over centrality (0-90\%) and in \snn\ = 8.16 TeV p+Pb collisions \cite{ALICE:2022wpn}. For details, see \cite{ALICE:2019lga} (Pb+Pb) and \cite{ALICE:2018mml} (p+Pb) for data in the top panel, and \cite{ALICE:2020wwx} (Pb+Pb) and \cite{ALICE:2019qie} (p+Pb) for data in the bottom panel.  Model calculations \cite{Albacete:2017qng} based on nuclear shadowing using EPS09-LO nuclear parton distributions are shown as the gray bands. The dashed band in the top panel represents a calculation based on the color glass condensate model and a non-relativistic QCD production mechanism for the J/$\psi$ \cite{Ma:2017rsu}.}
\label{figures:ALICE_JPsi_pPb_PbPb.png}
\end{figure}

ALICE data on \Jpsi\ and \Upsi(1s) suppression for p+Pb and Pb+Pb, shown in Fig.~\ref{figures:ALICE_JPsi_pPb_PbPb.png}, paint a similar picture regarding possible CNM effects at LHC energies. \cite{ALICE:2022wpn} The measured \RAA(\Jpsi) and \RAA(\Upsi(1s)) in Pb+Pb collisions at forward and backward rapidities exhibit strong suppression. By contrast, both the forward and backward \RpPb(\Jpsi) and \RpPb(\Upsi) are consistent with unity at high $p_T$, but exhibit systematic suppression at low p$_T$. The observed behavior is consistent with the expectation of nuclear shadowing as indicated by the grey bands in the figure. ATLAS \cite{ATLAS:2017prf} and CMS \cite{CMS:2017exb,CMS:2022wfi} have measured \RpPb\ and \RPbPb\ for \Jpsi\ and \Upsi\ in p+Pb and Pb+Pb collisions out to $p_T = 30$ GeV/c, and LHCb has measured \RpPb\ up to $p_T = 15$ GeV/c \cite{LHCb:2013gmv,LHCb:2014rku}, with all observing similar trends.

Summarizing this subsection, the light-particle induced reactions clearly exhibit the presence of suppression effects in the lower range of the p$_T$ measured, especially for the \Jpsi. As discussed, effects that could cause this suppression include nuclear shadowing of the nuclear gluon distributions. Momentum broadening of the initial-state partons in the light projectile, and final state absorption may also contribute, especially for the $\psi'$. The fact that these effects are small and generally understood reinforces the conclusion that quarkonium suppression in the Au+Au and Pb+Pb collisions at RHIC and the LHC, respectively, is a signature of sequential quarkonium melting and quark deconfinement in the QGP. The observation of charmonium regeneration in Pb+Pb collisions at the LHC further consolidates this conclusion.

\section{Temperature}

A main goal of temperature measurements as a function of the deposited energy was to determine the equation of state of QCD matter. A change in the number of effective degrees of freedom changes the entropy density $s(T)$ at a given temperature, which is closely related to the energy density $\varepsilon$ by the relation $s=(\varepsilon+P)/T$. The change of slope in the curve shown in the ``temperature'' panel of Fig.~\ref{figures:Signatures1996} around the critical energy density $\varepsilon_c$ reflects the expectation at the time that QCD matter would undergo a sharp, perhaps first-order, phase transition from hadronic matter to a quark-gluon plasma with the associated liberation of color-nonsinglet degrees of freedom carried by deconfined quarks and gluons. We will return to the equation of state in Section \ref{sec:EOS}; here we will focus on the status of temperature measurements.

There are few model-independent ways to measure the temperature in a relativistic heavy-ion collision. Thermal slopes deduced from the transverse momentum spectra of emitted particles are ``corrupted'' by the blue-shift caused by the transverse expansion of the fireball. In order to avoid this influence of collective flow, one needs to deduce the temperature from the measurement of a Lorentz invariant quantity that is independent from the frame of reference. The two measurements that satisfy this constraint are yields of particles with different masses, $dN_i/d\eta \propto e^{-m_i/T}$, and the invariant mass spectrum of lepton pairs. The former enable a frame-independent measurement of the temperature at which the hadrons are produced, commonly called the {\em chemical freeze-out temperature}, the latter provides for a measurement of the time-averaged temperature of the medium that emits the lepton pairs. 

Because the dilepton invariant mass spectrum is distorted by the decay of vector mesons, the most promising region for a temperature measurement is the intermediate mass region (IMR) of invariant masses between the $\phi$-meson and the \Jpsi :  $1.1~\mathrm{GeV}/c^2 < M_{\ell^+\ell^-} < 3~\mathrm{GeV}/c^2$. An experimental challenge is that dileptons in this mass range have a potentially large background contribution from semi-leptonic charm decays, especially at collision energies well above the charm threshold. 

The first and still most accurate measurement of the slope of the di-muon invariant mass spectrum was made by NA60 in $\sqrt{s_\mathrm{NN}} = 17.3$ GeV fixed-traget In+In collisions \cite{NA60:2008dcb}. The experiment reported an ``excess'' contribution with a spectral slope $T_\mathrm{IMR} \approx 193 \pm 16$ MeV, somewhat dependent on the chosen mass window and $p_T$-cut. The rather strong dependence of this slope parameter on the upper limit of the invariant mass window suggests contributions to lepton-pair production in the higher mass range from the very early (and therefore very hot) thermal or even pre-equilibrium stages.\footnote{R. Rapp, private communication. The argument is motivated by the observation that a fit of the form $M^{3/2}\exp(-M/T)$ to the intermediate mass region $1.2~\mathrm{GeV}/c^2 < M_{\mu\mu} < 2.5~\mathrm{GeV}/c^2$ yields $T_\mathrm{IMR}^{17.3~\mathrm{GeV}} = 246 \pm 15$ MeV, substantially larger than the apparent temperature reported in \cite{NA60:2008dcb} for a narrower mass window.}

\begin{figure}[ht]
	\centering
	\includegraphics[width=0.95\linewidth]{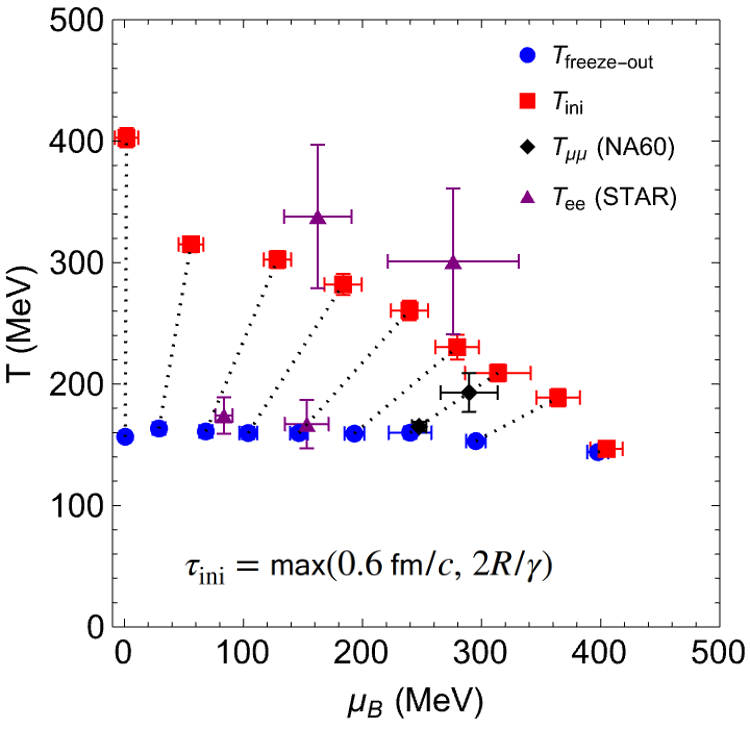}
	\caption{QCD phase diagram showing: chemical freeze-out points (blue dots), average initial temperatures and chemical potential (red squares) and effective temperatures obtained by thermal fits to the intermediate and low mass regions in dilepton invariant mass spectra. The dotted lines indicate lines of constant $T/\mu_B$, corresponding to approximately constant entropy per baryon in the QGP phase. (See text for literature references.)}
	\label{figures:phase_diagram}
\end{figure}

STAR recently reported invariant mass electron-pair spectra for Au+Au collisions at $\sqrt{s_{NN}} = 27, 54.4$ GeV \cite{STAR:2022Ye}, with thermal fits of the form $M^{3/2}\exp(-M/T)$ to the intermediate mass region (IMR) yielding $T_\mathrm{IMR}^{27~\mathrm{GeV}} = 301 \pm 60$ MeV and $T_\mathrm{IMR}^{54.4~\mathrm{GeV}} = 338 \pm 59$ MeV. These results are shown in Fig.~\ref{figures:phase_diagram} together with the values of $(T_c,\mu_{B,c})$ at chemical freeze-out (blue dots) and the initial thermalization conditions $(T_\mathrm{ini},\mu_{B,\mathrm{ini}})$, where $T_\mathrm{ini}$ is given by (\ref{eq:Tini}). Note the apparent temperatures deduced from the dilepton invariant mass spectra lie above the estimated initial temperatures at which the QGP thermalizes, again suggesting contributions from pre-equilibrium production in the measured invariant mass range. Thermal fits to the mass region around the $\rho$-meson, the low-mass region (LMR), on the other hand, yield temperatures consistent with those deduced from chemical freeze-out analyses \cite{NA60:2008dcb,STAR:2022Ye}. These are also shown in Fig.~\ref{figures:phase_diagram}.

A more model-dependent measurement of the temperature can be obtained from blast-wave fits to transverse momentum spectra of identified particles \cite{Schnedermann:1993ws}. There are many blast-wave fits of the temperature and expansion velocity at kinetic freeze-out \cite{STAR:2017sal,ALICE:2013mez,ALICE:2019hno}. Most of these show kinetic freeze-out temperatures $T_f$ that are too low to be associated with the QGP. Exceptions are \cite{Rybczynski:2012ee}, where the authors consider anisotropic momentum distributions at freeze-out, which allows them to describe the final spectra with $T_f = 165.6$ MeV, and \cite{Mazeliauskas:2019ifr}, where the authors determine the freeze-out parameters of the blast-wave fit from the fully-decayed hadron spectra and yields rather than from the spectra of primary hadrons. This method yields a common freeze-out temperature $T_{\rm fo} = (150 \pm 2)$ MeV for Pb+Pb collisions at $\sqrt{s_{\rm NN}} = 2.76$ TeV over the entire centrality range with an average transverse expansion velocity that varies with centrality.

It would be interesting to perform similar fits at lower collision energies. If the concept is correct that hadron formation occurs always at the same temperature, and the temperature reached initially is reflected in the transverse expansion velocity, the dependence of $\langle v_T/c\rangle$ on collision energy could reflect the amount of time the fireball spends in the QGP phase. The average transverse momentum $\langle p_T \rangle$ reflects both, $T_{\rm fo}$ and $\langle v_T/c\rangle$, as well as the particle mass. A direct comparison with data again requires taking resonance decays into account.

\section{Radiation from the plasma}

In principle, direct photons carry information about the temperature of the emitting QGP. In practice, the analysis is complicated by the fact that the QGP temperature changes with time during the collision and the photon spectrum is blue-shifted owing to the transverse expansion velocity of the emitting matter. Finally, there can be contributions from photons radiated by the final-stage hadron gas. Any interpretation of measured photon spectra is therefore model dependent. The PHENIX collaboration has compiled data from RHIC and LHC on the collision energy and system size dependence of the direct photon yield (see Figs.~7, 8 in  \cite{Khachatryan:2017dqo}) over a wide range. 

\begin{figure}[htb]
\centering
\includegraphics[width=0.95\linewidth]{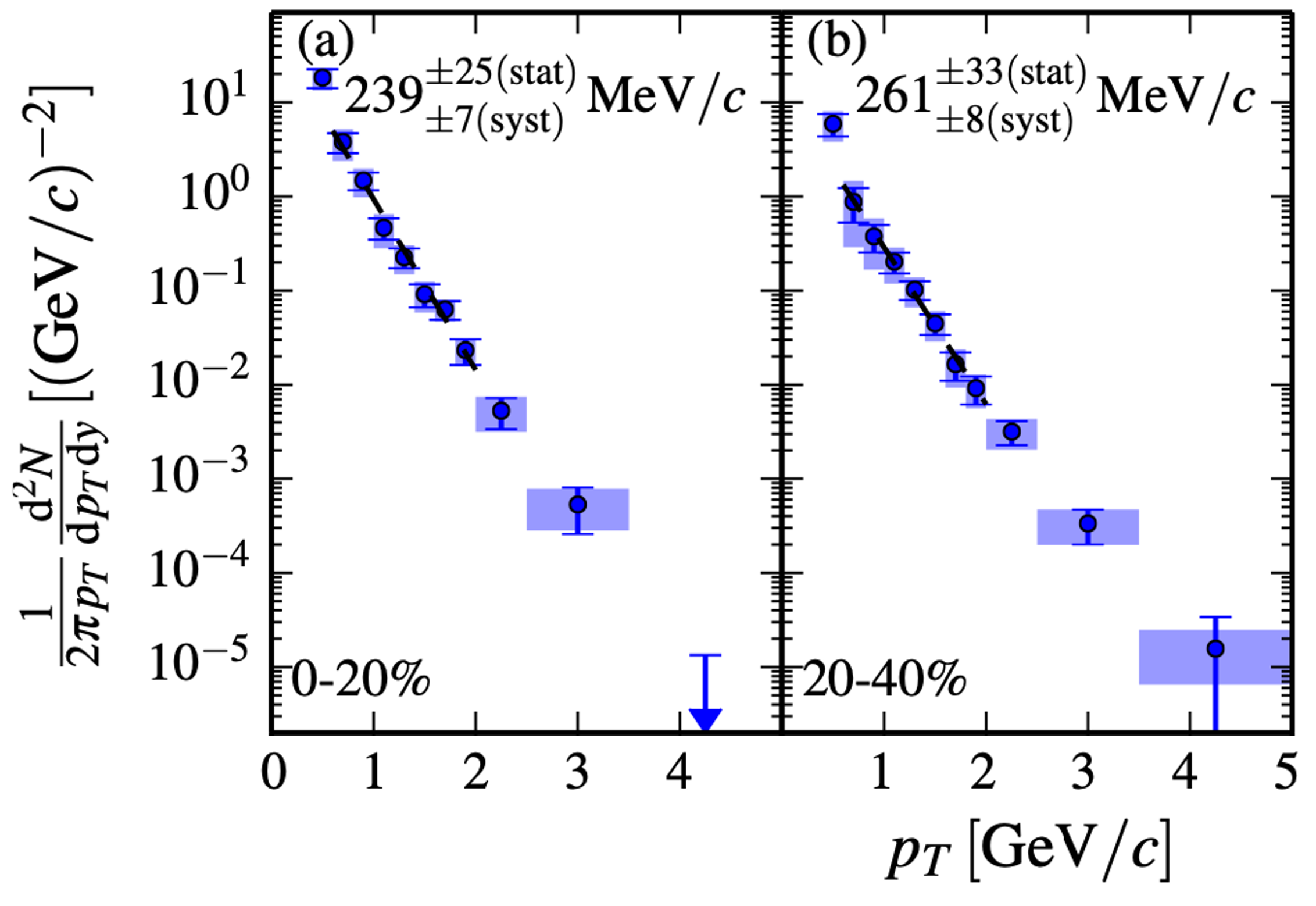}
\caption{Direct photon $p_T$-spectra from PHENIX for \snn\, = 200 GeV Au+Au collisions after subtraction of the $N_{\rm coll}$ scaled p+p contribution in centrality bins 0–20\% and 20–40\%. Dashed lines are fits to an exponential function in the range $0.6\ {\rm GeV/c} < p_T < 2.0\ {\rm GeV/c}$. [From \cite{PHENIX:2014nkk}]}
\label{figures:Photons_PHENIX}
\end{figure}

Here we present figures of low-energy direct photons for \snn\, = 200 GeV Au+Au collisions from PHENIX \cite{PHENIX:2014nkk} and for \snn\, = 2.76 TeV Pb+Pb collisions from ALICE \cite{ALICE:2015xmh}. The PHENIX data shown in Fig.~\ref{figures:Photons_PHENIX} are already background substracted and only show the spectrum of photons attributed to thermal radiation from the hot medium. The subtraction uses a power-law fit to the spectrum measured in p+p collisions, which is scaled by the average binary collision number in the selected Au+Au centrality window. As indicated in the figure, the resultant fits give $T_{\rm eff} = (239 \pm 25 \pm 7)$ for the most central 0-20\% window and $T_{\rm eff} = (261 \pm 33 \pm 8)$ MeV for the 20-40\% centrality window, and does not rely on theoretical prediction for the photon spectrum emitted in p+p collisions. 

\begin{figure}[htb]
\centering
\includegraphics[width=0.95\linewidth]{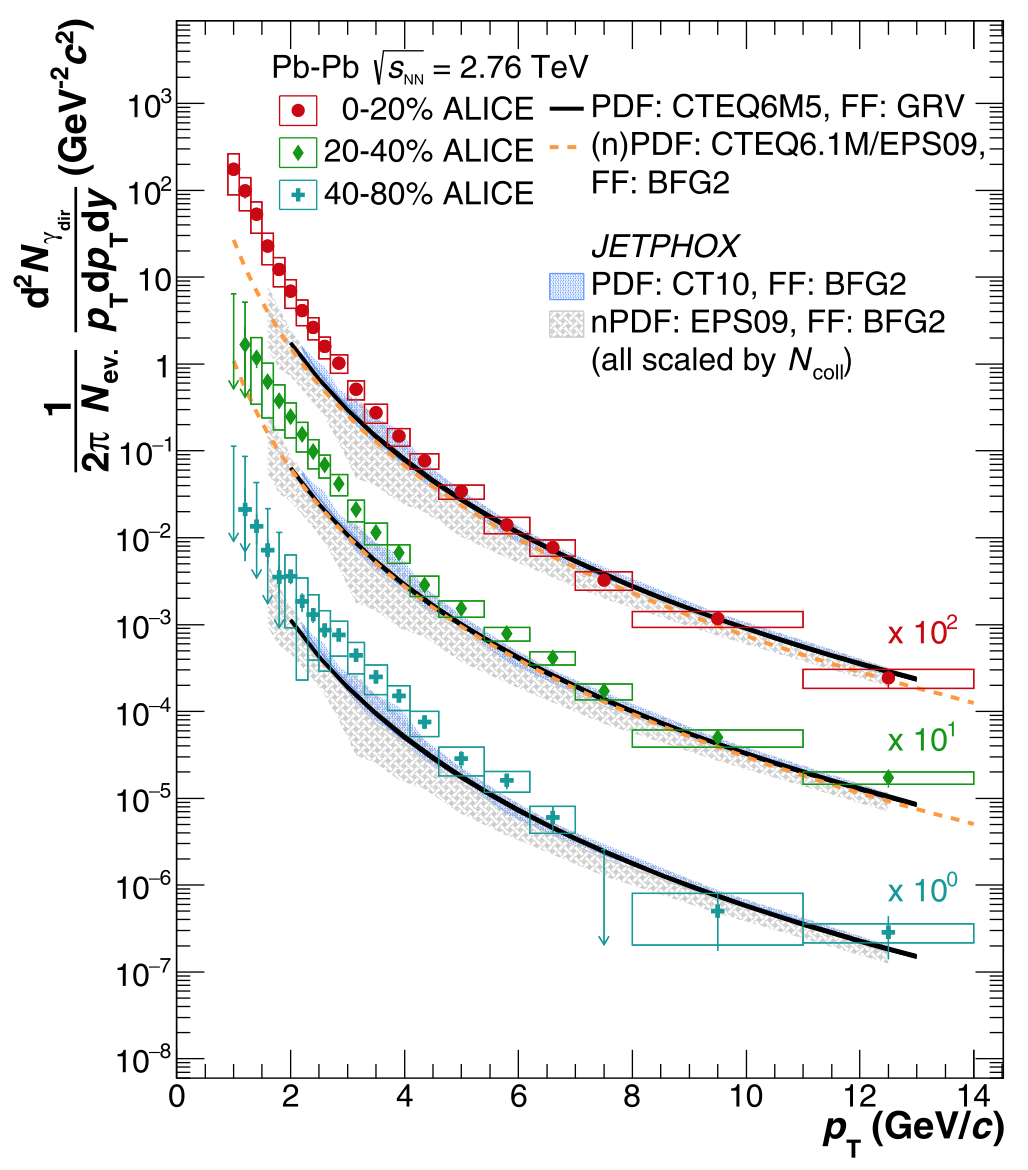}
\caption{Direct photon $p_T$-spectra in Pb+Pb collisions at \snn\, = 2.76 TeV for the 0–20\% (scaled by a factor 100), the 20–40\% (scaled by a factor 10) and 40–80\% centrality windows compared to next-to-leading order pQCD predictions for the direct photon yield in p+p collisions at the same energy, scaled by the number of binary nucleon collisions for each centrality window (from \cite{ALICE:2015xmh}). See text and \cite{ALICE:2015xmh} for details. }
\label{figures:Photons_ALICE}
\end{figure}

Figure~\ref{figures:Photons_ALICE} shows the unsubtracted ALICE data for \snn\, = 2.76 TeV Pb+Pb in three centrality windows. The figure also shows the scaled background of direct photons in p+p collisions, calculated at next-to-leading order in perturbative QCD and scaled with the average $N_{\rm coll}$ for each centrality window. Exponential fits to the low-$p_T$ spectrum for $p_T < 2.1$ GeV/c, after subtraction of the pQCD background, give thermal slopes of $T_{\rm eff} = (297 \pm 12 \pm 41)$ MeV for the 0-20\% centrality window and $T_{\rm eff} = (410 \pm 84 \pm 140)$ MeV for the 20-40\% window. It is not clear why the slope parameter is so much larger for the less central window; one reason may be that the data used in the fit start at a slightly larger value of $p_T$.

The ALICE publication \cite{ALICE:2015xmh} also contains a comparison with model calculations of the fireball evolution using boost invariant hydrodynamics and lists the initial temperatures for several of these models, which depend on the start time $\tau_{\rm ini}$ and the way the temperature is determined (at the center or averaged over the transverse profile). We here list results for the most central window (0-20\%). For the ideal hydrodynamics model of Hees, He, and Rapp \cite{vanHees:2014ida} using $\tau_{\rm ini} = 0.2$ fm/c the initial temperature at the center is $T_{\rm ini} = 682$ MeV; for the viscous hydrodynamics model of Paquet {\it et al.} \cite{Paquet:2015lta} the initial volume average temperature at $\tau_{\rm ini} = 0.4$ fm/c is $T_{\rm ini} = 385$ MeV. (A rough estimate based on boost-invariant ideal hydrodynamics scaling suggests that the two temperatures should be related by a factor $(4/3)(0.4/0.2)^{1/3} \approx 1.68$, which is close to the actual ratio $682/385 \approx 1.77$.)

Recently Paquet and Bass \cite{Paquet:2022wgu} showed in the context of an analytical model how the measured photon spectrum and yield can be related to the initial temperature $T_\mathrm{ini}$ of the QGP at the center of the fireball. The Bayesian fit most tightly constrains the combination  $\tau_\mathrm{ini}^{1/3}T_\mathrm{ini}$, which is found to have the value $450^{+100}_{-70}$ fm$^{1/3}$MeV for Pb+Pb collisions at $\sqrt{s_{\rm NN}} = 2.76$ TeV and $350^{+130}_{-60}$ fm$^{1/3}$MeV for Au+Au collisions at $\sqrt{s_{\rm NN}} = 200$ GeV.

\section{Event-by-event fluctuations}

There are three main sources of event-by-event fluctuations in relativistic heavy ion collisions: (1) Quantum mechanical density fluctuations in the colliding nuclei, (2) statistical fluctuations around thermal equilibrium, and (3) large fluctuations caused by instabilities during the dynamical evolution of the fireball. The first source is the origin of higher-order anisotropic flow, which will be discussed in Section \ref{sec:elliptic}. The second source is always present in finite-size thermal systems and creates fluctuation observables that can probe the thermodynamic properties of the fireball. The third source requires dynamical evolution of the system far off equilibrium, which can occur, e.~g., in a  system that undergoes a first-order phase transition. In this section we focus on the second and third sources of fluctuations.

If in the process of cooling through the critical temperature $T_c$ the fireball makes a sudden transition between supercooled phase without chiral symmetry breaking, implying a vanishing quark condensate, to a broken phase with a large quark condensate, extended domains with random orientation of the chiral quark condensate could be formed \cite{Rajagopal:1993ah}. The formation and decay of domains of disoriented chiral condensate (DCC) would reveal itself by non-Poissonian fluctuations of the neutral-to-charged pion ratio $N(\pi^0)/N(\pi^\pm)$ \cite{Anselm:1991pi,Anselm:1996vm}. The precondition for such a scenario is that the fireball evolves far out of equilibrium during the chiral transition.

Many searches have been carried out for signals from such DCC domains, but none of the searches have shown any sign of this effect \cite{Mohanty:2005mv,STAR:2014xli}. The most likely explanation for the absence of a DCC signal is that the expanding fireball never deviates far from thermal equilibrium, which is consistent with the smooth cross-over between the phases with broken and unbroken chiral symmetry at $T_c$ found in lattice QCD. Another possibility is that some off-equilibrium evolution occurs but that domains of disoriented chiral condensate produced in relativistic heavy-ion collisions are too small to be distinguishable from thermal fluctuations of the chiral order parameter.

Isospin fluctuations characteristic of DCC can also show up as anomalous charge fluctuations among kaons $N(K^0_s)/N(K^\pm))$ \cite{Gavin:2001uk,Nayak:2019qzd}. Measurements of cumulants of the neutral and charged kaon yields in Pb+Pb collisions at $\sqrt{s_{\rm NN}} = 2.76$ TeV by ALICE \cite{ALICE:2021fpb} revealed that $K^0_s-K^\pm$ correlations differ from charged and neutral kaon correlations. However, various kinematic aspects of the observed difference do not support the interpretation as a DCC signal.

As stated at the beginning of this section, event-by-event fluctuations can also reflect statistical fluctuations around thermal equilibrium. These fluctuations may have a chance to survive up to the final state, if they involve locally conserved quantum numbers, such as electric charge $Q$, baryon number $B$, and strangeness $S$. Thermodynamics relates these fluctuations to the corresponding susceptibilities $\chi_2^{(X)}$, where $X$ stands for the considered quantum number and the index 2 denotes the order of the fluctuation. Higher-order susceptibilities are related to higher-order event-by-event fluctuations. The thermal fluctuations of these quantities differ quite characteristically between a QGP and a hadron gas \cite{Asakawa:2000wh,Jeon:2000wg}, as do correlations, such as those between strangeness and baryon number \cite{Koch:2005vg}.

Transport theory predicts that locally conserved quantum number fluctuations adjust quickly to the changing thermodynamic conditions as the QGP cools down, but to change much more slowly after hadronization \cite{Shuryak:2000pd}. Thus, the experimentally measured event-by-event fluctuations and correlations of conserved quantum numbers are expected to reflect the conditions that are prevalent at the quark-hadron transition. This insight can be used for an independent experimental determination of the quark-hadron phase boundary \cite{Ratti:2018ksb,Ratti:2021ubw}.

\begin{figure}[htb]
\centering
\includegraphics[width=0.95\linewidth]{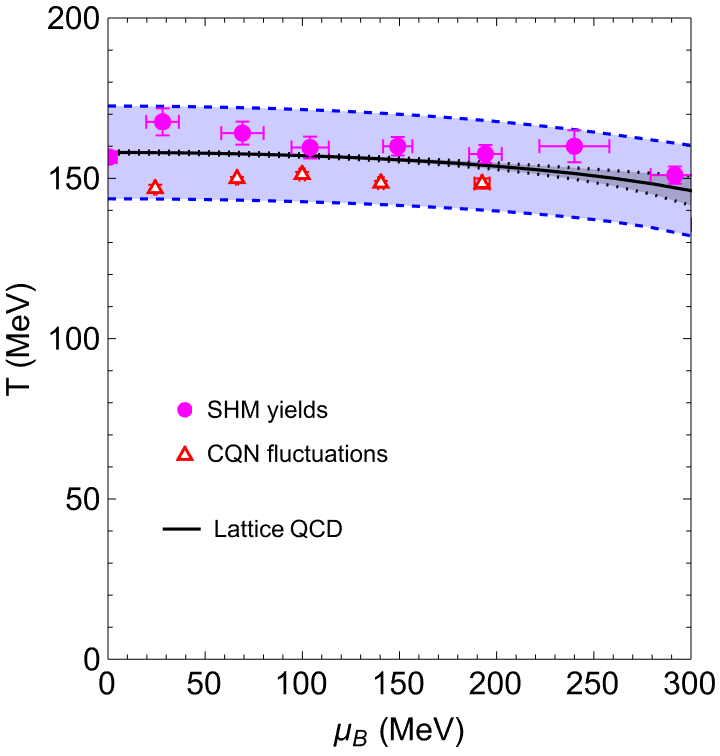}
\caption{Phase boundary between hot hadronic matter and QGP. The black line shows $T_c(\mu_B)$ calculated by lattice QCD; the blue shaded region indicates the width of the transition region \cite{Borsanyi:2020fev}. The results derived from hadron yields using the statistical hadronization model are shown as magenta dots \cite{STAR:2017sal,Andronic:2005yp,Andronic:2018qqt}, those deduced using the experimentally measured net-electric charge and net-proton number fluctuations are shown as red triangles \cite{Alba:2014eba}.}
\label{figures:PhaseBoundary}
\end{figure}

Figure \ref{figures:PhaseBoundary} compares the phase boundary between hot hadronic matter and QGP determined by lattice QCD simulations \cite{Borsanyi:2020fev} with results obtained from experimentally measured net-electric charge and net-proton number fluctuations \cite{Alba:2014eba} (red triangles) and those obtained from hadron yields using the statistical hadronization model \cite{STAR:2017sal,Andronic:2005yp,Andronic:2018qqt} (magenta dots). The black line and the grey shaded region show the pseudo-critical line $T_c(\mu_B)$; the blue shaded region represents the width of the transition derived from the width of the peak in the chiral susceptibility \cite{Borsanyi:2020fev}.

\begin{figure*}[htb]
\centering
\includegraphics[width=0.725\linewidth]{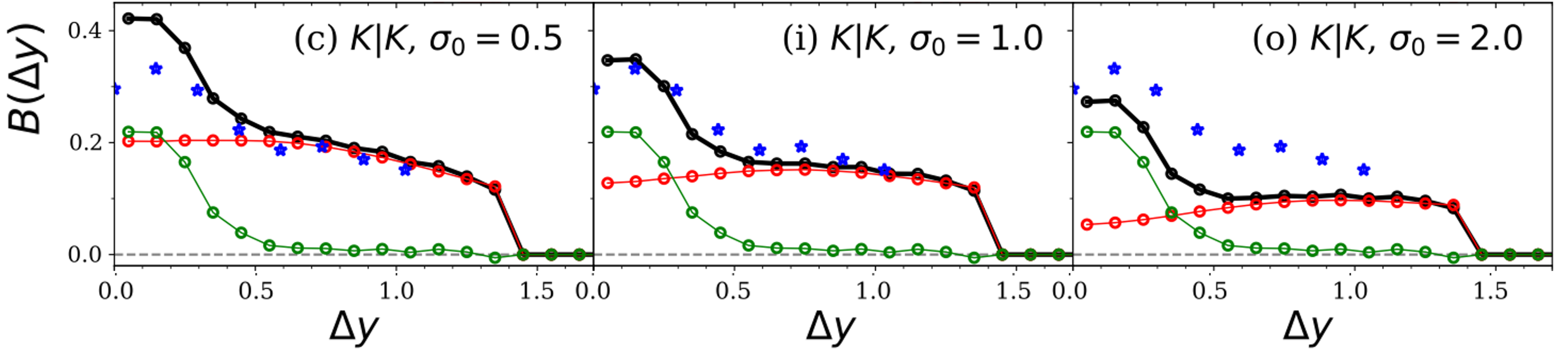}
\includegraphics[width=0.95\linewidth]{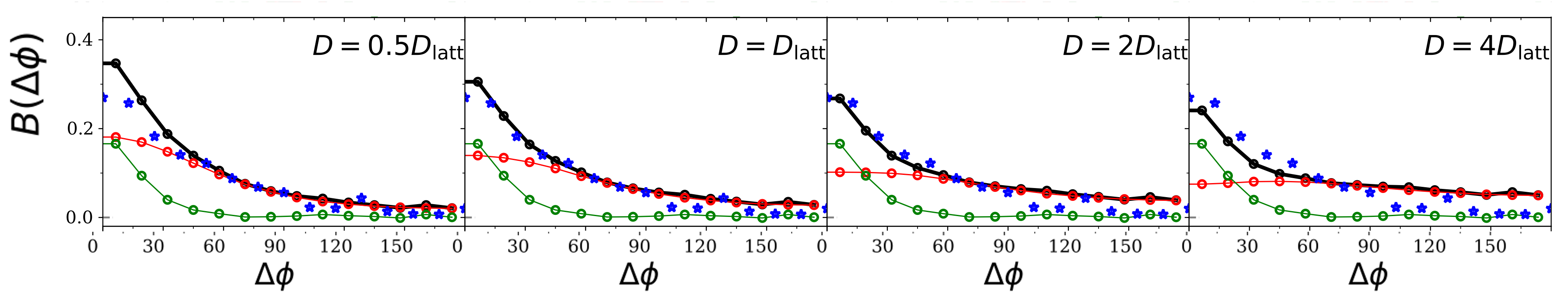}
\caption{Charged kaon balance functions $B_{K|K}$ for $0-5\%$ central Pb+Pb collisions at \snn\, = 2.76 TeV measured by ALICE \cite{ALICE:2021hjb} (blue dots) in comparison with theoretical simulations \cite{Pratt:2021xvg} (connected black dots). Upper panel: Rapidity-dependent balance function $B_{K|K}(\Delta y)$ for three values of the width $\sigma_0$ of the initial balance function at the moment of hydrodynamization (0.6 fm/c). Lower panel: Azimuth-dependent balance function $B_{K|K}(\Delta\phi)$ for four different values of the charge diffusion constant $D$ in the QGP. The red dots/lines account for diffusion in the QGP; green dots/lines account for hadron decays and rescattering; black dots/lines show the sum of both contributions.}
\label{figures:Balance}
\end{figure*}

A special case of such fluctuations are critical fluctuations in the vicinity of a critical point ($\mu_{B.c},T_c$) in the QCD phase diagram \cite{Stephanov:1999zu}. Because the critical mode has a component associated with the net baryon density, the critical fluctuations are expected to be manifested in net baryon number fluctuations, especially in the existence of a region with negative kurtosis \cite{Stephanov:2011pb}. After intriguing hints of such fourth-order fluctuations were observed in an exploratory beam energy scan at RHIC \cite{STAR:2013gus}, an extensive campaign of measurements (RHIC Beam Energy Scan II) was conducted \cite{Liu:2022wme}. We are currently awaiting a full analysis of these data. See also Section \ref{sec:EOS} for discussion of the critical point in the context of the QCD equation of state.

Another important application of event-by-event fluctuations of conserved quantum numbers are balance functions. In a closed system, such as the fireball created in a nuclear collision, any local fluctuation of a conserved quantity (``charge'') in a certain region of phase space must be compensated (``balanced'') by an equal but opposite fluctuation in the complementary part of phase space. The distribution of this compensating charge is called the balance function. The balance function is usually projected onto relative rapidity, $B(\Delta y)$, or relative azimuthal emission angle, $B(\Delta\phi)$. A wide separation of observables in (pseudo-)rapidity implies that they are established early in the collision; the separation in emission angle is sensitive to the diffusivity of the quanta carrying the observed charge, which then gets imprinted with the radial flow profile of the QGP.

Figure \ref{figures:Balance} shows the rapidity-dependence (upper panel) and angle-dependence (lower panel) of the kaon charge balance function, $B_{K|K}(\Delta y)$ and $B_{K|K}(\Delta\phi)$. $B_{K|K}(\Delta y)$ is shown for three different values of the space-time rapidity width $\sigma_0$ of the balance function at the hydrodynamization moment ($\tau_{\rm ini} = 0.6$ fm/c); $B_{K|K}(\Delta\phi)$ is shown for four different values of the charge diffusion constant $D$ \cite{Pratt:2021xvg}. The theoretical predictions are compared with data from ALICE in the 5\% most central Pb+Pb collisions at \snn\, = 2.76 TeV \cite{ALICE:2021hjb}. The conclusion is ({\it i}) that the chemical composition of the QGP is equilibrated at the time of hydrodynamization and ({\it ii}) that the charge diffusion constant $D$ agrees with values obtained on the lattice \cite{Aarts:2014nba} within a factor of two.

\section{Chiral symmetry restoration}

One of the defining characteristics of the QGP is the restoration of chiral symmetry. Lattice QCD calculations identify the crossover transition between the hadronic gas phase and the QGP phase by the location $T_c$ of the inflection point in the temperature dependence of the renormalized chiral condensate $\langle\bar{\psi}\psi\rangle_{\rm ren}$, or equivalently, by the location of the maximum of the chiral susceptibility. For $T < T_c$ the chiral condensate approaches its vacuum value; for $T > T_c$ the condensate rapidly tends to zero signalling restoration of the spontaneously broken chiral symmetry. 

A direct consequence of chiral symmetry restoration above $T_c$ is that excitation modes that differ only by parity must become degenerate. A prime example for this behavior are the vector and axial vector modes. In the vacuum, the lowest hadronic modes in these channels belong to the $\rho$-meson and the $a_1$-meson, respectively, which are separated in mass by approximately 500 MeV. It is predicted that the two modes become degenerate above $T_c$ \cite{Kapusta:1993hq}. The axial vector channel is difficult to access, but the vector channel can be probed by measuring the spectrum of emitted lepton pairs, either $e^+e^-$ or $\mu^+\mu^-$, which can be related to the photon spectral function. The restoration of chiral symmetry manifests itself in rather subtle changes in the continuum at masses above $m_\rho$ \cite{Holt:2012wr}. The $\rho$-meson peak in the spectral function, which is already collision broadened in hot or dense hadronic matter, completely disappears in the QGP phase. This is a signature of quark deconfinement and the associated disappearance of well-defined hadron states above $T_c$ \cite{Bochkarev:1985ex,Dominguez:1989bz}.

\begin{figure}[ht]
	\centering
	\includegraphics[width=0.95\linewidth]{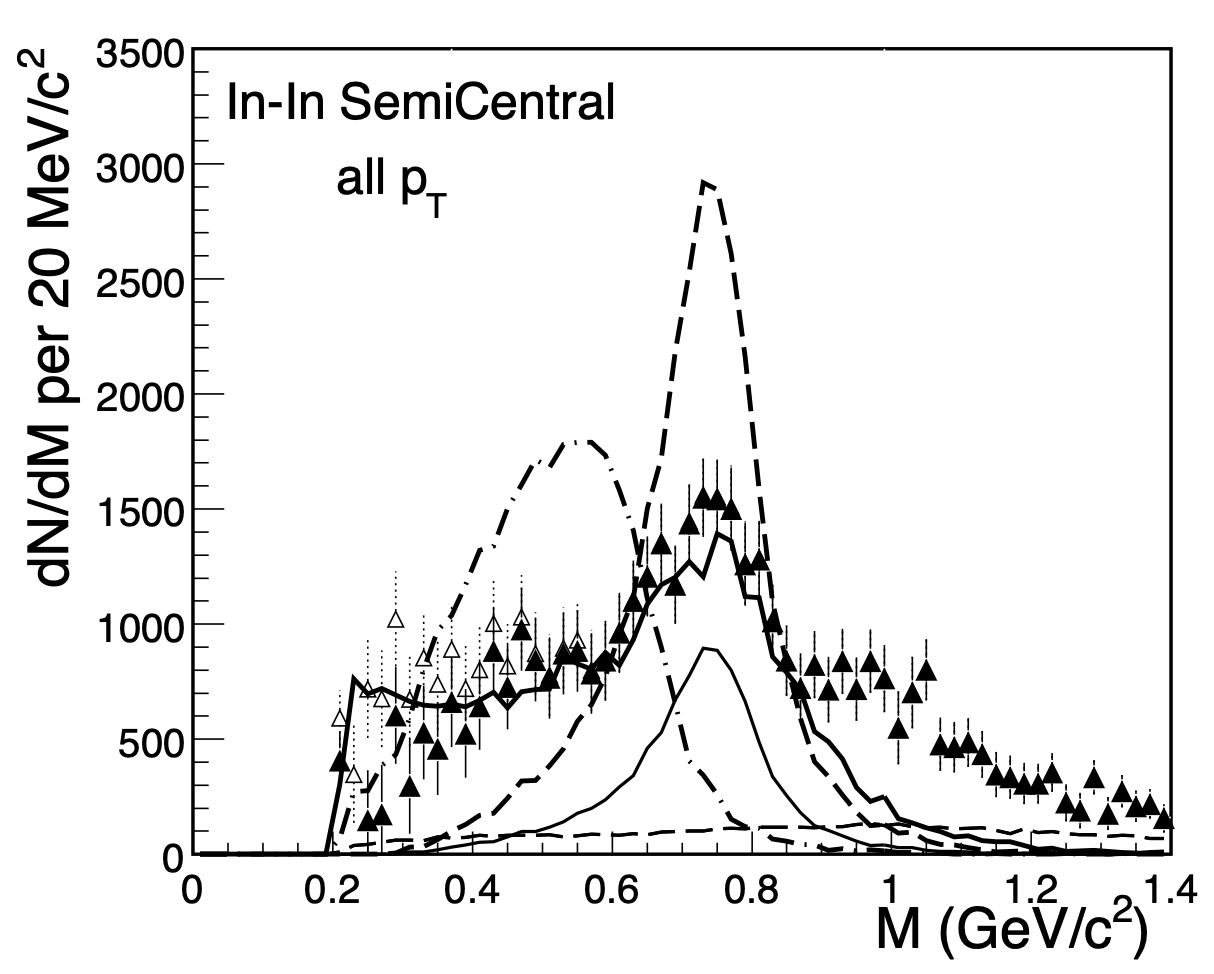}
	\caption{Excess $\mu^+\mu^-$ mass spectrum for the semicentral bin in 158 GeV/c In+In collisions in comparison with model predictions. The curves show: ``Cocktail'' $\rho$ (thin solid), unmodified (``vacuum'') $\rho$ (dashed), in-medium broadening $\rho$ (thick solid), in-medium shifted $\rho$ (dashed-dotted). The errors are purely statistical [from \cite{NA60:2006ymb}].}
 \label{figures:NA60_rho}
 \end{figure}

The most precise measurement of the lepton pair spectrum was carried out by the NA60 experiment for \snn\, 17.3 GeV In+In collisions at CERN-SPS in the $\mu^+\mu^-$ channel \cite{NA60:2006ymb,NA60:2007lzy,NA60:2008dcb,NA60:2008ctj,NA60:2008iqj}. The di-muon mass spectrum shows a much reduced peak at the $\rho$-meson mass corresponding to final-state decays of $\rho$-mesons in a dilute hadronic medium, as shown in Fig.~\ref{figures:NA60_rho}, superimposed on a broad background that is compatible with expectations from models of in-medium resonance broadening \cite{Rapp:1999ej}. There is no evidence of a mass shift that is predicted by some models of chiral symmetry restoration in dense, baryon-rich hadron matter \cite{Brown:2001nh}.

 Further analysis of the $\mu^+\mu^-$ spectrum revealed that the spectrum below $M_{\mu\mu} = 1$ GeV is azimuthally isotropic \cite{NA60:2008iqj} and its $p_T$-distribution is compatible with thermal emission from a collectively flowing hot hadronic medium \cite{NA60:2007lzy,NA60:2008ctj}. The spectrum for $M_{\mu\mu} > 1$ GeV shows a different $p_T$-dependence without indication of transverse flow, which is consistent with an origin from an early deconfined partonic phase \cite{NA60:2008dcb}.

\begin{figure}[ht]
	\centering
	\includegraphics[width=0.95\linewidth]{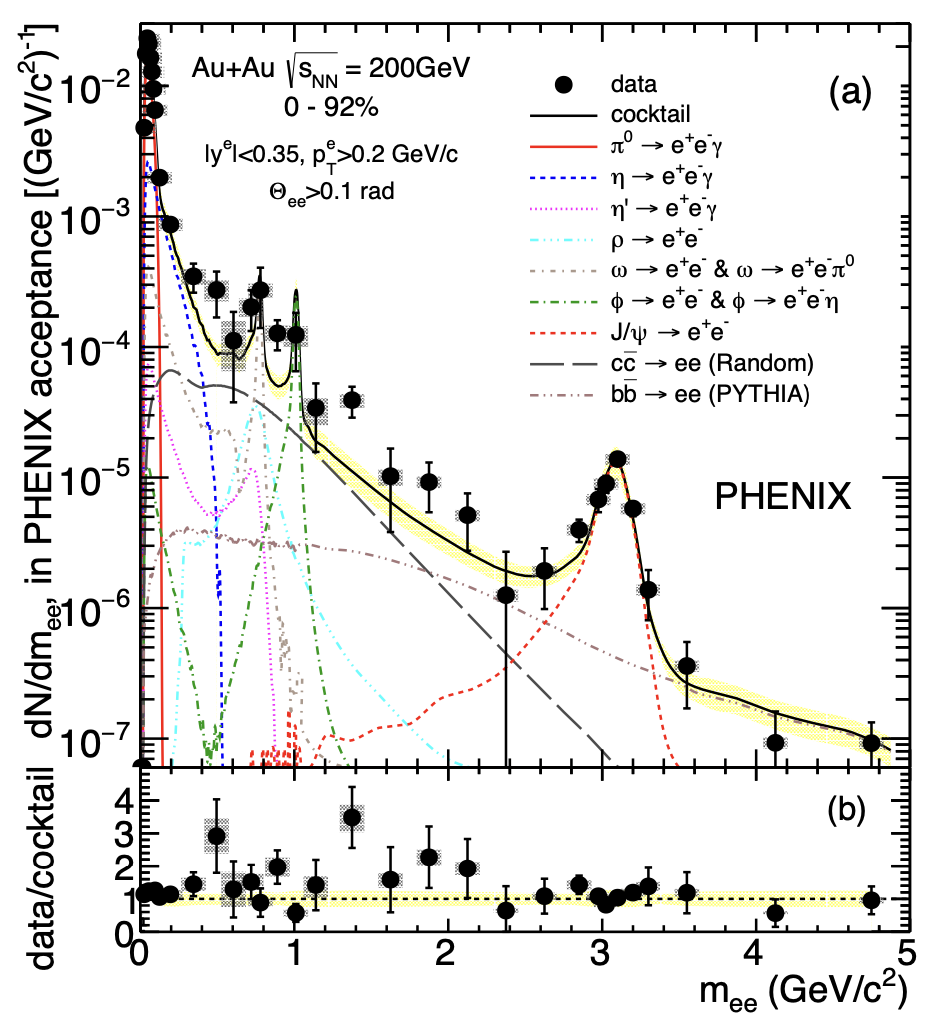}
	\caption{Dielectron mass spectrum for several centrality bins in 200 GeV/c Au+Au collisions measured by PHENIX \cite{PHENIX:2015vek}. The solid line shows the hadronic ``cocktail'' contribution; the various other curves represent specific contributing decay channels. A statistically significant excess is observed in the mass regions below and above the $\rho$ peak.}
 \label{figures:PHENIX_ee_AuAu}
 \end{figure}

\begin{figure}[ht]
	\centering
	\includegraphics[width=0.95\linewidth]{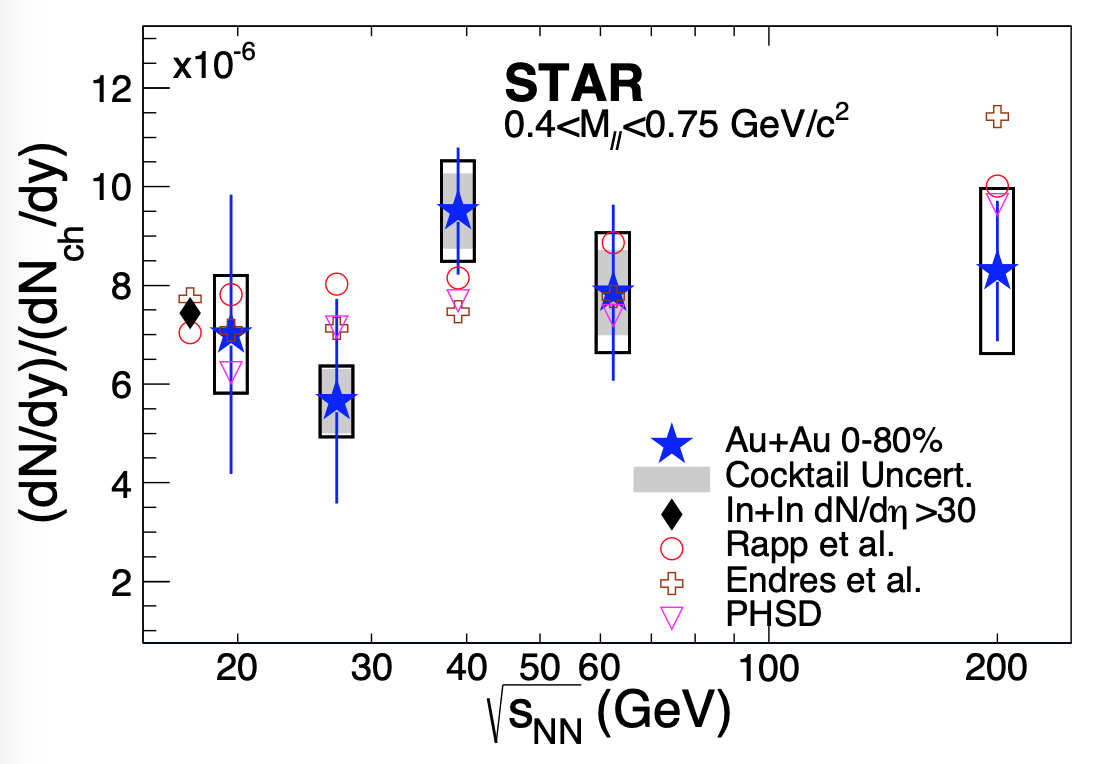}
	\caption{Dielectron excess over the hadronic ``cocktail'' contribution in $0-80\%$ central Au+Au collisions over a wide range of collision energies measured by STAR \cite{STAR:2023wta}. The blue stars show the STAR data; open symbols represent various theoretical model calculations (for details see \cite{STAR:2023wta}).}
 \label{figures:STAR_ee_AuAu}
 \end{figure}

 Low-mass electron pair production in $\sqrt{s_{\rm NN}} = 200$ GeV Au+Au collisions at RHIC energies has been measured by PHENIX \cite{PHENIX:2015vek} and STAR \cite{STAR:2015tnn}. The data exhibit similar features as those measured at SPS energies in the In+In system, albeit with lower statistical significance. The invariant mass spectrum shown in Fig.~\ref{figures:PHENIX_ee_AuAu} exhibits a broad excess over the ``cocktail'' from hadronic decays, especially in the region below the $\rho$ peak, which is compatible with predictions from models of resonance broadening in a hot hadron gas. Data from STAR shown in Fig.~\ref{figures:STAR_ee_AuAu} taken at lower collision energies are consistent with a linear scaling of the di-electron excess with the charged multiplicity \cite{STAR:2023wta}. \jh{Dielectron data from Pb+Pb collisions at LHC were reported in peripheral and semi-peripheral collisions with limited statistics \cite{ALICE:2022hvk}. Low-mass electron pair production data with high statistics in 0-10\% central Pb+Pb collisions have recently been presented, and exhibit similar effects as seen at lower energies \cite{ALICE:2023jef}}.

\section{Femtoscopy and Other Correlations}

Identical two-particle correlations are sensitive to the spatial extent and the life-time of the emitting source. This method of experimentally constraining the source geometry is called Hanbury-Brown--Twiss (HBT) interferometry, density interferometry, or femtoscopy (see \cite{Gyulassy:1979yi} for a detailed exposition of the theoretical foundations and \cite{Zajc:1992sz} for a pedagogical introduction). Experimental results for identical charged pions, kaons, and protons have been extensively published for a wide range of collision energies at AGS, SPS, RHIC, and LHC (see Fig.~20 in \cite{STAR:2020dav} and  Fig.~18 in \cite{ALICE:2022wpn}). 

Most analyses are based on a source distribution that uses a Gaussian profile with radius parameters that are aligned along the collision axis ($R_\mathrm{long}$), the combined momentum of the observed particle pairs ($R_\mathrm{out}$), and the axis perpendicular to these two directions ($R_\mathrm{side}$). The value of $R_\mathrm{out}$ is sensitive to the duration of the emission process and thus can serve as a probe of the late-stage expansion dynamics. A first-order phase transition involving the formation of a long-lived mixed phase is expected to increase the emission duration and to result in a (much) larger value of $R_\mathrm{out} > R_\mathrm{side}$. A steep drop in the compressibility of the expanding matter during hadron emission, corresponding to a drop in the sound velocity, would have a similar, albeit less pronounced effect.

\begin{figure}[ht]
	\centering
	\includegraphics[width=0.9\linewidth]{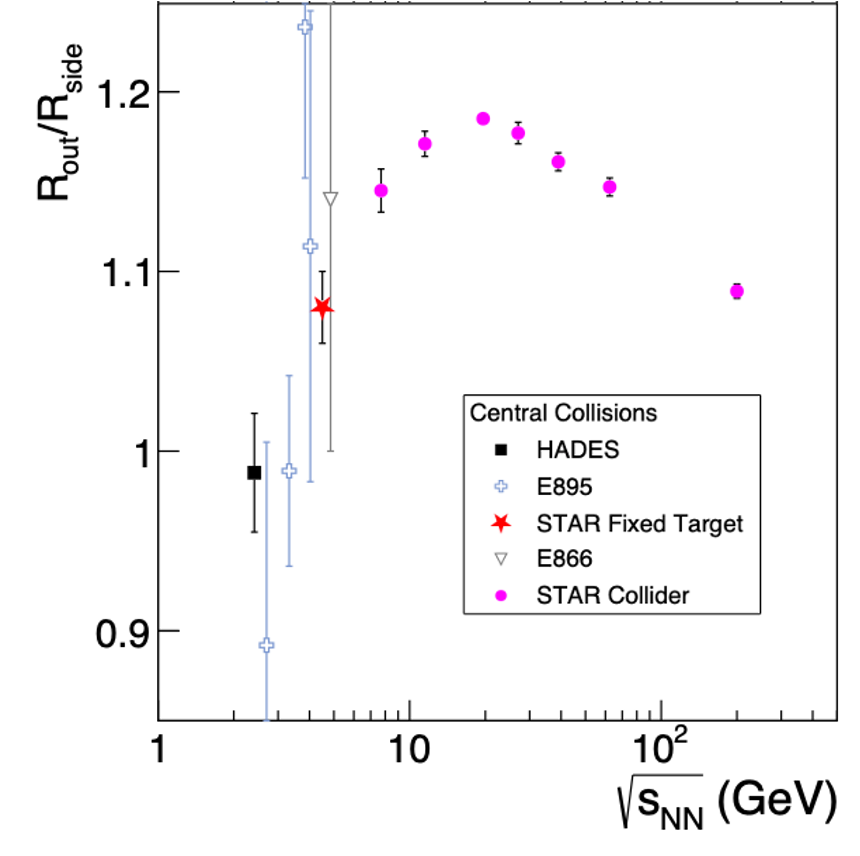}
	\caption{Data for the ratio $R_\mathrm{out}/R_\mathrm{side}$ over the energy range of the RHIC beam energy scan. The symbols refer to results from the different experiments as shown in the legend. For further details see \cite{STAR:2020dav}.}
 \label{figures:HBT_RoutRsideRatio}
 \end{figure}

The data for Au+Au collisions over the energy range of the RHIC Beam Energy Scan from STAR exhibit a rise in $R_\mathrm{out}/R_\mathrm{side}$ with increasing collisions energy up to $\sqrt{s_{\rm NN}} \approx 20$ GeV followed by a smooth fall-off for higher energies as seen in  Fig.~\ref{figures:HBT_RoutRsideRatio}. This behavior appears to be consistent with the interpretation of a minimum of the compressibility around $T_c$ during hadron emission, but a firm conclusion will require a detailed theoretical analysis, which is not yet available. 

\begin{figure}[ht]
	\centering
	\includegraphics[width=1.\linewidth]{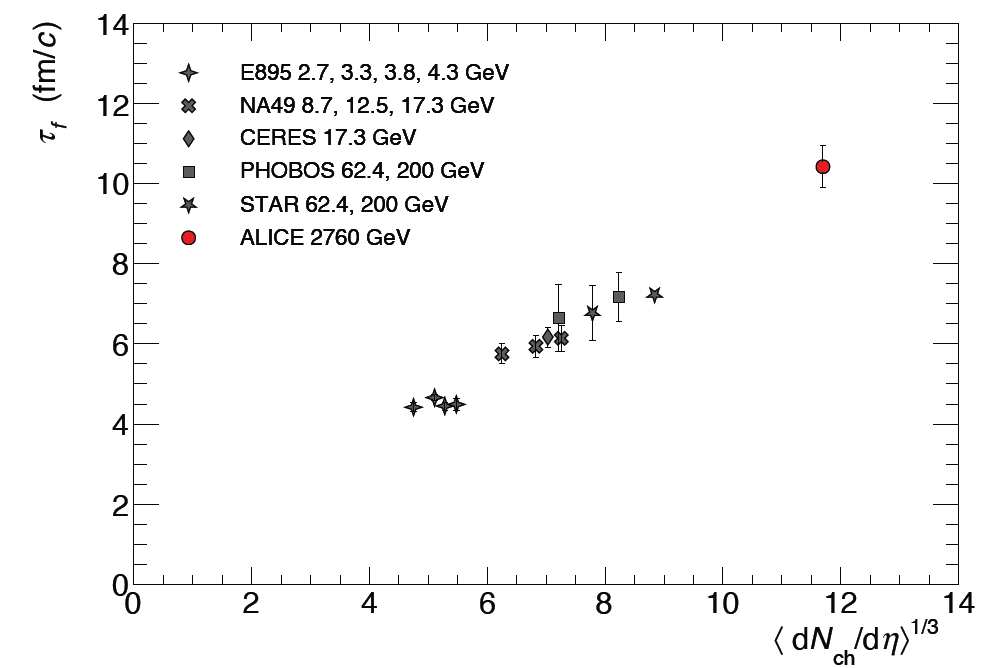}
	\caption{Life-time parameter $\tau_f$ as a function of the cube-root of the charged-particle multiplicity density. Data are from femtoscopy measurements of various experiments covering the center-of-mass energies labeled in the legend. \cite{ALICE:2022wpn}}
    \label{figures:HBT_formation_time_figure}
\end{figure}

The three radius parameters are sometimes combined to estimate the volume of a homogeneously flowing emission region at the moment of freeze-out. However, regions of the fireball that flow in different directions or are shielded from each other by opaque matter do not contribute to the HBT interference pattern. Therefore, the product $V_\mathrm{hom} = R_\mathrm{out}R_\mathrm{side}R_\mathrm{long}$, called the homogeneity volume, cannot be interpreted directly as the total volume of the fireball during the hadron emission process. The Gaussian life-time parameter $\tau_f$ measures the average duration of the stage during which hadrons freeze out from the fireball, or their emission time. The $\tau_f$ can be derived from the $R_\mathrm{long}$ and the kinetic freeze-out temperature \cite{Herrmann:1994rr}. The life-time $\tau_f$ increases smoothly with charged-particle multiplicity from around 4 to 10 fm/c, as seen in Fig.~\ref{figures:HBT_formation_time_figure}. This is also the case for the quantity $V_\mathrm{hom}$.

Momentum correlations of non-identical particles have been measured providing information about interactions among hadrons that cannot be easily measured in scattering experiments because the hadrons are unstable or beams are unavailable. For example, (p$\Lambda$) correlations have been measured in Au+Au collisions by STAR \cite{STAR:2005rpl} and (K$^-$p) correlations in collisions of p+p, p+Pb, and Pb+Pb by ALICE \cite{ALICE:2021szj,ALICE:2022yyh}. These are sensitive to the asymptotic form of the two-particle ${\overline{\textrm{K}}}{\textrm{N}}$ wave function at distances of several fm and are able to provide details of the coupling strength in various inelastic channels of exotic nuclear resonance states. When measured as a function of the source size can help understand the internal structure of these exotic states. 

Another example where heavy-ion collisions can help elucidate the structure of hadronic resonance states is the exotic $\chi$(3872) particle, which was first observed in p+p collisions \cite{LHCb:2020sey} collisions. The decay channel $\chi$(3872) $\to$ \Jpsi\,$\pi^+\pi^-$ was recently measured in inclusive Pb+Pb collisions \cite{CMS:2021znk}. The prompt $\chi$(3872)$/\psi$(2s) is observed to increase as a function of multiplicity in p+Pb and Pb+Pb, but to decrease with underlying event multiplicity in p+p reactions. This suggests very different dynamics, such as quark coalescence, for the exotic $\chi$(3872) particle at high density compared to the $\psi$(2s). Future measurements will aim to determine whether the $\chi$(3872) is a ($q\bar{q}$) molecule, a tetraquark state or some mixture of both.

The production of light anti-nuclei is enhanced in heavy-ion collisions \cite{STAR:2011eej} by the formation and rapid expansion of a QGP, as it allows anti-nuclei to escape more easily without annihilation. This is also true for production of light anti-hypernuclei \cite{STAR:2010gyg}. The relative yields of light nuclei and their antiparticles can be used to test their production mechanisms, such as statistical hadronization and final-state coalescence by comparing production yields in p+p, p+A and A+A collisions. 

Measurements of light anti-nuclei and anti-hypernuclei have potential impact in other realms of physics. Precision measurements \cite{ALICE:2015rey} of the mass differences between light nuclei and their antiparticles allow for unique tests of CPT invariance. Experimental results for light anti-nuclei are also important for better modeling of the particle composition of cosmic rays as well as the propagation of light anti-nuclei in the interstellar medium \cite{ALICE:2022veq}, which is an important ingredient of certain dark matter searches. The significantly enhanced yield of $^3_{\Lambda}$H measured at the lowest RHIC energies \cite{STAR:2021orx} favors low-energy heavy-ion collisions as a tool for the study of strange quark-doped nuclear matter, which is of relevance to the interior of neutron stars.

\section{Parton propagation}

The last diagram in Fig.~\ref{figures:Signatures1996} labeled ``parton propagation'' was a placeholder for a multitude of possible observables, comprehensively called jet quenching or jet modification, that were not well understood at the time. The simplest observable sensitive to the propagation of hard-scattered partons in the QGP is the inclusive yield of high-$p_T$ hadrons. An energy loss of partons in the QGP results in the suppression of the hadron yields. The combined energy loss of all partons in the jet shower manifests itself in the suppression of the overall jet yield. Both phenomena are usually expressed in terms of a suppression factor \RAA\ (defined in  Eq. \eqref{equation:RAA}) with respect to the yields measured in appropriately scaled p+p collisions.

The initial measurements of the charged-particle \RAA\ at RHIC \cite{BRAHMS:2004adc,PHENIX:2004vcz,PHOBOS:2004zne,STAR:2005gfr} revealed suppression in central collisions of heavy ions \cite{PHENIX:2001hpc,STAR:2002ggv}. Various approaches have since evolved to investigate the influence of the QGP on the propagation of partons through the medium, with experiments focusing on less inclusive observables that could be sensitive to the pathlength dependence of parton energy loss in the QGP.

Correlations between two back-to-back high-$p_T$ hadrons revealed the attenuation of hadrons on the opposite side (``away-side'') of a trigger hadron in the most central collisions \cite{STAR:2002svs}.  The interpretation is that the interactions of the away-side parton in the QGP degrade its momentum and thereby reduce the number of hadrons that escape on the away-side. In order to understand quantitatively the parton energy loss mechanisms in the QGP, experiments have sought to determine the pathlength dependence of partons traversing the QGP by measuring various correlations. Studies of high-p$_T$ hadron correlations \cite{STAR:2005ryu,PHENIX:2006kkn,PHENIX:2008osq,STAR:2009ngv,CMS:2011cqy} include short- and long-range correlations in azimuth and pseudo-rapidity. The results of these studies have led to tests of possible collectivity in high multiplicity events in smaller collision systems \cite{ALICE:2021nir}. Since jet measurements have become prevalent at the LHC and with upgrades at RHIC, correlations of hadrons with a trigger jet \cite{STAR:2013thw,ALICE:2019sqi}, of jets with a trigger hadron \cite{ALICE:2015mdb,STAR:2017hhs}, and between two back-to-back jets (dijets) \cite{ATLAS:2010isq,CMS:2012ulu} have been investigated. Such observables  represent semi-inclusive  measurements that are more complicated to interpret. 

Most recently, there has been a focus on jet measurements and flavor dependence of various energy-loss observables. They include investigations of the dijet asymmetry (or imbalance) \cite{ATLAS:2010isq,CMS:2011iwn,CMS:2015hkr,ATLAS:2022zbu} and acoplanarity \cite{Norman:2020grk,Anderson:2022nxb}, which are considered to be sensitive to the parton rescattering in the medium.  A larger di-jet imbalance between opposite jets of a dijet pair is observed in Pb+Pb compared to p+p collisions \cite{CMS:2015hkr}. The $p_T$ imbalance in the Pb+Pb dijets is compensated for by an enhanced multiplicity of low-$p_T$ (0.5 – 2.0 GeV/c) particles on the side of the less energetic (subleading) jet, indicating a softening of the radiation responsible for the imbalance in $p_T$. The dijet imbalance in Pb+Pb compared to p+p is greater for more central Pb+Pb collisions. Furthermore, the subleading jets are found to be more suppressed than leading jets, reaching up to 20\% stronger suppression in central collisions \cite{ATLAS:2022zbu}. These measurements can be used to constrain models of the path-length dependence of jet energy loss and its fluctuations.

The results of these investigations thus far have not yielded definite conclusions nor straight-forward interpretations regarding QGP medium properties beyond the jet quenching parameter $\hat{q}$. However, there appears to be some consistency developing between the longtime prediction \cite{Appel:1985dq,Blaizot:1986ma} of a broadening of the acoplanarity distribution and what has recently been observed in hadron-recoil jet measurements at the LHC \cite{Norman:2020grk} and RHIC \cite{Anderson:2022nxb}. The acoplanarity measurements exhibit a broadening of the recoil jet distribution in Pb+Pb relative to p+p collisions at low recoil jet $p_T$ indicating enhanced jet-medium interactions of low-$p_T$ jets opposite the trigger, presumably due to its longer path through the QGP.

To study the pathlength dependence of the interactions of partons traversing the QGP in detail \cite{Shibata:2022fyb,Beattie:2022ojg} event shape engineering has been implemented \cite{Beattie:2023mcz} in order to have better control of the initial geometrical event shapes for more precise path-length determination. The overall goal of the various jet asymmetry measurements is to provide additional insight into the pathlength dependence of jet modification and provide more rigorous tests of the energy-loss mechanisms in the QGP. Although several intriguing observations have been made, more theoretical work and incisive experimental results are needed to reach this goal.

More detailed information about the dynamics of parton propagation in the QGP can be gleaned from studies of the modification of the substructure of jets. The two simplest observables in this domain are fragmentation functions and jet shapes, which characterize the longitudinal and transverse momentum structure of jets, respectively. The interactions of showering partons with the QGP modify the gluon radiation pattern that imprints itself on the parton shower, which makes the momentum space structure of the shower a promising probe of the elementary nature of the parton interactions with the QGP. Increasing experimental capabilities combined with improved jet shower simulations are pushing the forefront of jet quenching studies in the direction of more exclusive studies of jet substructure modifications, on the one hand, and the search for globally defined observables that allow for rigorous QCD-based calculations.

In the following we discuss some of these findings in detail, focusing on high-$p_T$ inclusive hadron and jet suppression and modifications of the internal structure of jets by the QGP.

\subsection{High-Momentum Hadron Suppression}

\subsubsection{Light Hadrons}

Jet quenching in relativistic heavy-ion collisions \cite{Gyulassy:1990ye,Wang:1992qdg} (see \cite{Baier:2000mf} for a review of the basic theory) probes the mechanisms for secondary scattering and energy loss of fast partons, i.~e.\ quarks or gluons, in the medium created during the collision. The observable that most directly connects jet quenching to parton energy loss is the suppression of the yield of inclusive high-$p_T$ hadrons \cite{Renk:2014lza}, expressed as the ratio $R_{\rm AA}(p_T)$ of the inclusive single-hadron yield in A+A collisions and the single-hadron yield in proton-proton collisions, scaled by the number of binary nucleon-nucleon collisions $N_{\rm coll}$, defined in (\ref{equation:RAA}).

Suppression of the charged-hadron spectra was initially observed in measurements of \RAA\ at RHIC \cite{PHENIX:2001hpc, STAR:2002ggv,BRAHMS:2003sns,PHOBOS:2004juu}. Since then, a wealth of data has been accumulated on the \RAA\ of inclusive charged hadrons from LHC \cite{ALICE:2018vuu,ALICE:2010yje,CMS:2012aa,ALICE:2012aqc,ATLAS:2015qmb} and RHIC, as well as the \RAA\ of identified hadrons (discussed below). Inclusive charged hadron data at lower collision energies were taken in the RHIC beam energy scan \cite{PHENIX:2012oed,Horvat:2013lza}. For some collision energies a p+p reference was not available; in those cases a binary collision-scaled hadron spectrum measured in peripheral A+A collisions was used. The resulting ratio $R_\mathrm{CP}(p_T)$ can serve as a proxy for \RAA. 

The general shape of the curve $R_{\rm AA}(p_T)$ can be divided into a low-$p_T$ region, roughly $p_T \lesssim 5$ GeV/c, and a high-$p_T$ region with $p_T \gtrsim 5$ GeV/c, each encompassing different dominant dynamical processes. At low $p_T$ there is a complex interplay between collective flow and quark recombination, while at high $p_T$ the hadron spectrum reflects the fragmentation spectrum of the hard-scattered partons, modified by their energy loss caused by passage through the QGP. 

The sketch in Fig.~\ref{figures:Signatures1996} entitled ``parton propagation'' was based on the expectation that the amount of energy loss in a QGP would be quite different (either much larger or much smaller) than that in a hadron gas. It could be larger because the number of active scattering centers (gluons) is much larger in a QGP; but it could also be smaller because the strong confining force is screened in the plasma. In the absence of a theoretical framework it was not possible to make a definite prediction.

\begin{figure}[ht]
	\centering
	\includegraphics[width=0.95\linewidth]{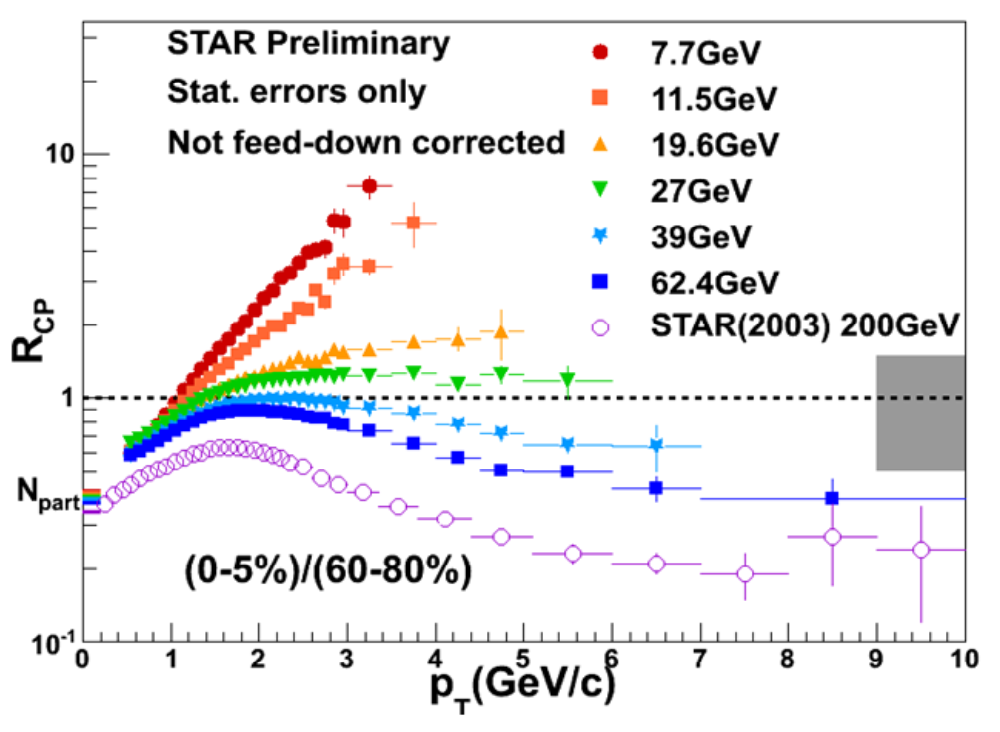}
	\caption{\RCP\ for inclusive charged hadrons measured by STAR in Au+Au collisions  \cite{Horvat:2013lza} over a wide range of collision energies as indicated in the legend.}
	\label{figures:STAR_BES_RCP}
\end{figure}

The most direct way of studying this question experimentally is to explore the dependence of \RAA\ (or $R_{\rm CP}$) on the collision energy and centrality.  The STAR data for \RCP\ of charged hadrons shown in Fig.~\ref{figures:STAR_BES_RCP} cover the energy range $\sqrt{s_\mathrm{NN}} = 7.7-200$ GeV. They exhibit suppression at large $p_T$ for collision energies greater than 27 GeV, the lowest collision energy for which $R_{\rm CP}(p_T)$ data for $p_T \gtrsim 5$ GeV/c exist. 
For lower collision energies an  enhancement ($R_\mathrm{CP} > 1$) is observed in the few GeV/c momentum range, which grows as the collision energy is lowered. This enhancement has been attributed to contributions from several mechanisms. These include the Cronin Effect  \cite{Cronin:1974zm,Kharzeev:2003wz}, the cumulative effect in nuclear parton distributions that extend into the region $x>1$ \cite{Braun:1994bf}, and collective transverse flow augmented by parton recombination \cite{Fries:2003kq}. All these effects have in common that multiple nucleon-nucleon collisions contribute to the transverse energy of the produced hadrons. Comparison with p+A data will be needed to sort out the relative importance of these mechanisms.

In nuclear collisions, recombination is enhanced at larger $p_T$ by the collective flow that blue-shifts the thermal parton spectrum. Fragmentation is depleted in the presence of a dense medium by the energy loss of the primary parton. The fragmentation mechanism generally dominates at sufficiently high $p_T$, because the primary parton spectrum from hard QCD scatterings has a power law tail, while the thermal parton spectrum falls off exponentially. The recombination contribution only weakly depends on \snn\, while the fragmentation contribution falls off steeply as \snn\ decreases. Thus, the relative magnitude of the two contributions depends on the collision energy. This means that the threshold value of $p_T$ beyond which jet quenching is visible shifts rapidly to higher $p_T$ as the collision energy is reduced and eventually becomes unobservable because sufficiently hard parton scatterings become rare.

The \RAA\ of identified protons and pions has been measured at midrapidity in d+Au collisions at \snn\ = 200 GeV and exhibits an enhancement
for 2 $< p_T <$ 7 GeV/c in central collisions \cite{STAR:2006xud}. NLO pQCD calculations are able to describe the data for pions at higher $p_T$ in both p+p and d+Au collisions indicating an emergence of effects outside pQCD at these lower $p_T$. Furthermore, the larger enhancement of protons than pions observed at low $p_T$ in the d+Au data reinforces the role of recombination and collective flow in the enhancement and possibly additional cold nuclear matter effects.

\begin{figure}[ht]
	\centering
	\includegraphics[width=0.95\linewidth]{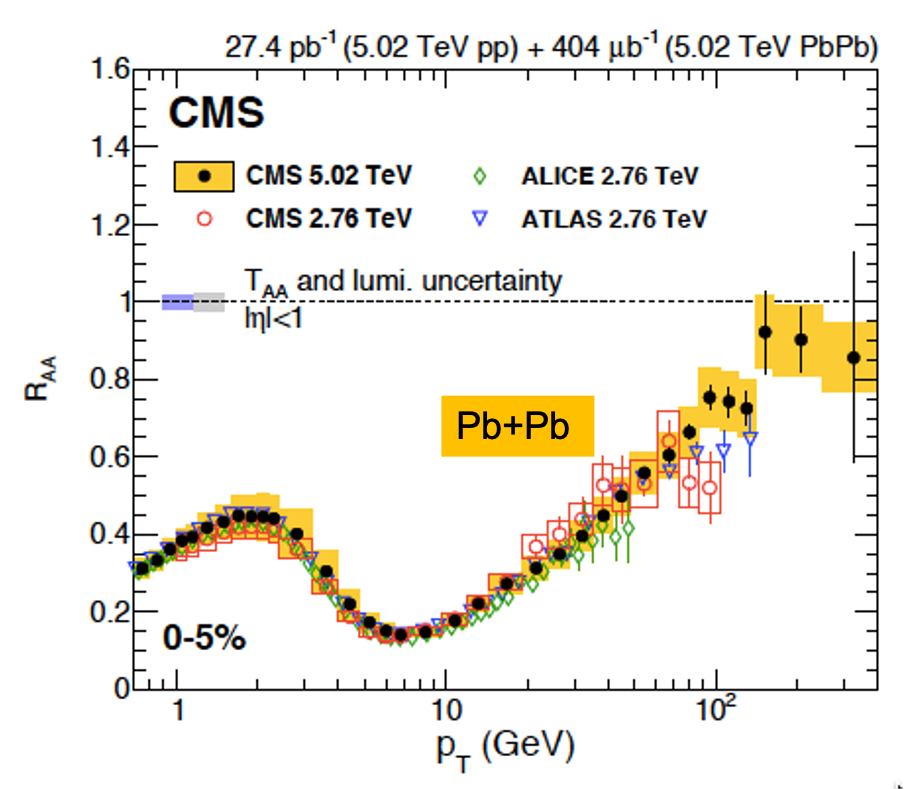}
	\caption{\RAA\ of charged particles in 0-5$\%$ central Pb-Pb collisions at $\sqrt{s_{NN}} =$ 2.76 and 5.02 TeV from CMS \cite{CMS:2012aa}. Also shown are $\sqrt{s_{NN}} =$ 2.76 TeV results from ALICE \cite{ALICE:2012aqc} and ATLAS \cite{ATLAS:2015qmb} as indicated in the legend. The boxes represent the systematic uncertainties of the 5.02 TeV CMS data.}   
	\label{figures:RAA_LHC_CMS_ATLAS_ALICE}
\end{figure}

The $p_T$ range covered by the data expands quickly with collision energy and reaches up to $p_T = 250$ GeV/c in Pb+Pb collisions at $\sqrt{s_{NN}} = 5.02$ TeV measured by ATLAS \cite{ATLAS:2017rmz}. For collision energies in the LHC range, as shown in Fig.~\ref{figures:RAA_LHC_CMS_ATLAS_ALICE}, one generally finds that $R_{\rm AA}(p_T)$ attains a minimum at $p_T \approx 6-8$ GeV/c, followed by a steady rise that extends up to the highest $p_T$ measured. This behavior indicates that the relative energy loss $\Delta E/p_T$ shrinks with increasing momentum $p_T$. 

\begin{figure}[ht]
	\centering
	\includegraphics[width=0.8\linewidth]{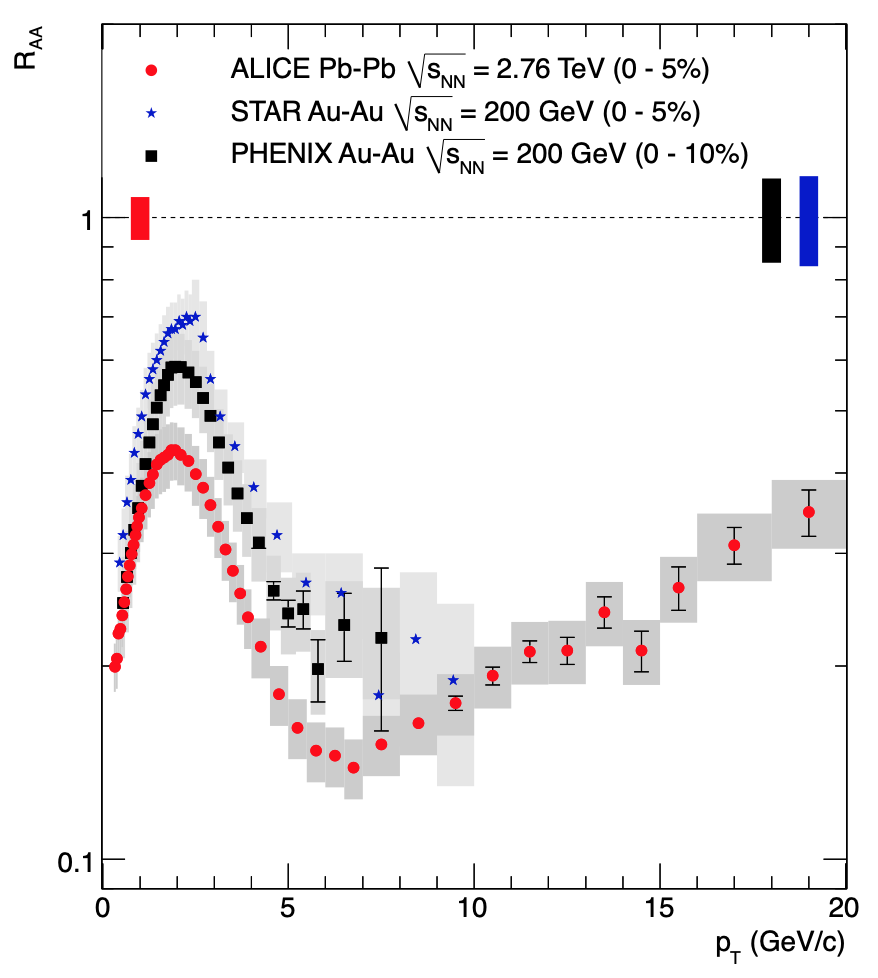}
	\caption{\RAA\ for inclusive charged hadrons measured by ALICE in central 2.76 TeV Pb+Pb collisions in comparison with \RAA\ for inclusive charged hadrons measured by STAR and PHENIX in central Au+Au collisions at 200 GeV \cite{ALICE:2010yje}.}
	\label{figures:ALICE-STAR-PHENIX_RAA}
\end{figure}

A comparison of the inclusive \RAA\ for central Pb+Pb collisions at LHC with that for central Au+Au collisions at the top RHIC energy in Fig.~\ref{figures:ALICE-STAR-PHENIX_RAA} shows that the suppression exhibits a similar pattern and appears only slightly stronger at LHC than at RHIC. This is somewhat of an illusion, because the charged-hadron spectrum falls off more steeply at RHIC, which means that a smaller energy loss $\Delta E$ is needed at RHIC to produce a comparably large suppression as that seen at LHC. 

\begin{figure}[ht]
	\centering
	\includegraphics[width=0.8\linewidth]{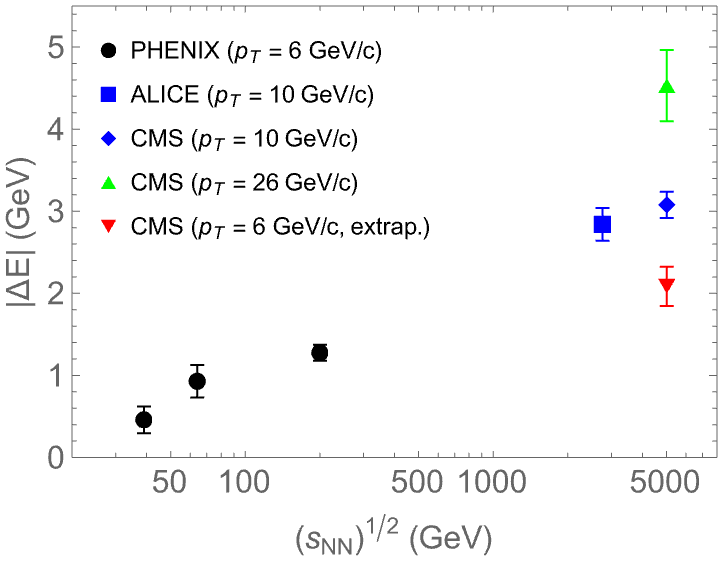}
    \caption{
	Energy loss $|\Delta E|$ at fixed p$_T$ for several different collision energies deduced from the nuclear suppression factor in central Au+Au collisions at RHIC \cite{PHENIX:2008saf,PHENIX:2012oed} and Pb+Pb collisions at LHC \cite{ALICE:2010yje,CMS:2016xef} using the relation (\ref{eq:DeltaE}). The $\Delta E$ for 5.02 TeV collisions has been extrapolated to $p_T = 6$ GeV/c for a visual comparison with the RHIC data.}
 \label{fig:DeltaE_vs_sNN}
\end{figure}

An estimate of the energy loss $\Delta E$ can be obtained as follows. Expressing the nuclear suppression factor as a downward (in $p_T$) shift of the hadron spectrum:
\begin{equation}
    R_{\rm AA}(p_T) = \frac{P_{\rm AA}(p_T)}{P_{\rm pp}(p_T)} = \frac{P_{\rm pp}(p_T-\Delta E)}{P_{\rm pp}(p_T)} .
\end{equation}
Expanding to first order in $\Delta E$ gives
\begin{equation}
    \Delta E = - \frac{\ln R_{\rm AA}(p_T)}{\frac{d}{dp_T}\ln P_{\rm pp}(p_T)} .
    \label{eq:DeltaE}
\end{equation}
Both PHENIX and STAR have published \RAA\ or \RCP\ data for pions at several collision energies from the RHIC beam energy scan \cite{PHENIX:2008saf,PHENIX:2012oed,Horvat:2013lza}. ALICE has published \RAA\ data for pions at the LHC collision energies of 2.76 and 5.02 TeV \cite{Sekihata:2018lwz}. The energy loss deduced from the measured \RAA\ for pions in $0-10$\% central Au+Au collisions at RHIC and for charged hadrons in $0-5\%$ central Pb+Pb collisions at LHC is shown in Fig.~\ref{fig:DeltaE_vs_sNN} for collision energies $\sqrt{s_{\rm NN}}$ ranging from 39 GeV to 5.02 TeV. The energy loss increases with both collision energy and the transverse momentum of the primary parton.

\begin{figure}[ht]
	\centering
	\includegraphics[width=0.95\linewidth]{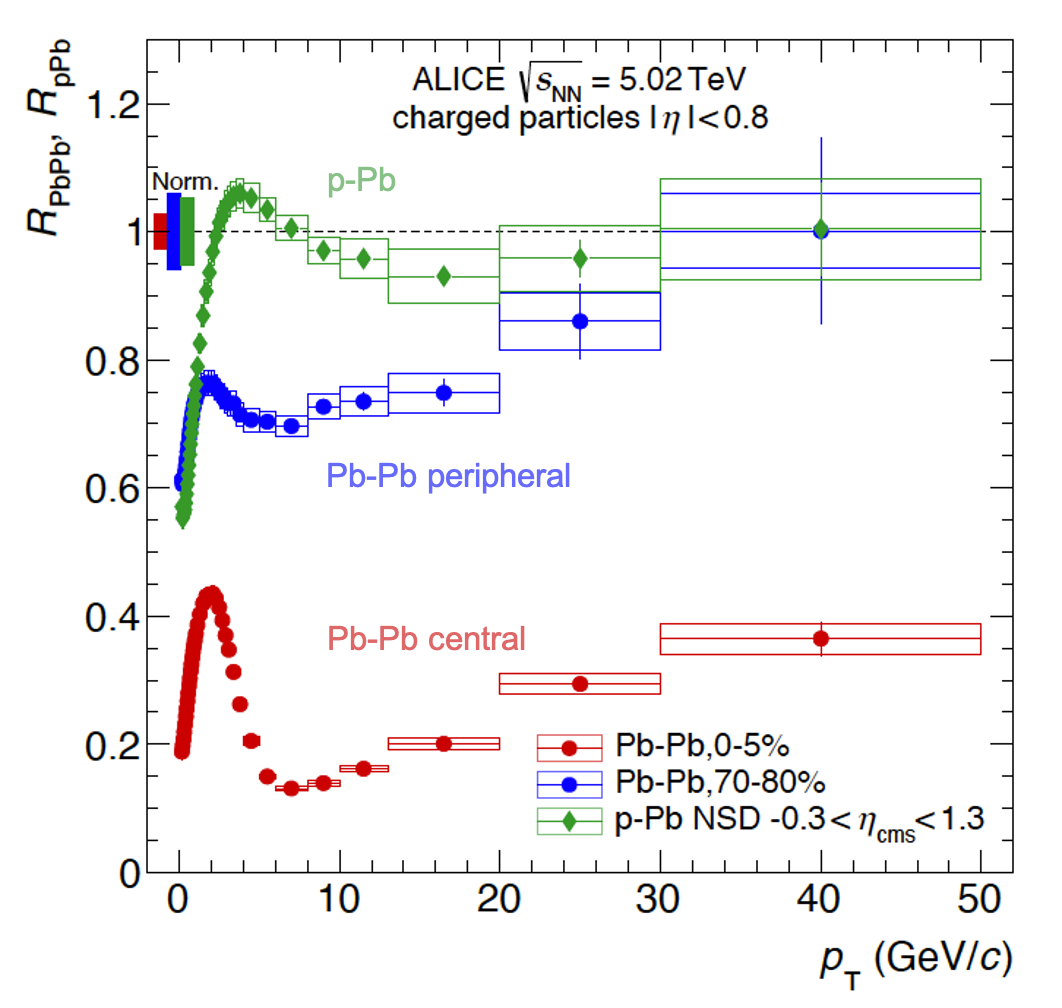}
	\caption{\RAA\ for central (0-5$\%$ centrality) and peripheral (70-80$\%$ centrality) Pb-Pb collisions, and \RpPb\ for non-single diffractive p–Pb collisions at $\sqrt{s_{NN}} = 5.02$ TeV \cite{ALICE:2022wpn}.}
	\label{figures:ALICE_RAA_pPb_PbPb}
\end{figure}

Figure \ref{figures:ALICE_RAA_pPb_PbPb} demonstrates that the nuclear suppression is a function of system size. Comparing \RAA\ measured in central Pb+Pb collisions with the \RAA\ measured in peripheral collisions and $R_{\rm pPb}$ measured in p+Pb collisions one sees that the suppression is much weaker in peripheral collisions, where hard partons have much less matter to traverse, and essentially absent in non-single diffractive (NSD) p+Pb collisions, where very little or no hot matter is produced. 

\begin{figure}[ht]
	\centering
	\includegraphics[width=0.9\linewidth]{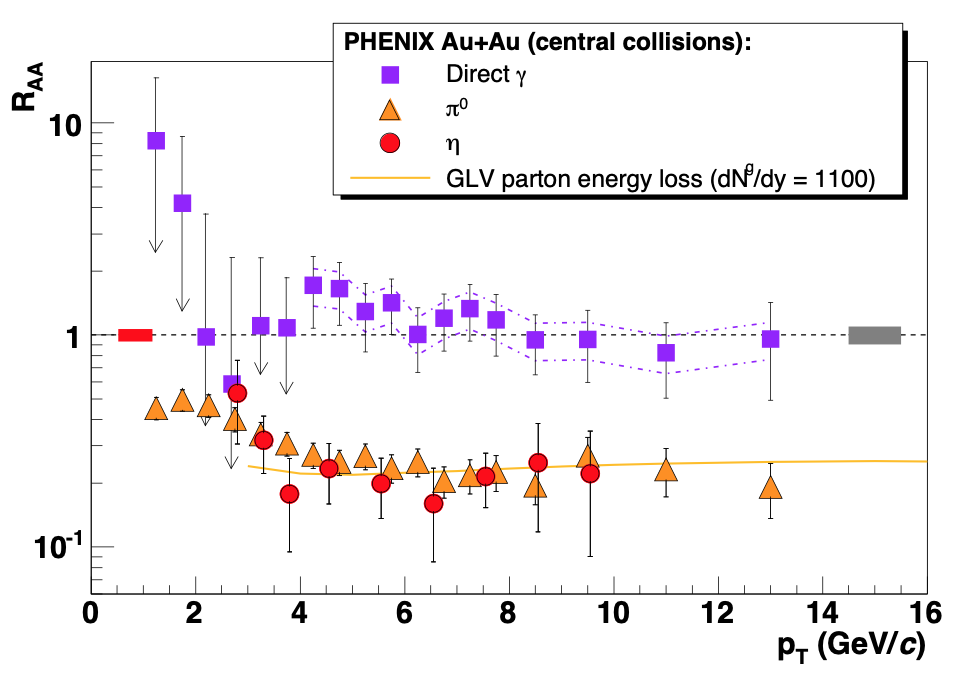}
	\caption{\RAA\ for identified $\pi^0$ and $\eta$-mesons measured by PHENIX in central 200 GeV Au+Au collisions in comparison with the \RAA\ for direct photons \cite{PHENIX:2006ujp}.}
	\label{figures:PHENIX_RAA_gamma_pi0_eta}
\end{figure}

One important question is whether the modification of the hadron spectrum is an initial-state effect, e.g.\ caused by nuclear modification of the parton distribution functions $f_i^{\rm (A)}(x)$, or a final-state effect. This question was answered with the initial results from RHIC by a comparison of the \RAA\ for direct photons with that for $\pi^0$ and $\eta$-mesons \cite{PHENIX:2006ujp}, which is shown in Fig.~\ref{figures:PHENIX_RAA_gamma_pi0_eta}. $\pi^0$ and $\eta$-mesons are almost identically suppressed while direct photons, which do not suffer significant final-state interactions in the QGP, are not suppressed. Further investigation into direct boson production in AA collisions at the LHC have confirmed that not only direct photons \cite{ALICE:2015xmh,ATLAS:2015rlt} but also W- and Z-bosons \cite{ATLAS:2019ibd,ATLAS:2019maq} are consistent with pQCD calculations and exhibit no signs of suppression.

\begin{figure}[ht]
	\centering
	\includegraphics[width=0.95\linewidth]{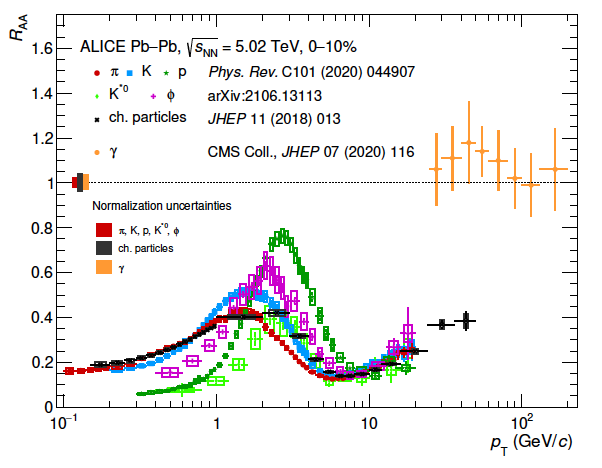}
	\caption{\RAA\ for charged particles, identified particles, and photons in central 5.02 TeV Pb+Pb collisions with particle species and references given in the legend. \cite{ALICE:2022wpn}}
	\label{figures:ALICE_particles_CMS_photons}
\end{figure}

Complementary data from the LHC on identified hadrons and photons extend this conclusion to larger $p_T$ as seen in Fig.~\ref{figures:ALICE_particles_CMS_photons}. The strong suppression of identified hadrons, combined with the lack of suppression of direct photons, singles out a final-state effect (parton energy loss) as the cause of the observed suppression and rules out any initial-state mechanism as the cause. Figure~\ref{figures:ALICE_particles_CMS_photons} also demonstrates that particle-specific effects, such as collective flow and recombination from the QGP, strongly affect the \RAA\ for various hadron species in the range $p_T < 10$ GeV/c. This is observed in heavier mass particles, e.g.\ protons, whose \RAA$(p_T)$ peaks at successively larger $p_T$. However, for $p_T > 10$ GeV/c one finds a universal behavior in \RAA\ for all hadrons composed of light ($u,d,s$) quarks, indicating that these particles are all created by the same mechanism, fragmentation of a hard-scattered primary parton.

\begin{figure}[ht]
	\centering
	\includegraphics[width=0.95\linewidth]{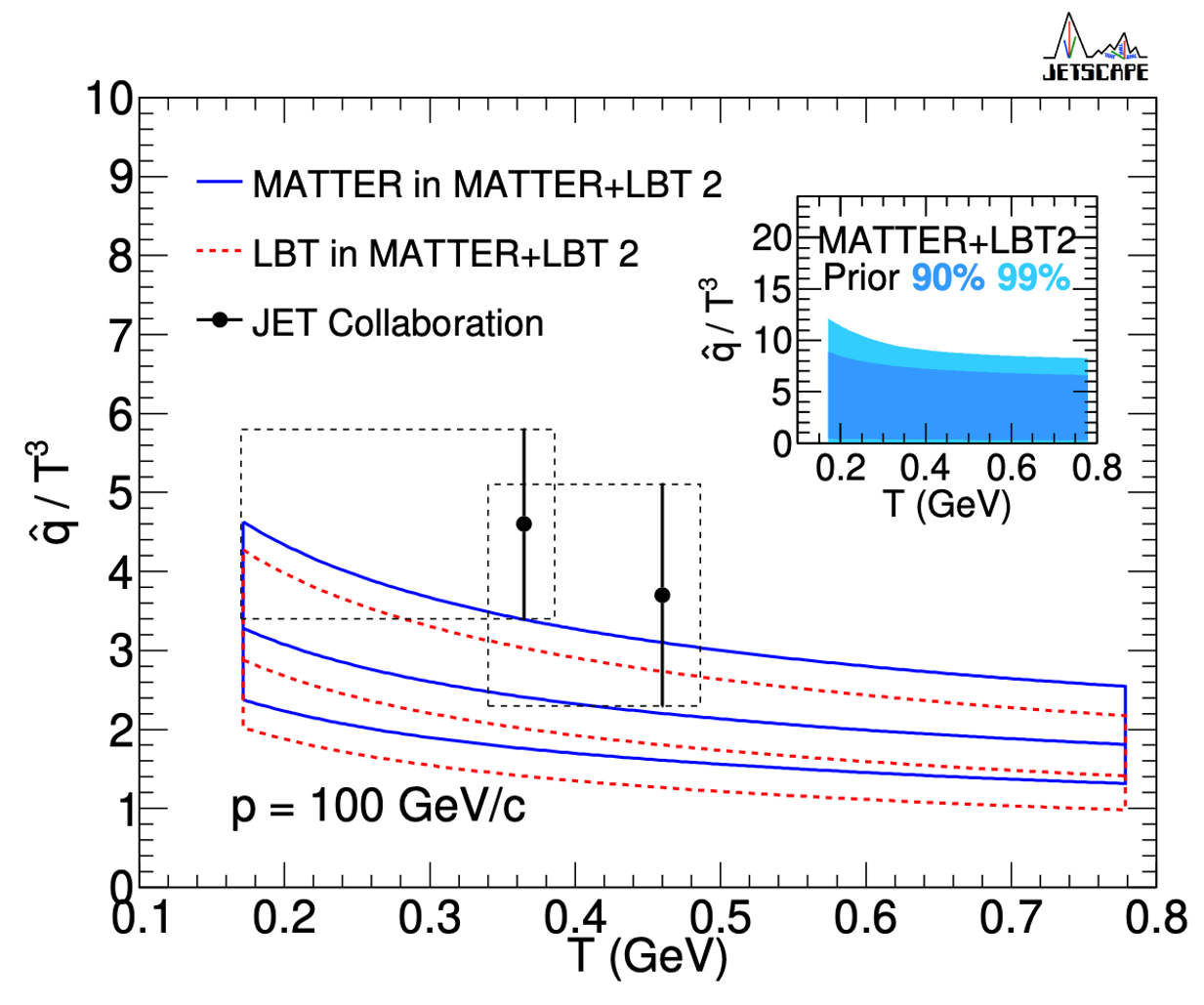}
	\caption{Bayesian parameter extraction of $\hat{q}/T^3$ from experimental measurements of inclusive hadron suppression in Au+Au collisions at RHIC and Pb+Pb collisions at LHC \cite{JETSCAPE:2021ehl}. The 90\% confidence regions for the MATTER+LBT2 model encompass the top and bottom curves of each color as a function of medium temperature $T$. The curves in the middle of  the bands indicate their median values. The solid black circles with error bars represent the results obtained by the JET Collaboration \cite{JET:2013cls}. The dotted boxes indicate the temperature ranges considered in that analysis. The insert shows the prior range of values for $\hat{q}/T^3$ used in the Bayesian analysis with the darker (lighter) area depicting the 90\% (99\%) likelihood range. [From \cite{JETSCAPE:2021ehl}] }
	\label{figures:qhatT3}
\end{figure}

This universal behavior justifies using experimental data to extract using Bayesian parameter estimation a value for the radiative jet quenching parameter $\hat{q}$, \jh{which represents the rate of
change of the \bm{square of the} transverse momentum of a hard parton as it traverses the medium}. Following early work by the JET Collaboration \cite{JET:2013cls}, the JETSCAPE Collaboration performed a systematic analysis to constrain the dependence of $\hat{q}$ on the jet energy, virtuality, and medium temperature from experimental measurements of inclusive hadron suppression in Au+Au collisions at RHIC and Pb+Pb collisions at LHC \cite{JETSCAPE:2021ehl}. The results, shown in Fig.~\ref{figures:qhatT3} favor a model in which the ratio $\hat{q}/T^3$ depends logarthmically on the primary parton virtuality and energy, and it scales quadratically with the color charge of the parton.

\subsubsection{Heavy Flavor Hadrons}

Heavy-flavor quarks are produced mainly in hard scattering of partons in the initial stage of a heavy-ion collision prior to formation of the QGP. Thus, they experience the entire history of the  collision process, interact with the QGP, and probe the flavor and mass dependence of parton energy loss in the evolution of the QGP.

Initial investigations into the possible suppression of heavy-flavor hadrons were carried out at RHIC with measurements of non-photonic electron spectra from semileptonic decays of open-charm and open-beauty hadrons. The \RAA(e$^{\pm}$) was found to be strongly suppressed at mid-rapidity in central \snn\ = 200 GeV Au+Au collisions, indicating significant energy loss of heavy quarks in the QGP \cite{PHENIX:2006iih, STAR:2006btx}. The suppression approaches that of the $\pi^{0}$ for p$_{T} > 4$ GeV/c. 

Later, a direct measurement of the \RAA(D$^{0}$) from $D^{0} \rightarrow K^{-} + \pi^{+}$ for $p_{T} > 2$ GeV/c in semi-central Au+Au collisions (\Npart\ $> 170$) confirmed that the open-charm hadrons are suppressed when traversing the QGP \cite{STAR:2014wif}. The D$^0$-meson yield integrated over $p_{T} < 8$ GeV/c is suppressed by a factor \RAA(D$^{0}$) $\approx 0.5$, while an enhancement by a factor \RAA(D$^{0}$) $\approx 1.3$ is observed over the narrower momentum range 0.7 GeV/c $< p_{T} <$ 2.2 GeV/c. The suppression is consistent with a charm quark energy loss similar to that of light quarks, while the enhancement at low $p_T$ for these most central collisions is a reflection of the chemical oversaturation of charm quarks and may suggest a coalescence mechanism for low-$p_T$ open-charm hadrons. Additional evidence for coalescence comes from the observed enhancement of the $\Lambda_c/{\rm D}^0$ ratio \cite{Vanek:2020sbq}.

\begin{figure}[ht]
	\centering
	\includegraphics[width=0.99\linewidth]{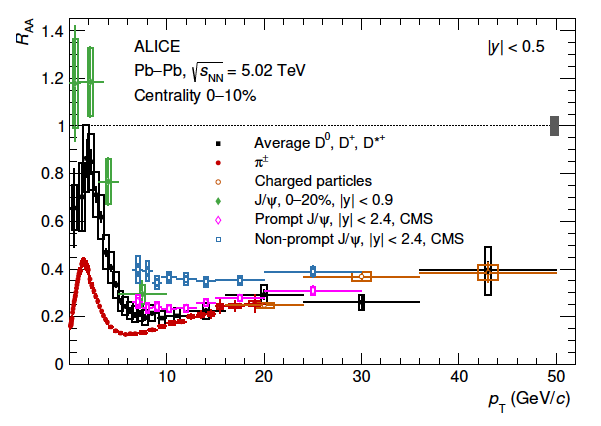}
	\caption{\RAA\ for prompt D-mesons, charged pions, charged particles, and \Jpsi\ from ALICE \cite{ALICE:2021rxa}. Also shown are \RAA\ results for prompt and non-prompt \Jpsi\ from CMS \cite{CMS:2017uuv}. All measurements are for $\sqrt{s_\mathrm{NN}} = 5.02$ TeV central Pb+Pb collisions at mid-rapidity with ranges stated in the legend.}
	\label{figures:ALICE_R_AA_particles}
\end{figure}

Better statistics at the higher energies of the LHC in Run 2 and refinement of experimental techniques enabled a more thorough investigation of the particle and quark mass dependence of the suppression. The \RAA\ of identified hadrons ($\pi^{\pm}$, D$^{0}$, D$^{+}$, D$^{*+}$, \Jpsi) are displayed in Fig.~\ref{figures:ALICE_R_AA_particles} for $\sqrt{s_\mathrm{NN}} = 5.02$ TeV central Pb+Pb collisions at mid-rapidity \cite{ALICE:2021rxa}. The data show that \RAA(D) $>$ \RAA($\pi$) for $p_{T} \lesssim 10$ GeV/c, indicating that effects due to radial flow and hadronization affect D-meson and light- and heavy-hadron yields differently as a function of $p_{T}$, which complicates the interpretation of their \RAA\ values at low to intermediate $p_{T}$. At $p_{T} \gtrsim 10$ GeV/c, the D-meson \RAA\ reaches values similar to that of pions. However, due to the harder $p_{T}$ spectrum and different fragmentation function of charm quarks compared to light quarks and gluons, the interpretation of the differences in the pion and D-meson \RAA\ requires detailed model calculations. 

The \RAA\ for prompt and non-prompt \Jpsi\ from CMS \cite{CMS:2017uuv} is also shown in Fig.~\ref{figures:ALICE_R_AA_particles}. The \RAA\ of prompt D-mesons is observed to be lower than that of non-prompt \Jpsi\ mesons from beauty decays indicating a quark mass dependence of parton energy loss, whereby heavier $b$-quarks lose less energy than lighter $c$-quarks when traversing the QGP. 
\jh{Larger values of \RAA\ for \Jpsi\ are observed than for D mesons at low p$_T$ . This reflects the predominance of recombination of thermal charm quarks in the Pb+Pb system compared to that in p+p. This is also seen in  larger \RAA\ values for D mesons compared to pions at low p$_T$.}
Additional measurements of the \RAA\ of light, open-charm, and open-beauty hadrons via non-photonic electrons at RHIC \cite{STAR:2021uzu,PHENIX:2022wim} and LHC \cite{ALICE:2019nuy}, and muons at LHC \cite{ATLAS:2021xtw,ALICE:2020sjb} confirm the flavor and mass ordering of the suppression of charm and beauty quarks. The investigation into the flavor and mass dependence of hadron suppression in Pb+Pb collisions at LHC continues with new measurements of mixed-quark hadrons such as the D$^{+}_{s}$ \cite{ALICE:2022xrg}, B$^{0}_{s}$ \cite{CMS:2021mzx}, and B$^{+}_{c}$ \cite{CMS:2022sxl}.

\subsection{Jets}

\subsubsection{Jet Suppression}

Understanding the parton energy loss processes in the QGP requires measurement of the resulting parton showers known as jets. Jets and their properties have been measured extensively in p+p collisions. In heavy-ion collisions the showering process becomes convoluted with the energy loss of the partons as they traverse the QGP. It is thus important to compare jet measurements in Pb+Pb collisions with those in p+p collisions to extract the jet energy and yield as a function of $p_{T}$ with the aim to better understand the parton energy loss mechanism. 

In addition to the total jet energy loss relative to the initial hard scatter it is important to distinguish as much as possible between the elastic interaction processes, i.~e.\ two-body scattering off medium constituents, and various inelastic ones, such as collisionally induced gluon radiation. For example, the analysis of inclusive hadron suppression \RAA\ in terms of the jet quenching parameter $\hat{q}$ \cite{JET:2013cls,JETSCAPE:2021ehl} assumes that the entire energy loss of a hard-scattered parton is caused by collisionally induced gluon radiation. One goal of studying jet modification by the medium is to determine whether the picture underpinning such energy loss analyses is correct.

The various parton-medium interaction processes will manifest themselves not only in longitudinal momentum loss but also in momentum broadening transverse to the jet axis. Thus, there is the need to determine differences between jets from heavy-ion collisions and parton showers in vacuum, represented in p+p collisions, and to identify the influence of the flavor and mass of partons on the jet structure. In turn, the medium responds differently to the elastic and inelastic interaction processes that contribute to the parton energy loss. By using jets and high-$p_T$ partons, we seek to understand not only the parton energy-loss mechanisms, but also to probe the QGP at various resolution scales with the ultimate goal of gleaning information about its microscopic structure.

It is important to note that high-$p_{T}$ hadrons are most likely to be produced downstream from the hardest splitting in the jet shower, which is calculable in pQCD, and are most sensitive to the energy loss in that branch of a parton shower. In contrast, jets are sensitive to the energy lost in the entire shower and the various energy loss processes down to the non-perturbative level, but the lost energy ends up outside the kinematic cuts that are used to define the jet. 

Because jets are not unambiguously defined states in QCD, they must be characterised by the experimental procedure by which they are identified. This procedure includes the resolution parameter (also called the jet cone opening angle) $R \leq 1$, the clustering algorithm, such as anti-$k_T$ \cite{Matteo Cacciari_2008}, and possibly a low-$p_T$ cutoff. Only data with the same selections of the clustering algorithm and cone parameter $R$ are comparable. The method for subtracting out the soft background underlying the jet in heavy-ion collisions is also important. 

The jet suppression measured at RHIC and the LHC has been analyzed using transport models, which have found the $\hat{q}/T^3$ transport coefficient for the energy loss distribution to be in the range $\hat{q} = 2 - 4$ GeV$^2/$fm for $300 < T < 500$ MeV over the range of temperatures of the QGP at RHIC and LHC \cite{Ke:2020clc}. The JETSCAPE analysis \cite{JETSCAPE:2021ehl} is only for inclusive hadron production! There is also the recent JETSCAPE analysis of jet substructure \cite{JETSCAPE:2023hqn}, but it does not attempt to extract values for $\hat{q}$.

\begin{figure}[ht]
	\centering
	\includegraphics[width=0.9\linewidth]{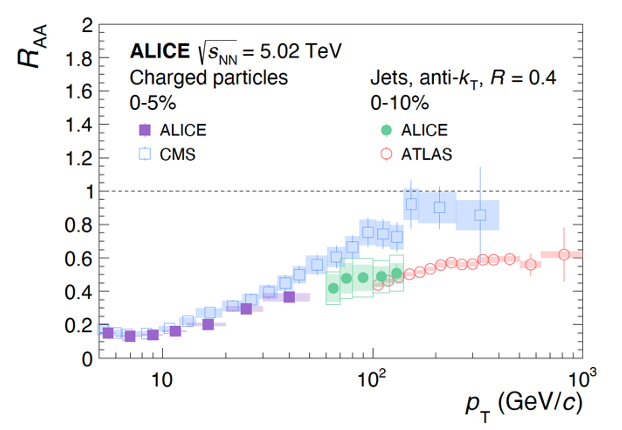}
	\caption{Comparison of \RAA\ for charged hadrons from ALICE \cite{ALICE:2018vuu} and CMS \cite{CMS:2016xef}, and jets from ALICE \cite{ALICE:2019qyj} and ATLAS \cite{ATLAS:2018gwx} in central Pb+Pb  collisions at \snn\ = 5.02 TeV. Compilation from \cite{ALICE:2022wpn}.}
	\label{figures:LHC_Particles_and_Jets.png}
\end{figure}

Figure~\ref{figures:LHC_Particles_and_Jets.png} shows the jet \RAA\ in central Pb+Pb collisions in comparison with the \RAA\ of charged hadrons. The hadron and jet \RAA\ are both found to be strongly suppressed, with the jet \RAA\ exhibiting stronger suppression than that of the inclusive hadrons at the same $p_{T}$. At higher $p_{T}$, jets are more suppressed than hadrons with the same $p_T$, since the inclusive hadrons at a given $p_{T}$ originate from energetic partons that fragment at late times and thus lose less energy in the medium than the combined energy loss of the entire parton shower that constitutes an average jet (see Section \ref{subsubsect:Fragmentation} for a more detailed discussion of jet fragmentation). The measurement of hadrons does not extend as high in $p_{T}$ as that of jets, since the jets encompass the entire shower from the parton rather than just one (leading) hadron.

\begin{figure}[ht]
	\centering
	\includegraphics[width=1.0\linewidth]{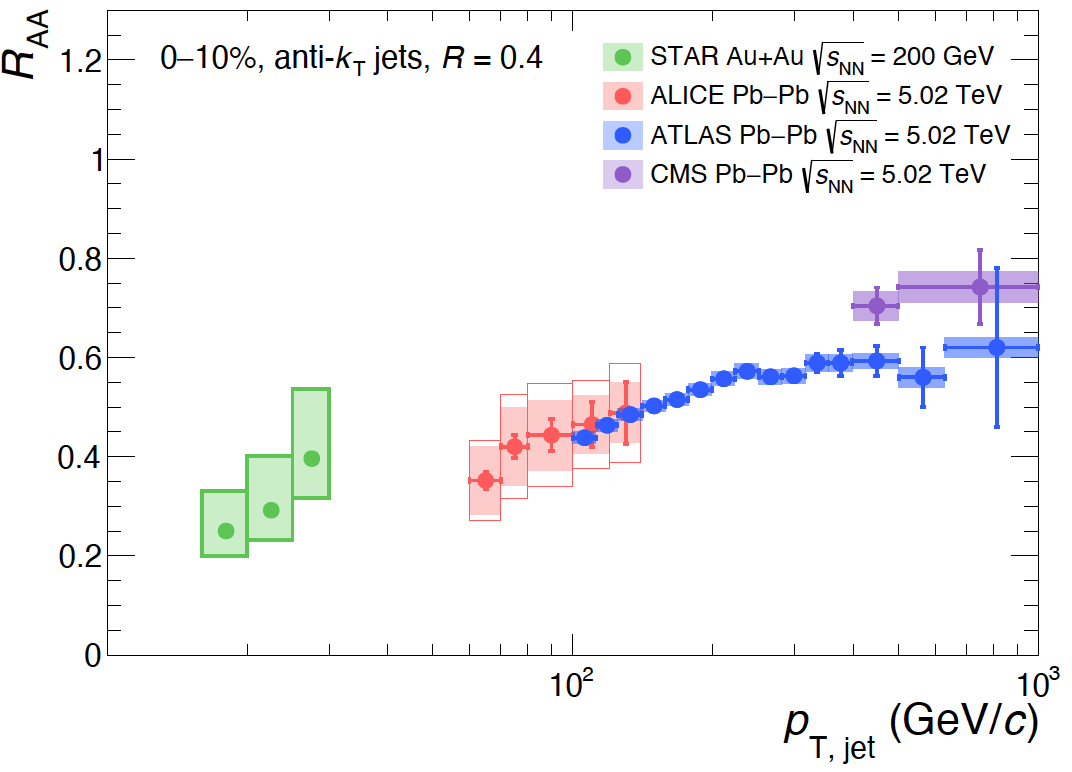}
	\caption{A compilation \cite{Bossi:2023nmu} of jet \RAA\ measurements at RHIC and LHC \cite{STAR:2020xiv,ALICE:2019qyj,ATLAS:2018gwx,CMS:2021vui}. Measurements are for full jets at LHC and charged-particle jets at RHIC. See text for more details.}
	\label{figures:Jet-Raa-Summary-Plot.png}
\end{figure}

A summary plot of current jet \RAA\ measurements from RHIC and LHC is shown in Fig.~\ref{figures:Jet-Raa-Summary-Plot.png} for central (0-10\%) Au+Au at RHIC and Pb+Pb at the LHC. \cite{Bossi:2023nmu} The ATLAS and CMS results represent full (electromagnetic and hadronic) calorimetric measurements of jets, ALICE comprises electromagnetic energy and charged particles, while STAR measurements are jets measured solely with charged particles, all with the same jet resolution parameter $R = 0.4$. The uncertainties are larger for the STAR and ALICE jet measurements and increase as the jet-$p_{T}$ decreases. Several effects contribute to the increased uncertainty at low jet-$p_{T}$: the dependence of the experiments on charged-particle tracking rather than calorimetry, the increased influence of the soft background at lower jet-$p_{T}$, and greater dependence on the low-$p_T$ cutoff. Also noticeable is the gap in $p_{T}$ between the RHIC and LHC data, which is partly due to the circumstance that only the energy by charged particles is detected in the STAR measurements. The entire region $p_{T} < 100$ GeV/c is important to theoretical comparisons in order to better understand jet energy loss mechanisms and the response of the medium. Therefore, it is a focus of new experimental background and jet-isolation techniques and continued higher statistics data-taking. 

The dependence of jet quenching on the color charge of the primary parton can also be derived from a comparison of jets initiated by a hard-scattered quark (quark jets) with those initiated by a gluon (gluon jets). Experimentally, this can be achieved statistically by comparing ensembles of photon-tagged jets (jets opposite in azimuth from an isolated photon) with inclusive jets. Event generators predict that the fraction of quark jets in a photon-tagged sample of jets in a typical kinematic range at the LHC is $0.7-0.8$ as compared to a quark fraction of $0.3-0.5$ for inclusive jets in the same range \cite{CMS:2018jco}. If the jet energy loss is proportional to the square of the color charge of the primary parton ($C_q/C_g = 4/9$) as predicted by theory, a smaller quark energy loss should be reflected in less suppression, i.~e.\ a larger \RAA\ for photon-tagged jets than for inclusive jets. 

\begin{figure}[ht]
	\centering
	\includegraphics[width=0.9\linewidth]{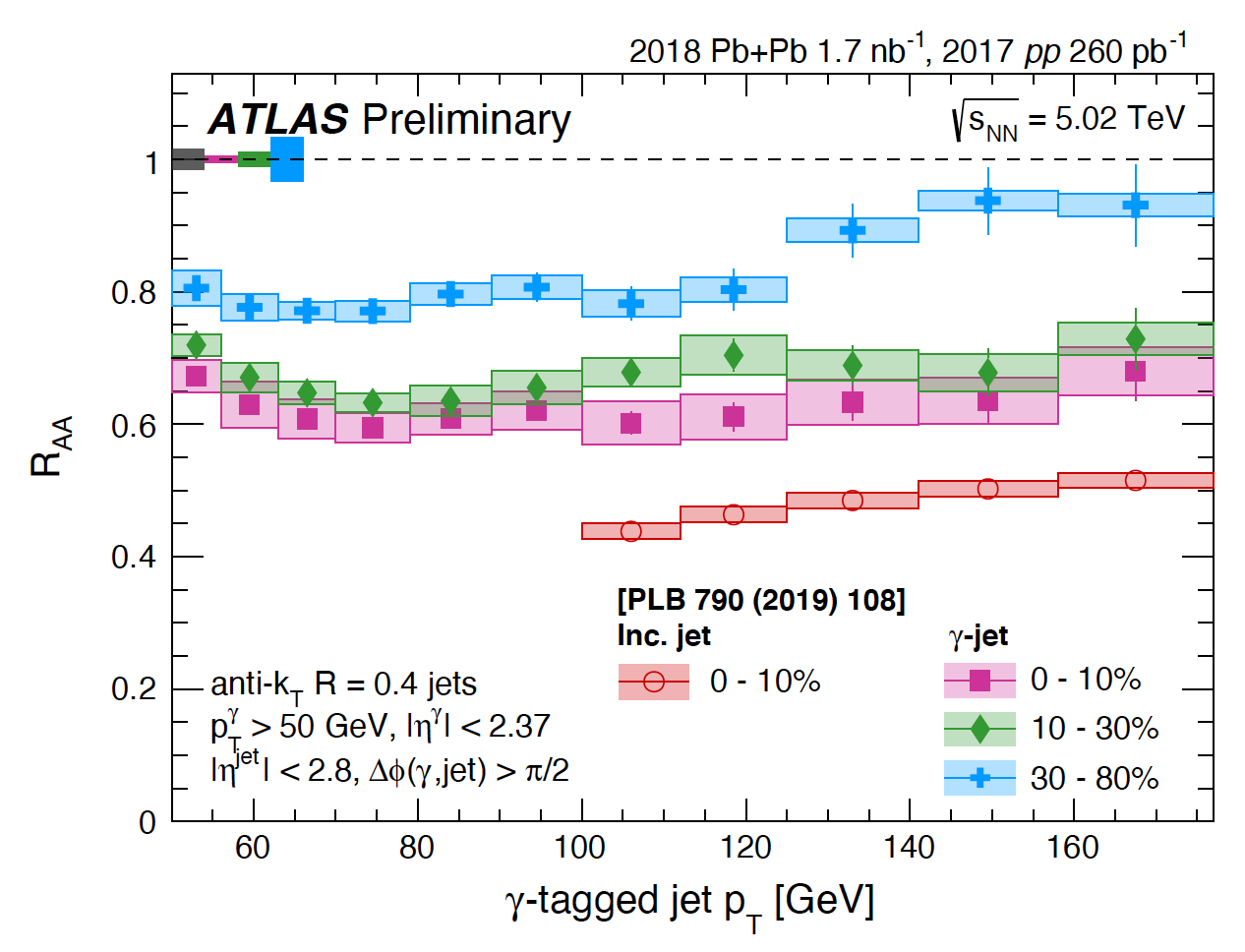}
	\caption{Jet \RAA\ compilation from ATLAS for $\gamma$-jet and inclusive jets in \snn\ = 5.02 TeV Pb+Pb collisions. Details in the legend and text. [from \cite{ATLAS:2022cim}]}
	\label{figures:ATLAS_gamma-tagged-jet_figure.png}
\end{figure}

The jets opposite isolated photons will consist predominantly of quark jets, enabling potential discrimination between the energy loss of a primary quark with the medium and that of a mixture of quarks and gluons that make up the inclusive jet sample. These events with jets opposite a photon (referred to as $\gamma$-jet) were investigated and compared to inclusive jet production \cite{ATLAS:2023iad} in p+p and Pb+Pb interactions at \snn\ = 5.02 TeV. Figure~\ref{figures:ATLAS_gamma-tagged-jet_figure.png}
displays a comparison of the \RAA\ for $\gamma$-jet measurements at three centralities with an inclusive jet measurement at 0-10\% centrality. As already seen for inclusive jets, the $\gamma$-jet measurements also exhibit increased suppression for more central collisions. However, as Fig.~\ref{figures:ATLAS_gamma-tagged-jet_figure.png} highlights, the most central $\gamma$-jet \RAA\ results show significantly less suppression than inclusive jets reflecting the enhanced presence of gluons with their larger energy loss in the inclusive sample.

\subsubsection{Jet Fragmentation and Jet Shape}
\label{subsubsect:Fragmentation}

It is important to note that a general difference between the jet and hadron $p_{T}$-spectra is that the hadron spectra result from fragmentation of the primary parton into a jet that contains a leading parton carrying above average momentum. Therefore, the fragmentation function plays an integral role in the difference between the hadron and jet $p_{T}$-spectra, and the jet spectrum is harder than that of inclusive hadrons. In fact, a hadron and a jet at a given $p_{T}$ do not originate from partons with the same $p_{T}$. The primary parton momentum, which is represented for the most part by the jet, must be convoluted with the fragmentation function in order to obtain the $p_T$ of an individual hadron. Clearly, this entails the need to measure the fragmentation function in p+p collisions and its modification in A+A collisions. Similarly, the desire to understand the transverse momentum broadening of the jet shower by its interaction with the medium requires a quantitative understanding of the transverse jet shapes in p+p collisions and their modification in A+A collisions.

\begin{figure}[ht]
	\centering
	\includegraphics[width=0.8\linewidth]{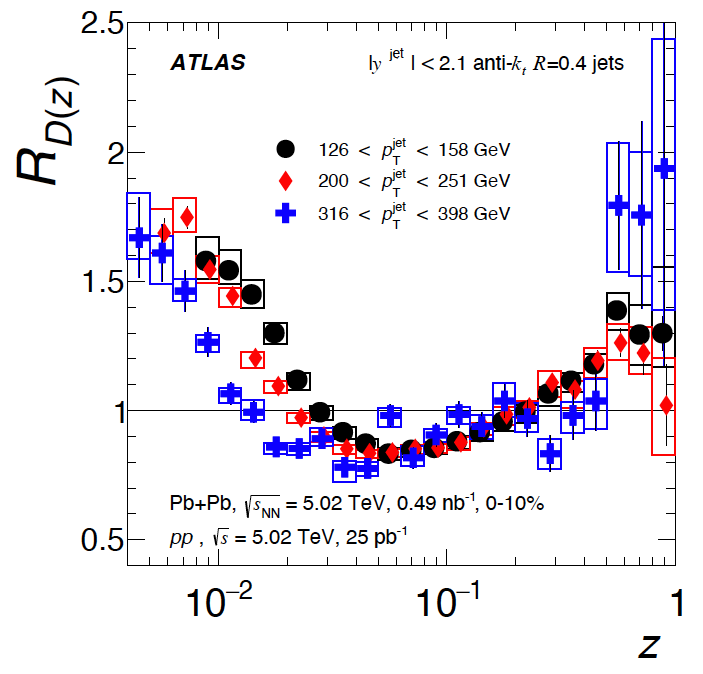}
	\caption{Fragmentation function ratio $R_{D}$ of fragmentation functions plotted as a function of $z$ in central Pb+Pb collisions relative to those in p+p collisions. Details of the jet selection are given in the legend.\cite{ATLAS:2018bvp}}
	\label{figures:ATLAS_Fragmentation_Function_Ratios.png}
\end{figure}

The fragmentation functions $D(z)$ for charged hadrons have been measured in p+p and Pb+Pb collisions \cite{ATLAS:2018gwx} for a variety of centralities \cite{ATLAS:2018bvp}. Figure~\ref{figures:ATLAS_Fragmentation_Function_Ratios.png} shows the measured ratios $R_{D(z)}$ of jet fragmentation into charged hadrons in central Pb+Pb collisions relative to p+p collisions as a function of $z = p_{T}/p_{\rm jet}$. A strong enhancement is observed for hadrons at low $z$, while a suppression is seen for hadrons in the intermediate region $0.03 < z < 0.1$. This is consistent with a scenario in which partons that would normally contribute in this intermediate region interact with the medium, lose energy, and form hadrons at lower $z$ resulting in the observed low-$z$ enhancement. The slight enhancement observed for hadrons with $z > 0.5$, a kinematic region typically dominated by leading hadrons, may reflect a selection bias in favor of narrow jets, which do not interact as strongly with the medium as wider jets.

\begin{figure*}[t]
    \centering
        \includegraphics[width=0.85\linewidth] {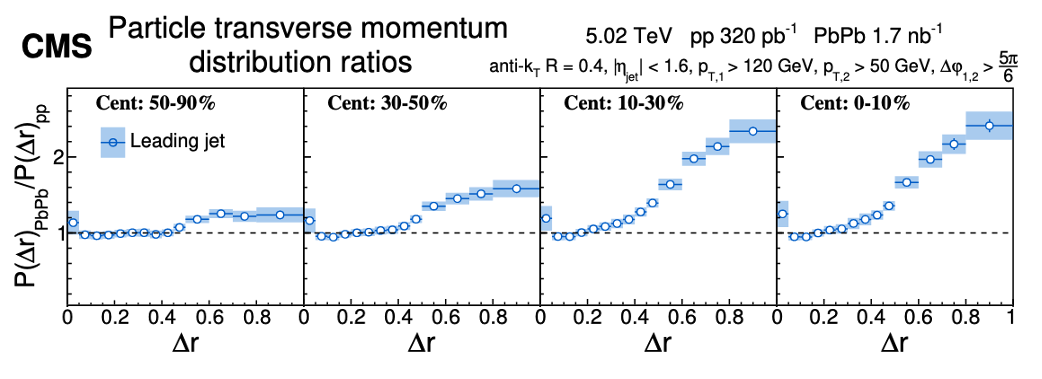}
    \caption{The ratio of the jet radial momentum distributions as a function of the angular distance $\Delta r$ from the jet axis in Pb+Pb for various centrality intervals relative to those measured in p+p collisions. The CMS data are for leading jets with $R = 0.4$ and $p_T > 120$ GeV/c, and for charged particles with 0.7 GeV/c $< p_T^{\rm (track)} <$ 300 GeV/c \cite{CMS:2013lhm}.}
    \label{figures:CMS_Leading_jet_figure.png}
\end{figure*}

Since $D(z)$  only provides a measure of the longitudinal fragmentation of jets, it is important to also measure the transverse structure of the jets to gain additional insight into the medium modification of the fragmentation process and the role of parton-medium interactions. This is commonly achieved by measuring the angular distribution of hadrons with respect to the jet axis within the jet cone. Figure~\ref{figures:CMS_Leading_jet_figure.png} displays the ratio of the jet radial momentum distributions as a function of the angular distance $\Delta r$ from the jet axis in Pb+Pb for various centrality intervals relative to that measured in p+p collisions for leading jets with $p_T >$ 120 GeV/c, $R = 0.4$ and 0.7 GeV/c $< p_T^{\rm track} < 300$ GeV/c \cite{CMS:2013lhm}. The Pb+Pb radial momentum distributions are enhanced over the p+p distribution for charged particles farther away from the jet axis and the enhancement increases with centrality primarily outside the jet cone ($\Delta r > 0.4$). 
This behavior indicates that there is significant out-of-cone radiation associated with the jet \cite{CMS:2018zze}. Thus, jets defined with a larger cone radius $R$ should recover more of this large-angle radiation than jets defined with a narrower cone and therefore should be expected to incorporate more sources of potential energy loss. The magnitude of the out-of-cone radiation will depend on the parton-medium interactions and also differences in the energy-loss mechanisms between quark and gluon jets.

Another promising probe of the mechanisms of jet-medium interactions are jets with a leading $b$-quark ($b$-jets. These jets overall are observed to be broader than inclusive jets \cite{CMS:2022btc}, with a broadening of the angular distribution of charged hadrons beyond $R = 0.2$ that increases significantly in Pb+Pb collisions for more central events and extends beyond the cone radius that defines the $b$-jet. Thus, the energy in $b$-jets is redistributed to larger angles in Pb+Pb collisions compared with p+p collisions. This finding is consistent with measurements of the \RAA\ for $R = 0.2$ $b$-jets compared to inclusive jets, where the \RAA\ appears larger for $b$-jets than that for inclusive jets in central Pb+Pb collisions \cite{ATLAS:2022fgb}. In general, the $b$-jet measurements are suggestive of mass and color-charge effects in the mechanisms of jet energy loss in heavy-ion collisions. Higher statistics data and new measurements will be required to disentangle the various sources of these effects.

\subsubsection{Jet Substructure}

We now turn to the emerging field of jet substructure measurements. As compared to  inclusive jet measurements, jet substructure measurements seek to elucidate the dynamical evolution of the internal structure of the jet as it propagates through the QGP medium and thus aim to provide information on the microscopic processes leading to parton energy loss in the QGP. There are two possible approaches to this goal. One, which can be called the microscopic approach, strives for the  complete reconstruction of the underlying parton propagation and kinematics in the QGP in the hope that this will permit one to distinguish and understand the energy loss processes and the response of the QGP to the evolving jet. The other, which can be called the global approach, aims at the precision measurement of semi-inclusive observables that are sensitive to the substructure of jets and can be rigorously calculated in QCD without the need for somewhat arbitrary kinematic cuts. We first discuss the microscopic approach.

\begin{figure}[ht]
	\centering
	\includegraphics[width=0.8\linewidth]{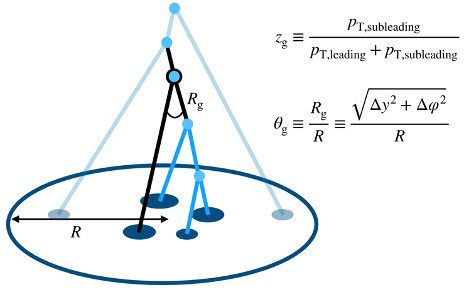}
	\caption{Diagram of angular-ordered re-clustering of constituents of a jet and the Soft Drop grooming procedure \cite{Dasgupta:2013ihk, Larkoski:2014wba}) to reduce background and then re-clustering \cite{Catani:1993hr}. The identified splitting is shown in black and the groomed-away splittings in light blue. From \cite{ALICE:2021mqf}.}
\label{figures:ALICE_Substructure}
\end{figure}

In order to reconstruct the evolution or shower history of a jet and determine its parton energy-loss mechanisms in the medium, the parton splittings and interactions must be derived from the final jet constituents. The splittings can be investigated using a technique that involves grooming of the jets (one popular approach is Soft Drop \cite{Dasgupta:2013ihk}) to reduce background and then reclustering (\cite{Larkoski:2014wba}) to determine the angular ordering in the QCD evolution of the jet. The jet substructure splittings can be characterized by the momentum fraction ($z_g$) and opening angle ($\theta_g$) of the first splitting after grooming, as shown in Fig.~\ref{figures:ALICE_Substructure}. This algorithm is well suited to analyze jet fragmentation in the vacuum, i.~e.\ in p+p collisions, where the branching tree obeys angular ordering. Within a medium the angular ordering can be destroyed by medium-induced interactions that change the color flow within the branching jet, and the usefulness of this method is less well established.

\begin{figure}[ht]
	\centering
 	\includegraphics[width=0.8\linewidth]{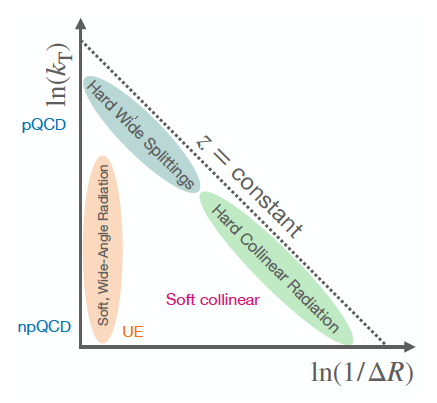}
	\caption{The Lund Plane representation \cite{Dreyer:2018nbf} of the kinematic regions available within a jet. The $\Delta R$ and $k_T$ are the angle and transverse momentum of a gluon emission with respect to its parent parton.}
	\label{figures:Lund-Diagram}
\end{figure}

Two variables that describe the splittings after grooming -- $z_g$ (the momentum fraction of first splitting) and $R_g$ (the angular opening of the first splitting) - can be derived in theory and extracted from experiment in jet analyses. These variables are typically plotted in a diagram, known as the Lund Plane \cite{Dreyer:2018nbf} (see Fig.~\ref{figures:Lund-Diagram}), where $k_{T} = p_{T,{\rm subleading}} \sin(R)$ and $\theta_{g} = R_{g}/R$, with $R$ being the jet cone angle. \cite{ATLAS:2020bbn, ATLAS:2022vii, ALICE:2021mqf}

The different regions in the Lund plane are populated by splittings ranging from the non-perturbative at low $\ln(k_T)$ to perturbative at high $\ln(k_T)$. Wider splittings and soft wide-angle radiation populate lower values of $\ln(1/\Delta R)$, where $\Delta R$ is the angle between the splitting and the jet axis. Splittings that are more collinear correspond to higher values of $\ln(1/\Delta R)$. The Lund Plane also provides insight into regions where coherence may take place.

Fully corrected measurements of $z_g$ distributions in Pb+Pb are found to be consistent with those measured in p+p collisions over the entire range of jets measured. However, the $\theta_g$ (and R$_g$) distributions are narrower for smaller-angle jet splittings in Pb+Pb collisions, and the wider-angle splittings are significantly more suppressed relative to those in p+p \cite{ALICE:2021mqf, ATLAS:2022vii}. In central collisions, the values of the jet suppression factor \RAA\ range between 0.75 for narrow jets and $\sim 0.3$ for the widest jets. We already speculated that this phenomenon is responsible for the rise of $R_{D(z)}$ for $z \to 1$ in Fig.~\ref{figures:ATLAS_Fragmentation_Function_Ratios.png}.

Presumably, the wider jets reflect incoherent interactions or larger gluon fractions and thus suffer more energy loss than narrow jets. These results are qualitatively in line with a recent JETSCAPE study of jet substructure modifications caused by jet-medium interactions \cite{JETSCAPE:2023hqn}, which confirms that parton scattering with the QGP at high virtuality is highly suppressed by coherence effects. The reduced interaction of highly virtual partons with the medium then leads to the enhancement of narrow jets relative to wide jets. Further studies along these lines could allow for a determination of the scale dependence of elastic parton scattering in the medium that goes beyond the jet quenching parameter $\hat{q}$ and thereby yield insight into the scale dependence of the microscopic structure of the QGP.

A more global approach to the study of jet substructure, which does not rely on the use of jet shower simulations is the measurement of energy-energy correlators (EEC) \cite{Gao:2019ojf,Dixon:2019uzg} and, more generally, correlators involving track functions \cite{Jaarsma:2022kdd}. Track functions are asymptotic expectation values of observables, such as energy flow or conserved currents, integrated along a given angular direction (the track) pointing away from the interaction vertex. Their usefulness derives from the fact that they can (a) be rigorously defined in quantum field theory \cite{Sveshnikov:1995vi} and (b) are the natural objects measured by calorimeters with or without particle identification.

\begin{figure}[ht]
	\centering
 	\includegraphics[width=0.95\linewidth]{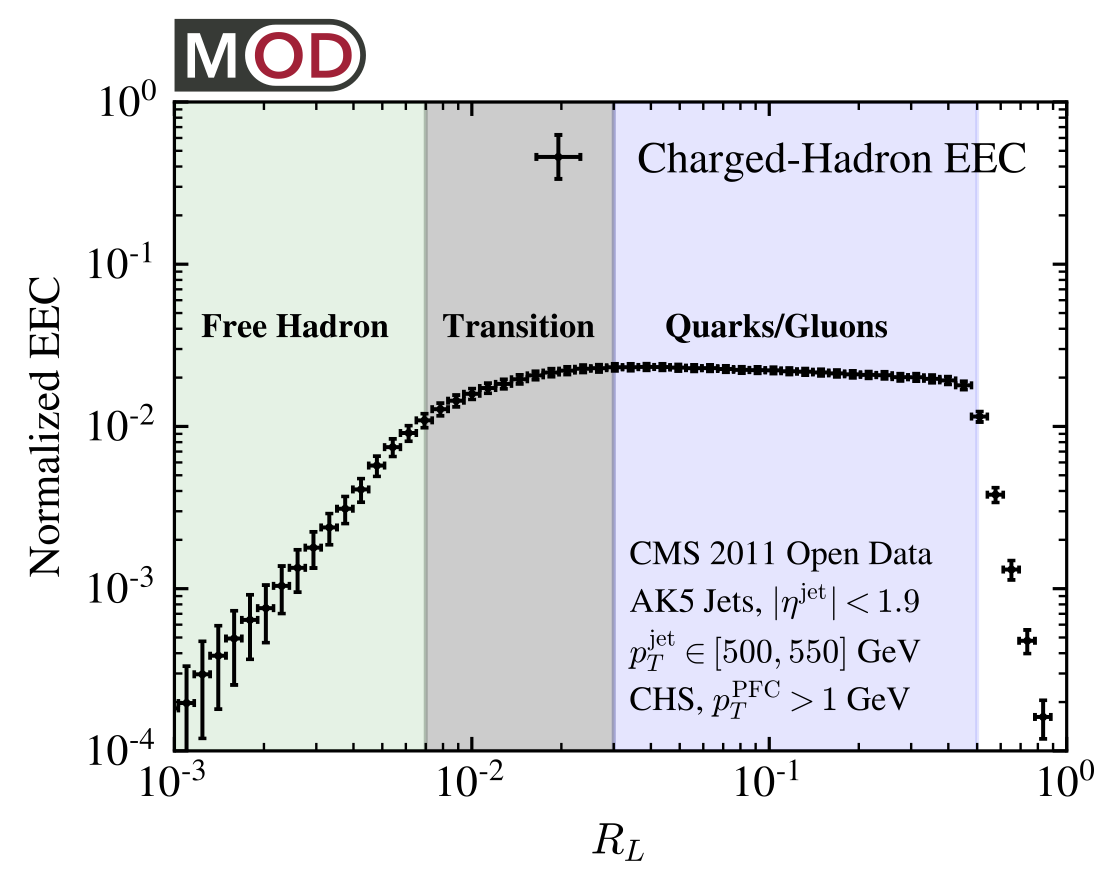}
	\caption{The two-point energy-energy correlator restricted to charged hadrons, evaluated from CMS Open Data for p+p collisions at LHC. The data, which are plotted as a function of the relative angle $R_L$ between the tracks, exhibit distinct scaling regimes associated with asymptotically free partons (at large $R_L$) and free hadrons (at small $R_L$). [From \cite{Komiske:2022enw}]}
	\label{figures:EEC}
\end{figure}

Recent progress in the calculation of the renormalization group flow for EECs \cite{Dixon:2019uzg} and moments of track functions \cite{Jaarsma:2022kdd,Chen:2022pdu} together with the demonstration of a universal scaling behavior of EECs in p+p data from LHC \cite{Komiske:2022enw,Lee:2022ige} have raised interest in using such global jet substructure observables for the study of jet quenching in A+A collisions. As an example of this behavior, Fig.~\ref{figures:EEC} shows the EEC restricted to charged hadrons for p+p collisions at LHC using CMS open data \cite{Komiske:2022enw}. The magenta shaded region labeled ``Quarks/Gluons'' is well described by next-to-next-to-leading QCD perturbation theory \cite{Lee:2022ige} indicating that it is governed by perturbative parton showers. Ongoing research focuses on the measurement of the modifications of EECs and track function moments in p+A and A+A collisions where characteristic changes due to jet-medium interactions are predicted, which are sensitive to the dynamics of color coherence in the parton shower \cite{Andres:2023xwr}.

\section{Collective flow}
\label{sec:elliptic}

Not all signatures of the QGP that are now understood to be relevant and important were recognized as such in our 1996 review and are thus absent from Fig.~\ref{figures:Signatures1996}. This section will be devoted to a brief discussion of those signatures that have had great phenomenological impact but were not fully appreciated before the advent of data from heavy-ion colliders. The most important and ubiquitous of these are the collective flow anisotropies $v_n$, most importantly, the elliptic flow coefficient $v_2$.

Many-body systems exhibit collective flow that can be described by viscous hydrodynamics if the mean-free path $\lambda_f$ of their constituents is short compared to the system size $L$, i.e.\ if the Knudsen number $Kn = \lambda_f/L \ll 1$. Before the advent of collider data, this condition was not expected to be satisfied by the QGP, because the strong long-range color force is screened in it, and lowest-order perturbative calculations of $\lambda_f$ yield rather large values. Although some theorists argued otherwise \cite{Danielewicz:1984ww}, the general consensus was that the specific shear viscosity $\eta/s$, where $s$ is the entropy density, of the QGP was of order unity or larger, prohibiting well developed collective flow for fireballs of nuclear size.

Features of collective flow were initially observed in fixed-target experiments at the BEVALAC in 400 MeV/u Ca+Ca and Nb+Nb collisions \cite{Gustafsson:1984ka} and 800 MeV/u Ar+Pb collisions \cite{Renfordt:1984et}. A detailed characterization of collective flow in terms of directed and elliptic flow was performed in 158 GeV/u fixed-target Pb+Pb collisions \cite{NA49:1997qey} at the SPS.  Data from Au+Au collisions at RHIC and later in Pb+Pb collisions at LHC clearly showed that the initial geometrical features of the QGP fireball are translated into characteristic collective flow patterns. For early summaries of these results and their interpretation see \cite{PHOBOS:2004zne,STAR:2005gfr,PHENIX:2004vcz,Muller:2006ee,Muller:2012zq}. 

The geometric features imprinted on the fireball during the initial collision can be expressed in terms of eccentricities $\varepsilon_n$ that measure the azimuthal anisotropies of the deposited energy density with respect to the beam axis. Hydrodynamics translates these geometric anisotropies into azimuthal anisotropies of the spectra of emitted particles, which are parameterized by flow coefficients $v_n$ in the form
\begin{multline}
E \frac{d^3N}{dp^3} = \frac{1}{2\pi} \frac {d^2N}{p_t dp_t dy} \times 
\\
\left[ 1 + \sum_{n=1}^\infty 2 v_n(p_T) \cos[n(\phi-\Psi_n)] \right] ,
\label{eq:vn}
\end{multline}
where $\Psi_n$ denotes the $n$-th order event plane.

\begin{figure}[ht]
	\centering
	\includegraphics[width=0.95\linewidth]{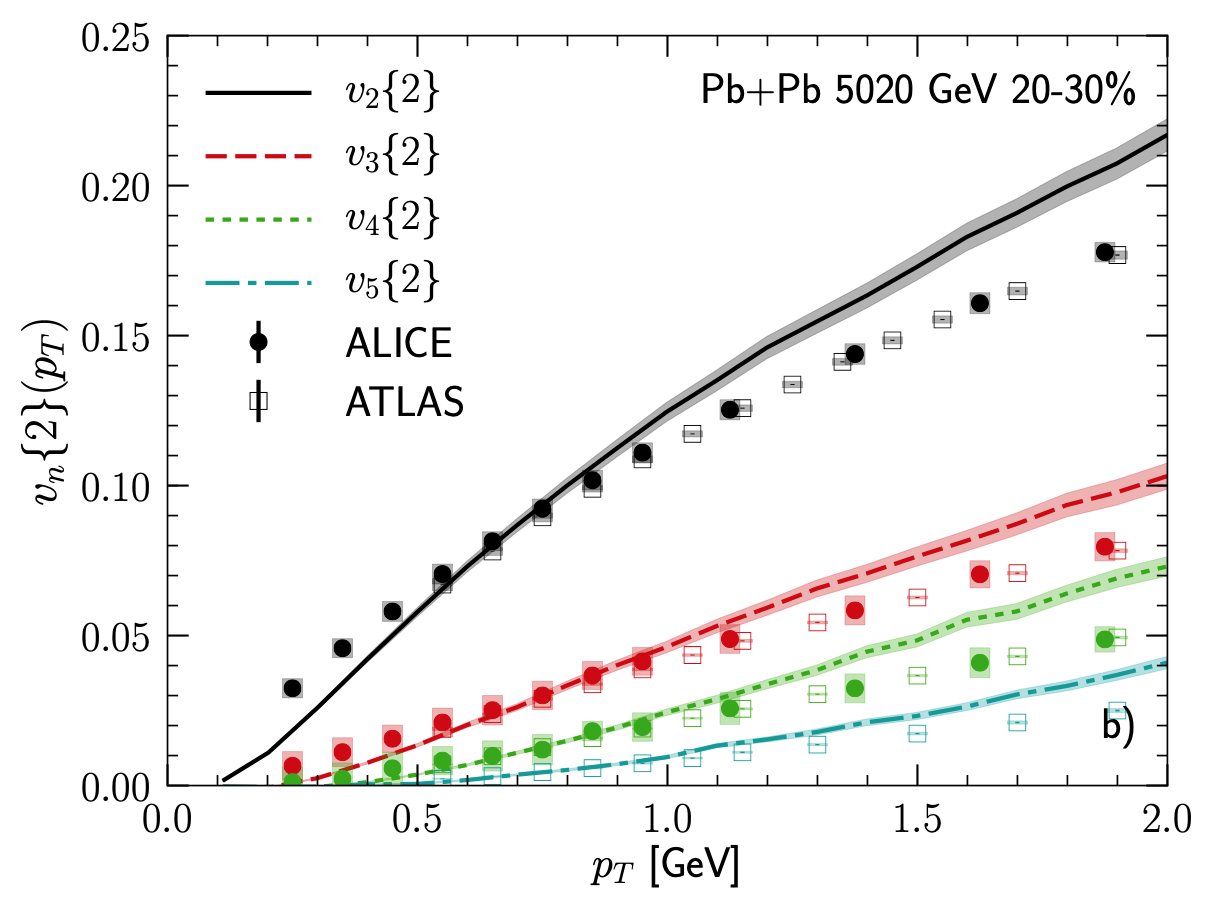}
	\caption{v$_n(p_T)$ ($n=2,3,4,5$) measured in $20-30\%$ central Pb+Pb collisions at \snn\ = 5.02 TeV by ALICE \cite{ALICE:2018rtz} and ATLAS \cite{ATLAS:2018ezv}. The data are compared with simulations in a hybrid collision model \cite{Schenke:2020mbo} based on viscous hydrodynamical evolution of the QGP phase. [From \cite{Schenke:2020mbo}]}
	\label{figures:LHC_vn}
\end{figure}

The magnitude of the observed $v_n(p_T)$ depends on the initial eccentricities $\varepsilon_n$ and the specific shear viscosity $\eta/s$. Since the $\varepsilon_n$ can be reliably modeled based on our knowledge of nuclear structure and elementary nucleon-nucleon collisions, the data for $v_n(p_T)$ can be used to deduce the value of $\eta/s$ from the data by means of a Bayesian model-data comparison. Here we can only present a few examples of the many published comparisons of viscous hydrodynamics simulations with experimental data. Figure~\ref{figures:LHC_vn} shows the flow coefficients $v_n(p_T)$ measured by ALICE \cite{ALICE:2018rtz} and ATLAS \cite{ATLAS:2018ezv} in \snn\ = 5.02 TeV Pb+Pb collisions compared with the results of hybrid model calculations using second-order viscous hydrodynamics with $\eta/s = 0.12$ to describe the QGP phase \cite{Schenke:2020mbo}.

The fact that all flow components $v_n(p_T)$ can be described by the same hydrodynamic equation with the need for fine-tuning of the initial eccentricities presents clear evidence for a rapid ``hydrodynamization'' of the QGP fireball. Theoretical studies of the approach to viscous hydrodynamic motion in the context of kinetic theory and holographic models have shown that the onset of hydrodynamics can occur when the system is still quite far from local thermal equilibrium because of the presence of large viscous effects (see e.~g.\ \cite{Kurkela:2015qoa,Heller:2016gbp}). Therefore, viscous deviations from thermal equilibrium must be taken into account in calculations of thermal quantities during the early collision stage even when the QGP is already expanding as a fluid.

\begin{figure}[ht]
	\centering
	\includegraphics[width=1.\linewidth]{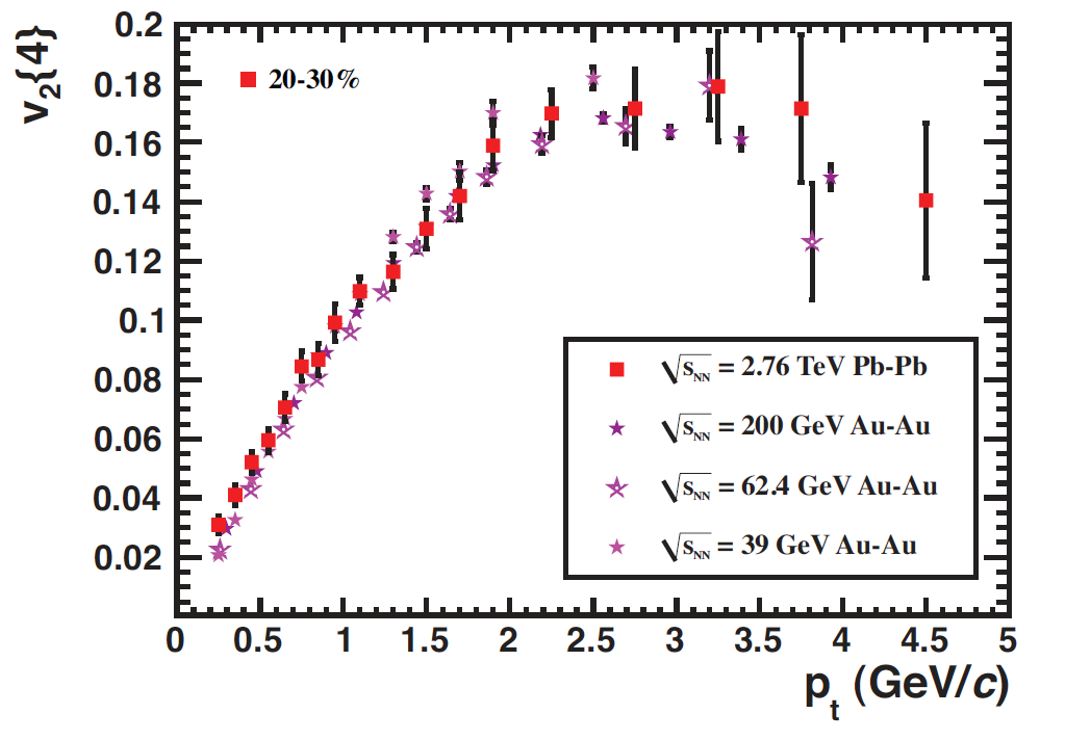}
	\caption{The $v_2(p_T)$ measured in $20-30\%$ central Au+Au (Pb+Pb) collisions over the collision energies \snn\ from 39 GeV to 2.76 TeV. The fact that the data all follow the same curve is indicative of elliptic flow that is driven by hydrodynamic expansion of a fireball with the initial geometric shape of the nuclear overlap associated with the impact parameter window. [From \cite{Kumar:2011de}]}
	\label{figures:v2_pT_sNN}
\end{figure}

Figure \ref{figures:v2_pT_sNN} indicates that the elliptic flow $v_2(p_T)$ of charged hadrons in Au+Au (Pb+Pb) collisions remains the same in a fixed centrality bin ($20-30\%$) over a large range of collision energies \snn\ from 39 GeV to 2.76 TeV. As Fig.~\ref{figures:eps-ini} suggests, the initial conditions of the fireball lie deep in the QGP regime over this energy range, and the colliding nuclei are sufficiently Lorentz contracted for the Bjorken model of a boost-invariant hydrodynamic expansion to be applicable at midrapidity. The observation that the $v_2(p_T)$ data all follow the same curve indicates that the elliptic flow is driven by the scale-invariant hydrodynamic expansion of a fireball whose initial geometric shape is the nuclear overlap region in the associated impact parameter window.

\begin{figure}[htb]
\includegraphics[width=0.85\linewidth]{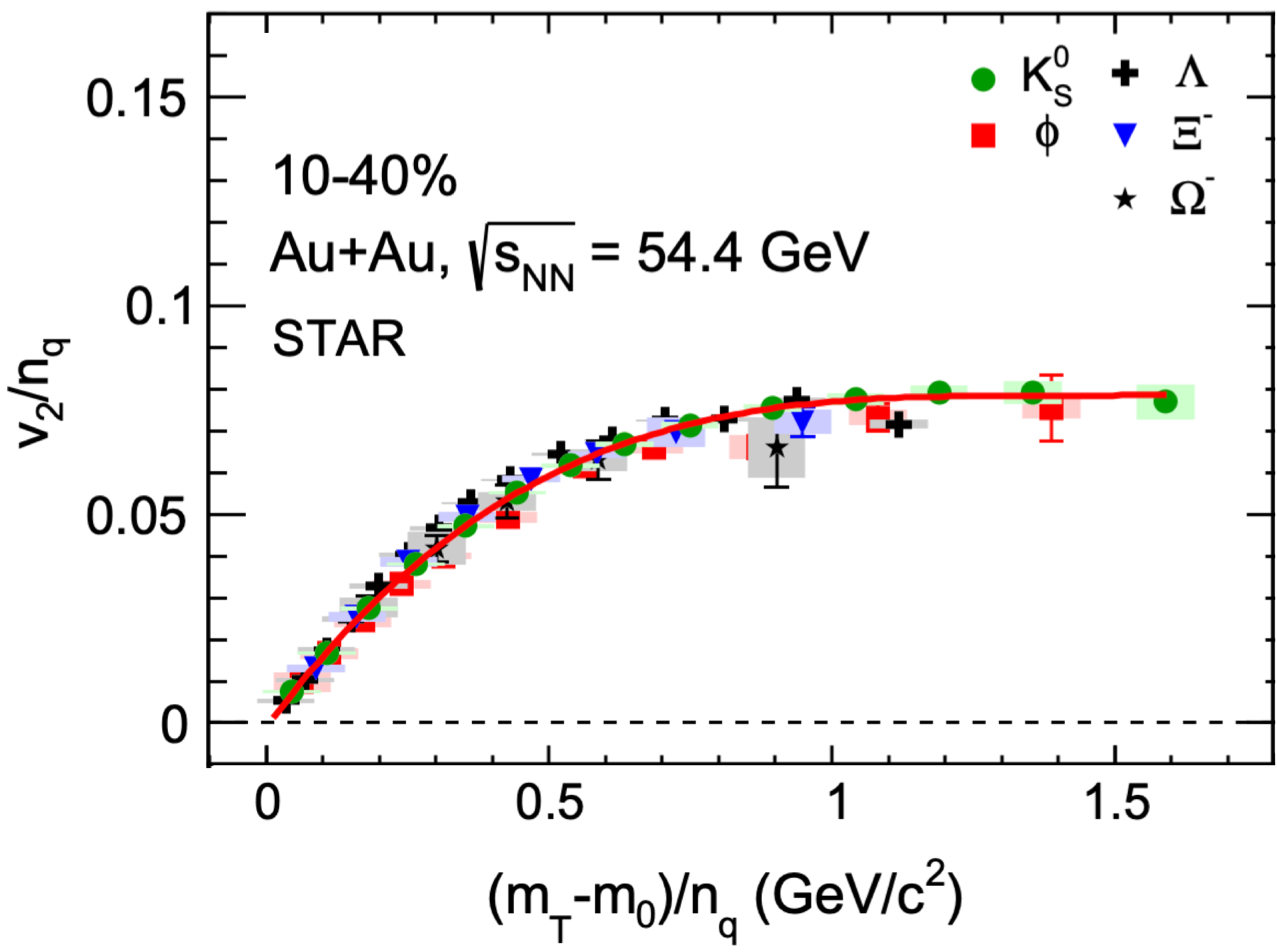}
\caption{Valence quark scaled elliptic flow coefficient $v_2/n_q$ for five different hadron species in $\sqrt{s_{\rm NN}} = 54.4$ GeV Au+Au collisions as a function of the scaling variable $(m_T-m_0)/n_q$. The solid red line indicates a fit to the $K^0_s$ data. [From \cite{STAR:2022tfp}]}
\label{fig:v2_scaled}
\end{figure}

While the strength of the observed elliptic flow of inclusive charged hadrons points to its early generation during the expansion phase, it does not directly indicate whether the flow is created at the (deconfined) quark level. This information comes from characteristic differences between the elliptic flow of mesons and baryons \cite{Fries:2003vb,Fries:2003kq}. If the flow is carried by the valence quarks of a hadron, the elliptic flow functions of different hadrons will satisfy the scaling law
\begin{equation}
v_2^{(i)}(p_T)/n_q^{(i)} = v_2^{\rm (q)}(p_T/n_q^{(i)}) \, ,
\end{equation}
where $n_q^{(i)}=2,3$ is the number of valence quarks of hadron species $i$, and $v_2^{\rm (q)}(p_T)$ is the elliptic flow function for quarks. Figure \ref{fig:v2_scaled} shows the valence quark scaled elliptic flow coefficient $v_2/n_q$ measured by STAR \cite{STAR:2022tfp} in $\sqrt{s_{\rm NN}} = 54.4$ GeV Au+Au collisions for five different hadron species containing strange quarks: the mesons $K_s^0, \phi$ and the baryons $\Lambda, \Xi^-, \Omega^-$. The flow coefficient $v_2$ is plotted as a function of the variable $(m_T-m_0)/n_q$, where $m_T = \sqrt{p_T^2 + m_0^2}$ is the transverse mass.\footnote{\bm{It is not well established whether the valence quark scaling works better in the variable $p_T$ or in $m_T$. It is noteworthy that there is no significant difference between the two scalings in the kinematic domain $p_T \gg m$ where the sudden recombination model applies.}} Similar results for $v_2, v_3, v_4$ have been obtained by ALICE in Pb+Pb collisions at LHC \cite{ALICE:2018yph}.

The $p_T$-integrated flow coefficients $v_n$ for $n \ge 2$ provide a good measure of the specific shear viscosity $\eta/s$, because the coefficients are increasingly sensitive to flow dissipation for growing values of $n$ \cite{Staig:2010pn}. These coefficients have been measured by several LHC experiments in Pb+Pb collisions at \snn\ = 5.02 TeV \cite{CMS:2017xnj,ATLAS:2018ezv,ALICE:2020sup}. The data are in good agreement with hybrid model calculations that use values $\eta/s \sim 0.1-0.2$ in the QGP phase.

\begin{figure}[ht]
	\centering
	\includegraphics[width=0.95\linewidth]{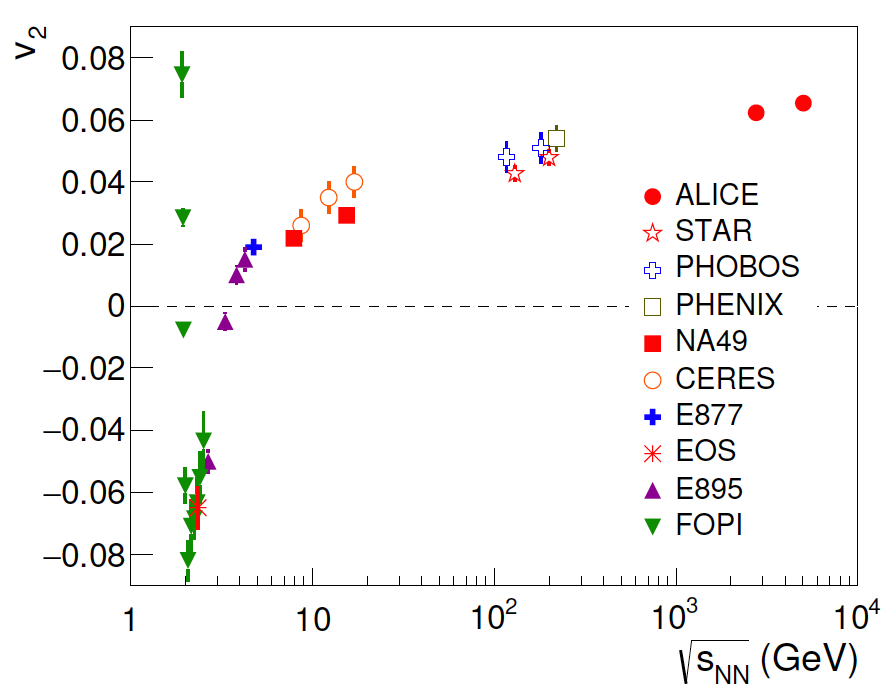}
	\caption{The $p_T$-integrated elliptic flow $v_2$ for Au+Au (Pb+Pb) collisions over the entire collision energy range covered by Au+Au (Pb+Pb) collisions at the GSI, AGS, SPS, RHIC, and LHC. See text for details. [From \cite{ALICE:2022wpn}]}
	\label{figures:v2_sNN}
\end{figure}

The collision energy dependence of $v_2$ of charged hadrons has been measured from \snn\ $\simeq$ 2 GeV to 5.02 TeV in Au+Au (Pb+Pb) collisions in experiments at GSI, AGS, SPS, RHIC, and LHC. The data collected in Fig.~\ref{figures:v2_sNN} show that the physical mechanism driving the elliptic flow changes for \snn\ $<$ 10 GeV. The slow increase of $v_2$ for \snn\ $>$ 10 GeV can be reconciled with the invariant behavior of $v_2(p_T)$ visible in Fig.~\ref{figures:v2_pT_sNN} \jh{and the gradual increase in the mean p$_T$ of the particle spectrum with increasing \snn\ .}

The numerical value for the QCD transport parameter $\eta/s$, \jh{the ratio of shear viscosity $\eta$ to entropy \bm{density} s of the system,} that can be extracted from the RHIC and LHC data has systematic uncertainties that derive from the need to simultaneously fix other parameters of the transport models, such as the initial energy density, the granularity of the density fluctuations, and the earliest time at which viscous hydrodynamics becomes a valid description. \jh{Fortunately certain observables exhibit primary sensitivity to a small number of model parameters. Specifically, the momentum anisotropy of emitted hadrons, characterized by the elliptic flow coefficient $v_2$, is most sensitive to the value of the shear viscosity $\eta$. The mean transverse momentum $\langle p_T \rangle$ is most sensitive to the equilibrium equation of state and the bulk viscosity $\zeta$. Assuming knowledge of the equation of state, obtained from lattice calculations, an estimate of the temperature dependent bulk viscosity can then be derived from the data.} 

\begin{figure}[ht]
\centering
\includegraphics[width=0.95\linewidth]{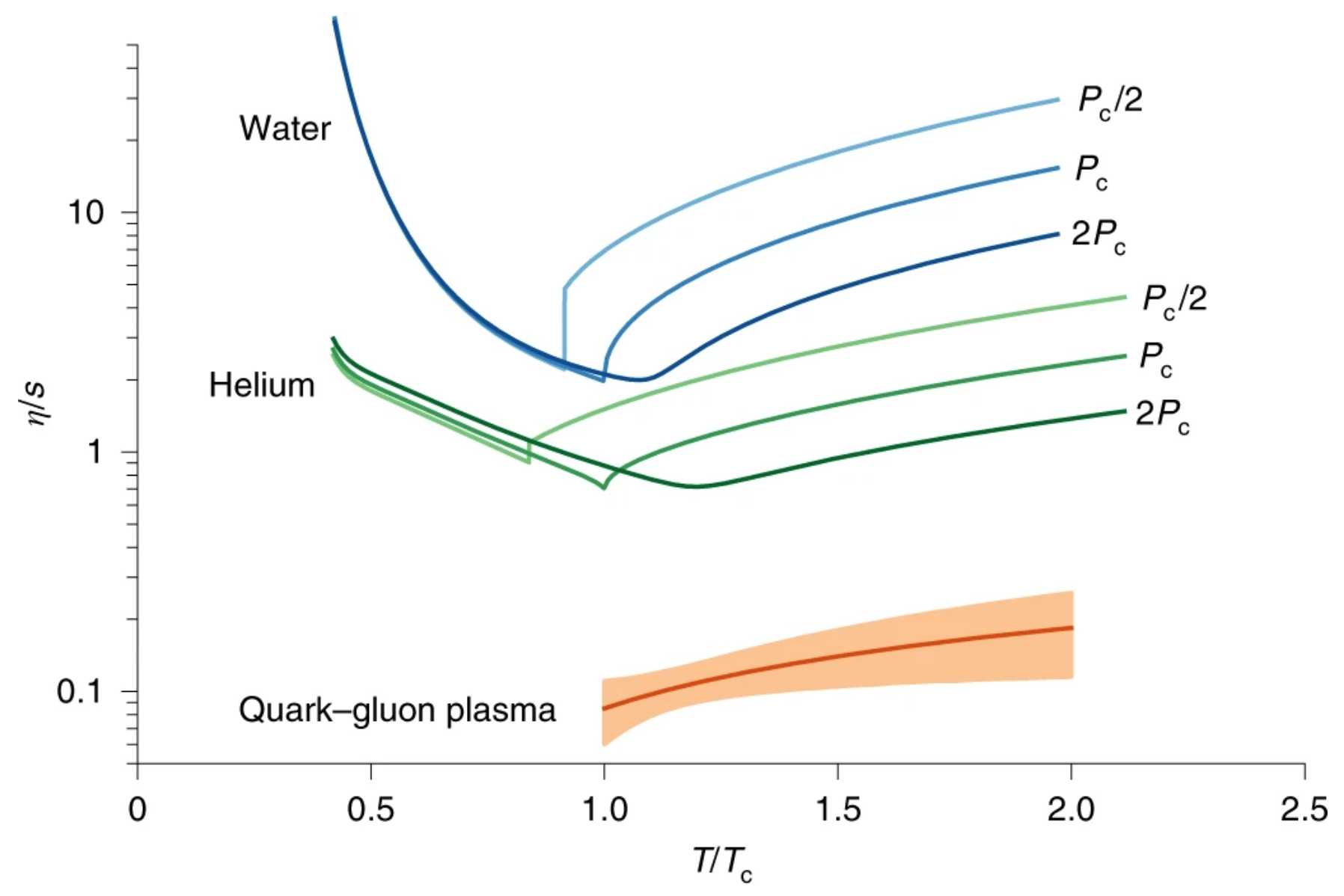}
\caption{Comparison of the specific shear viscosity $\eta/s$ of the QGP extracted from heavy-ion collision data with the values measured for helium and water. [From \cite{Bernhard:2019bmu}]}
\label{figures:etas_QGP_He_H2O}
\end{figure}

Comprehensive model-data analyses using Bayesian methodology that take many of these uncertainties into account have been conducted in recent years. \jh{Results of a recent analysis \cite{Bernhard:2019bmu} that allows for a temperature-dependent extraction of the specific shear viscosity $\eta/s$ is displayed in Fig.~\ref{figures:etas_QGP_He_H2O}. The red curve represents the most probable value for a QGP and the orange area covers the 90\% likely region. The figure indicates clearly that the $\eta/s$ of the QGP is approximately one and two orders of magnitude lower than that of helium and water, respectively.}

The importance of this result derives from the insight that values of $\eta/s \sim 0.1-0.2$ require the QGP to be a strongly coupled fluid \cite{Gyulassy:2004zy,Shuryak:2004kh}. In fact, this value establishes an exceptional role of the QGP as a nearly ``perfect'' fluid (see Fig.~\ref{figures:etas_QGP_He_H2O} for a comparison with other ``good'' fluids) with a sound dissipation coefficient that is near the quantum bound $(4\eta/3+\zeta)/s = (3\pi)^{-1}$ \cite{Kovtun:2004de}.\\

The information with respect to the initial azimuthal shape of the fireball that is gleaned from the collective flow measurements can be used to study the pathlength dependence of parton energy loss by measuring properties of the jet as a function of its angle relative to the flow anisotropy. Jets that are emitted along the major axis of the initial elliptic shape, created from the geometrical overlap of the colliding nuclei, must traverse a longer distance through the QGP and lose more energy than those emitted along the minor axis. Radiative energy loss of partons is predicted to grow quadratically with the pathlength \cite{Baier:1996kr}, whereas collisional energy loss would depend linearly on the pathlength \cite{Baier:1996sk}. Measurements of the azimuthal anisotropy of the jet yield relative to the event plane can thus provide information on the mechanism by which partons lose energy. Such studies have been implemented using event shape engineering methods \cite{Beattie:2023mcz} to have better control of the initial geometrical event shapes for more precise pathlength determination. Results to date are consistent with the assumption of a dominance of radiative energy loss for light partons.

\begin{figure}[ht]
\centering
\includegraphics[width=0.92\linewidth]{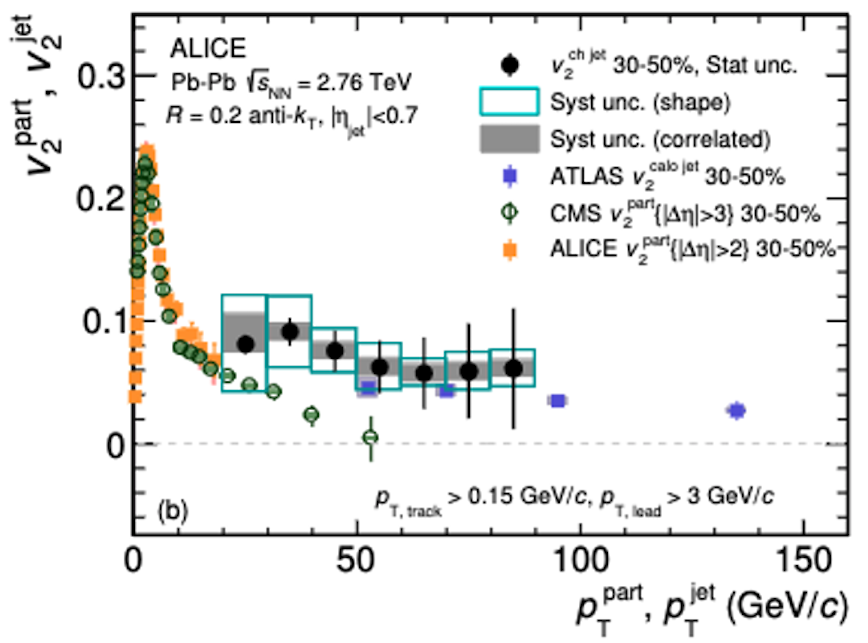}
\caption{The jet $v_2^{\rm jet}$ and particle $v_2^{\rm part}$ in 30-50\% central \snn\ = 2.76 TeV Pb+Pb collisions as a function of $p_T^{\rm jet}$ and $p_T^{\rm part}$, respectively. The particle $v_2^{\rm part}$ \cite{CMS:2012tqw, ALICE:2015efi}, charged jet $v_2^{\rm ch jet}$ \cite{ALICE:2015efi} and calorimetric jet $v_2^{\rm calo jet}$ \cite{ATLAS:2021ktw} results are shown. See legend for more details. [Compilation from \cite{ALICE:2015efi}]}
\label{figures:ALICE_Jet_v2_central.png}
\end{figure}

Analogous to the azimuthal correlation measurements of soft particles in an event, the azimuthal anisotropy of jets $v_n^{\rm jet}$  can be measured with respect to the second harmonic event plane, after separating jets from the underlying event background. Displayed in Fig.~\ref{figures:ALICE_Jet_v2_central.png} is a compilation \cite{ALICE:2015efi} of results on the jet $v_2^{\rm jet}$ and particle $v_2^{\rm part}$ for semi-central collisions. The ATLAS calorimetric jet $v_2^{\rm calo jet}$ and ALICE charged jet $v_2^{\rm ch jet}$ are consistent with each other and exhibit a significant $v_2^{\rm jet}$ up to large $p_T$.\footnote{Note that for any initial parton $p_T$, the particle $p_T$ will be less than that of a charged jet and it is likewise less than that for a calorimetric jet due to the missing initial parton energy in the particles and charged jets.} Also shown are the $v_2^{\rm part}$ of charged particles for comparison. The ALICE and ATLAS $v_2^{\rm jet}$ measurements are indicative of pathlength-dependent parton energy loss. A recent measurement by CMS \cite{CMS:2023lgq} in pp collisions of the two-particle $v_2$ (defined with respect to the jet axis) of constituent particles within a jet as a function of \Npart ~is consistent with theoretical model predictions up to \Npart\ $\sim 80$. There is an intriguing increase in the two-particle $v_2$ within jets for \Npart\ from 80 to 110 suggesting possible collective or other effects within jets in high-\Npart\ pp collisions. If this is, indeed, an effect of collective particle transport, it could alter our overall view of the emergence of collectivity in AA collisions.

While the elliptic and higher flow harmonics are especially sensitive to the shear viscosity of the QGP, the radially symmetric component of the transverse flow, as expressed in the mean transverse momentum $\langle p_T \rangle$, is primarily sensitive to the speed of sound of the QGP and the bulk viscosity. The bulk viscosity is thought to be most important at lower collision energies, where the initial temperature does not much exceed $T_c$, the speed of sound is predicted to dominate at higher collision energies. One way to avoid complications from changes in the overall geometry is to measure the $\langle p_T \rangle$ in ultra-central collisions, where variations of the particle multiplicity are thought to be indicative of different values of the initial temperature $T_{\rm ini}$ caused by quantum fluctuations in the energy deposition within the same maximal nuclear cross section. The proposal \cite{Gardim:2019xjs,Gardim:2019brr} is to measure the speed of sound $c_s$ via the thermodynamic relation 
\begin{equation}
c_s^2 = \frac{dP}{d\varepsilon} 
= \frac{d\ln T}{d\ln s} 
=\frac{d\ln\langle p_T \rangle}{d\ln N_{\rm ch}},
\end{equation}
where the mean $p_T$ is used as a proxy for the average temperature and the charged multiplicity $N_{\rm ch}$ as proxy for the average entropy density. For comparison, the speed of sound $c_s(T)$ can be easily calculated by lattice-QCD. Recent preliminary results from CMS \cite{CMS:2023byu} in ultra-central Pb+Pb collisions at LHC yield a value $c_s^2/c^2 = 0.241 \pm 0.002 \pm 0.016$ at the average temperature $T = 219 \pm 8$~MeV, which is in excellent agreement with lattice-QCD results. It will be interesting to see whether the extracted value for collisions at RHIC energies is lower as predicted by lattice-QCD.

An attribute of the QGP fluid that was not anticipated at the time of our 1996 review \cite{Harris:1996zx} is vorticity. Because of the very low specific shear viscosity of the QGP any vorticity that is seeded into the fluid at early times can survive for an extended period of time as Kelvin's theorem states that circulation is strictly conserved in an ideal fluid. The seeding of vorticity in non-central heavy-ion collisions was first recognized in \cite{Liang:2004ph} where also global hyperon polarization with respect to the collision plane was identified as an experimental signature. Global $\Lambda$-hyperon polarization in the percent range was subsequently observed in Au+Au collisions at \snn\ $= 7-200$ GeV \cite{STAR:2017ckg}. The magnitude of the polarization can be related to the average vorticity of the QGP at the moment of hadronization and gives an average value $|\vec\omega| = (9\pm 1)\times 10^{-21}\, {\rm s}^{-1}$ for Au+Au collisions within the energy range studied in \cite{STAR:2017ckg}. The observed magnitude can be explained as the transfer of vorticity into the QGP from the initial orbital angular momentum of the colliding nuclei that results in a spin polarization of the QGP fluid \cite{Karpenko:2016jyx,Florkowski:2017ruc}. The detailed vorticity pattern of the QGP fluid and the microscopic mechanisms of spin transfer into the QGP and its equilibration are areas of active research. In addition to spin polarization of hyperons, STAR and ALICE have also reported a nonzero spin alignment of several vector mesons ($K^*$, $\phi$) \cite{STAR:2022fan,Kundu:2021lra} the origins of which are not yet well understood.

\section{Equation of State}
\label{sec:EOS}

Interest in the equation of state of nuclear matter was the primary motivation for our field of research and the inception of experiments utilizing collisions of energetic heavy ions \cite{Chapline:1973kkq}. After initial studies of baryon-rich nuclear matter in the GeV range \cite{Harris:1984up}, the interest became focused on understanding the equation of state of excited QCD matter, which was originally a centerpiece of the RHIC experimental program as exemplified by the panel in Fig.~\ref{figures:Signatures1996} entitled  ``Temperature.''  \bm{This interest has waned somewhat in recent years as the definitive results of lattice gauge theory for zero to moderate net baryon densities (see \cite{Ratti:2022qgf} for a recent comprehensive review) have been shown to be in good agreement with hydrodynamic simulations for many observables \cite{Moreland:2015dvc}.}\footnote{\bm{This statement does not apply to the equation of state at high net baryon density, which is relevant to astrophysical phenomena such as neutron stars and neutron star mergers, where reliable lattice-QCD simulations are still unavailable (see e.~g.\ \cite{Annala:2019puf}). In this domain the equation of state still remains a primary focus of research (see also below).}}
Instead of the equation of state, the analyses of relativistic heavy ion collision data have recently focused mainly on dynamical properties, such as the effects of viscosity on the collective flow.

Experimental interest in the equation of state of nuclear matter has now shifted back to much lower collision energies, in the few-GeV range as explored in the second RHIC beam energy scan, where the net baryon density of the matter created is above that covered by reliable lattice calculations. A main focus of this investigation is to determine whether the smooth crossover between hadronic matter and QGP at low net baryon density gives way to a first-order phase transition with a threshold critical point. The primary experimental probes for a first- or second-order phase transition are large-scale spinodal density fluctuations and critical net-baryon number fluctuations, respectively. Hints of such critical behavior were observed in net-proton number fluctuations in the first RHIC beam energy scan \cite{STAR:2013gus} but require confirmation with much higher statistics data \cite{Liu:2022wme}.

The recent detection of gravitational waves from binary neutron-star mergers \cite{LIGOScientific:2017ync,LIGOScientific:2017zic} has sparked interest in connecting the equation of state governing the collapse of binary-neutron star systems to the equation of state of matter probed in heavy-ion collisions in the few-GeV energy range \cite{Oechslin:2004yj,Huth:2021bsp}. The shape of the gravitational wave signal is expected to be sensitive to the degrees of freedom in the core of neutron stars.
Calculations are currently focused on exploring connections to the dynamical evolution of few-GeV heavy-ion collisions in terms of the pressure, temperature, entropy, and isospin \cite{Most:2018eaw,Bauswein:2018bma,Tan:2020ics,Most:2022wgo}. A first-order phase transition to quark matter is expected to look very different than a smooth crossover, and the next generation of gravitational wave observatories may be able to distinguish between the two. Furthermore, the lowest energy probes in the second RHIC beam energy scan and the future Compressed Baryonic Matter (CBM) experiment at FAIR \cite{CBM:2016kpk} are expected to provide the data necessary for a quantitative comparison with neutron-star merger observations.

\section{Small Systems}

The motivation for colliding ultra-relativistic heavy ions at RHIC and the LHC was that at such high energies large nuclei would be most likely to create hot QCD matter in the thermodynamic limit. Notwithstanding this argument, there was also an old idea that even high-energy proton-proton collisions could produce a statistical system that might exhibit aspects of hydrodynamic behavior \cite{Fermi:1950jd,Belenkij:1955pgn,Hagedorn:1965st}. After the advent of QCD, the question remained as to whether a statistical system composed of locally deconfined quarks and gluons could be produced in sufficiently energetic p+p collisions and behave as a hot fluid, i.e. a QGP. However, attempts to find evidence for QGP formation in high-multiplicity p+$\bar{\rm p}$ collisions at the TEVATRON remained inconclusive \cite{E735:1994hbl}. 

The general consensus remains that minimum-bias p+p collisions do not involve the formation of a QGP, and such events are commonly used as a baseline against which nuclear modifications of hard probes are measured. This does not rule out that a QGP fireball can be produced in rare high-multiplicity p+p events. The first clear evidence for behavior that resembles a collective flow pattern was observed by CMS in p+p events at \snn\ = 7 TeV with more than 90 charged tracks \cite{CMS:2010ifv}. Angular correlation measurements at \snn\ = 2.76 and 13 TeV \cite{ATLAS:2015hzw,CMS:2015fgy} confirmed this observation. Similar observations of collective flow patterns have been made for p+Pb collisions at LHC \cite{ALICE:2012eyl,ATLAS:2012cix,CMS:2012qk,CMS:2015yux} and in p+Au, d+Au, and $^3$He+Au collisions at RHIC \cite{PHENIX:2018lia} (see \cite{Nagle:2018nvi} for a review). The similarity of the collective behavior seen in p+p, p+A, and A+A systems can be explained if a strongly coupled QGP is formed in all these systems \cite{Weller:2017tsr}.

Surprisingly, on the other hand, no evidence has been found for the formation of a QGP in p+Pb collisions at \snn\ = 5.02 TeV in modifications of hard probes, such as jets \cite{ALICE:2017svf}. It is presently unclear how the finding of apparent collectivity in soft particle emission can be reconciled with the absence of evidence for jet quenching. One possibility is that the soft collective behavior observed in p+p and p+A collisions is generated without hydrodynamic flow (see e.~g.\ \cite{Bierlich:2017vhg}). It is well known in other fields, e.~g.\ plasma physics, that collective motion of particles can be created by non-hydrodynamical mechanisms, such as the action of coherent fields \cite{Davidson:2016mhd,Hernandez:2017mch}. If the origin of collective behavior in p+p and p+A collisions were found to have an alternative explanation, our current understanding of the origin of flow patterns in A+A collisions would have to be revisited.

\section{Summary and Outlook}

Nearly three decades of experimental and theoretical research have affirmed the scientific strategy aimed at the discovery and characterization of the quark-gluon plasma that was described in \cite{Harris:1996zx}. Extensive measurements have converted the qualitative expectations for the quark-gluon plasma signatures summarized in Fig.~\ref{figures:Signatures1996} into quantitative knowledge. As with any preconceived strategy, adjustments were made in reaction to new insights gathered along the way. Some signatures have been found to be less useful or more difficult to measure than originally thought. Others have proven to be immensely valuable including several that were unanticipated or some that were known in principle but underappreciated.

The average initial energy density reached in the most central heavy-ion collisions in Fig.~\ref{figures:eps-ini} exceeds the threshold for QGP formation above \snn\ $\sim$ 10 GeV. In the high energy range, \snn\ $>$ 50 GeV, this can be deduced from the measured charged-particle multiplicity $dN_{\rm ch}/dy$ and the short hydrodynamization time deduced from elliptic flow. At lower energies, it requires some assumptions about the dynamics of energy deposition, which is no longer quasi-instantaneous. The argument here is based in part on the continuity of the valence quark number scaling of elliptic flow that is observed down to \snn\ $=$ 11.5 GeV, although increasing deviations from the scaling show up for \snn\ $<$ 39 GeV indicating a growing contribution to flow from the hadronic phase \cite{Shi:2013jha,Parfenov:2020fuo}.

Identical particle (HBT) interferometry has revealed that a fireball of nuclear size and a lifetime of $4-10$ fm/c\jh{, depending on the collision energy,} acts as the common source of the hadrons that are emitted. As already mentioned above, the composition of the emitted hadrons and the fluctuations of conserved quantities have been used to map the chemical properties of the hadronizing fireball. Future experiments with extended pseudorapidity coverage will allow balance functions of conserved quantities to reach farther back into the history of the evolution of the fireball and track when chemical equilibrium is first established.

The intense investigation of the collective flow patterns in experiments has made it possible to quantitatively determine fluid properties of the QGP. The specific shear viscosity of the QGP has been found to lie in the range $0.05 < \eta/s < 0.2$ depending on $T/T_c$, establishing this novel QGP state of matter as the most ``perfect'' fluid known. Furthermore, the valence quark scaling of the flow pattern has provided strong evidence that the collective flow is generated at the quark level in a fluid in which quarks are not confined as hadrons. The spin polarization of hyperons adds a new dimension to the exploration of the flow pattern by its sensitivity to the vorticity and thermal shear of the fluid. In the future, more precise measurements of the interaction of heavy quarks with this fluid will further probe the strongly-coupled nature of the QGP by yielding quantitative determinations of its diffusion constants.

Among soft signatures, the enhancement of strange hadron production \jh{with respect to pp collisions} and, more generally, the complete chemical equilibration of all light hadron species at common thermodynamic conditions have provided strong evidence for the transition from hadronic matter to a deconfined state -- the QGP -- at a temperature $T_c \approx 155$ MeV, in excellent agreement with lattice-QCD simulations. As shown in Fig.\ref{figures:PhaseBoundary}, the boundary between hadronic matter and the QGP has been mapped by two different methods over a range of baryon chemical potentials $\mu_B$ up to at least 300 MeV and agrees well with expectations from lattice gauge theory. 

The measured suppression pattern of heavy quarkonium states, especially the \Upsi\ states, and their observed sequential melting provide further confirmation for the deconfinement of quarks and gluons in the QGP, but the mechanisms responsible for the suppression pattern are more complex than originally thought. In particular, the reduced suppression, by regeneration at the phase boundary, of the \Jpsi\ in A+A collisions at LHC compared to that at RHIC energies provides clear evidence that charm quarks are deconfined in the QGP.

Electromagnetically interacting and hard QCD signatures provide complementary information about the properties of the QGP. Measurements of the spectrum of direct photons and the invariant mass spectrum of dileptons have yielded lower bounds for the temperature at which the QGP initially thermalizes. These spectra exhibit thermal temperatures substantially above the transition temperature $T_c$. The spectrum of dileptons in the mass region of the $\rho$-meson confirm the hadronization (chemical freeze-out) temperature deduced from the hadron yields. 

An unambiguous detection of chiral symmetry restoration will require high-precision measurements of the lepton pair spectrum in the mass region 1 GeV $< M_{\ell^+\ell^-} < 2$ GeV. Theoretical predictions indicate a difference of approximately 15\% between models that involve chiral symmetry restoration in the QGP phase and models that do not. Measurements of this level of precision require very precise knowledge of the background from semi-leptonic charm decays and are out of reach for the existing detectors. The proposed ALICE 3 \cite{ALICE:2022wwr} and NA60+ \cite{NA60:2022sze} experiments aim at reaching the required precision to be able to detect the enhancement of the dielectron spectrum at invariant masses above the $\phi$-meson peak characteristic of $\rho-a_1$ mixing that is the signature of chiral symmetry restoration.

The most versatile, but also the most complex probes of the QGP are energetic quarks and gluons, created by hard scatterings during the first moments of the nuclear collision. Such hard-scattered partons materialize as jets, in which the initial momentum of the primary parton is shared among many hadrons. A number of different observables have been found that encode the energy loss of the primary parton on its path through the QGP, beginning with the suppression of the inclusive yield of high p$_T$ hadrons in A+A collisions observed from the mid-range of RHIC energies to those of the LHC and corroborated by the observation of a strong suppression of the high-$p_T$ hadrons opposite in azimuth to a high-$p_T$ trigger hadron.

These measurements involving individual hadrons were subsequently extended to jets and di-jets, where a similar quenching of jets attributable to parton energy loss was observed. More recently, differential measurements of jets and their substructure have emerged as tools to investigate the mechanism that causes parton energy loss and help determine the conditions under which energy loss is primarily radiative or when elastic processes dominate. In parallel, flavor tagging of jets has given evidence for a mass and color charge dependence of the parton energy loss in the QGP.

According to our current understanding, the energy loss of the primary parton and the redistribution of its momentum within the jet is controlled by just a few parameters characterizing the medium. In a dilute or thin medium, they are the density of scattering centers and the range of the color force in the medium. In a dense, thick medium, the jet quenching parameter $\hat{q}$ encodes the transverse scattering power per unit length of the medium.  The suppression factor \RAA\ of inclusive hadrons provides a direct measurement of $\hat{q}$ under the assumption that the energy loss of the primary parton is predominantly caused by gluon radiation induced by scattering in the medium. The dimensionless parameter $\hat{q}/T^3$ is found to lie in the ($\pm 1\sigma$) range $3.4 < \hat{q}/T^3 < 5.8$ at RHIC and $2.4 < \hat{q}/T^3 < 5.0$ at LHC \cite{JETSCAPE:2021ehl}, which is consistent with values for $\hat{q}/T^3$ required to describe the inclusive jet suppression measured at RHIC and LHC.

The values of $\eta/s$ and $\hat{q}/T^3$ deduced from the heavy ion data by Bayesian model-data comparison are two examples where experimental data have helped bracket fundamental transport coefficients of the QGP that cannot (yet) be reliably calculated in QCD. A fundamental question that is still to be resolved, is to what extent it is possible to probe the dynamical evolution of the matter created in heavy-ion collisions from partons in the initial state to the thermal quarks and gluons of the QGP and, finally, into hadrons. This quest involves the investigation and understanding of the parton structure of the initial state, of the energy sharing mechanisms that produce a thermal plasma, and the response of the QGP to hard probes that are sensitive to a range of different scales.

Future measurements with better resolution and higher statistics will probe more deeply to reveal the various scales involved in the interactions of jets with the QGP. Investigation of coherence effects, both theoretically and through jet substructure measurements, will determine the extent to which the medium is able to resolve the interactions of the parton as it propagates through the QGP. By constraining the dependence on the color charge and mass of the parton they can further confirm the scattering dynamics underpinning parton energy loss. At the same time, these differential measurements become effective probes of the shower evolution inside a jet and contribute to our understanding of QCD. 

Over the next few years, the new sPHENIX detector \cite{PHENIX:2015siv} at RHIC and the existing RHIC and LHC experiments with upgraded detectors will make precision measurements of jet modifications in heavy-ion collisions. In the future, a newly proposed ALICE 3 \cite{ALICE:2022wwr}
experiment is expected to join in that endeavor at the LHC. Parallel advances in the theory of jet interactions with the QGP medium will be required to turn the wealth of expected data into firm insights into the structure and properties of the QGP and the internal dynamics of jet formation. The remarkable success achieved for soft QGP probes, where data--theory comparisons within well-defined frameworks have enabled quantitative measurements of QGP bulk properties, can serve as a guide for the scientific approach aimed at elucidating the microscopic structure of the QGP over the wider range of scales that is accessible with hard QCD probes.

Another increasingly central direction of investigation is research into the parton structure of cold nuclear matter. A better understanding of the structure of the colliding nuclei is important as one attempts to understand the initial conditions of a high-energy collision of nuclei. An example of such investigations is the monitoring of sub-nucleonic proton shape fluctuations by studying \Jpsi\ production in diffractive e+p collisions \cite{Mantysaari:2022ffw}. Alternative experimental approaches utilize \Jpsi\ photo-production in ultra-peripheral d+Au collisions \cite{STAR:2021wwq} and coherent \Jpsi\ production in ultra-peripheral Pb+Pb collisions \cite{ALICE:2019tqa}. 

Understanding the interaction of cold nuclear matter with hard probes is also an essential aspect in the interpretation of the nuclear modification factor \RAA\ as already discussed in conjunction with the physics of quarkonium suppression and jet quenching. Phenomena that will benefit from additional experimental investigations in p+A collisions include nuclear suppression or enhancement effects at relatively low $p_T$ that are alternatively attributed to shadowing of nuclear parton distributions, momentum broadening of incident partons, or final-state absorption.

In the more distant future precision studies of the parton structure of nucleons and complex nuclei will be the scientific focus of the electron-ion collider (EIC) \cite{Accardi:2012qut}. Generalized parton distributions and transverse momentum dependent parton distributions will ve used to map the transverse parton structure of the proton, while diffractive e+p and e+A collisions will provide precise quantitative constraints on the saturation of gluon distributions at small Bjorken-$x$. Besides being valuable in their own right, these results will help reduce the model dependence of the initial state of relativistic heavy-ion collisions.

In conclusion, the strategy for the investigation of hot QCD matter outlined in \cite{Harris:1996zx} has been successful beyond expectations. As in any field of physics, experimental and theoretical progress have gone hand-in-hand, leading to changes in research emphasis and readjustments of the strategy. Many questions at the core of the initial RHIC research program have been answered and given way to new ones \cite{Busza:2018rrf}. Among those most important are the following. How does the partonic microscopic structure of the QGP evolve into a ``perfect'' fluid at longer distance scales? How small can a QGP that behaves fluid-like be? What is the structure of the QCD phase diagram at high net baryon density? We can be optimistic that improved experimental techniques, supported by theoretical advances, and combined with creative and novel approaches will provide information over the next decade that will help answer these questions.

\bmhead{Acknowledgments}

We thank Roberta Arnaldi, Steffen Bass, Hannah Bossi, Helen Caines, Charles Gale, Marek Gazdzicki, Laura Havener, Joseph Kapusta, Raghav Kunnawalkam Elayavalli, Andras Laszlo, Yen-Jie Lee, Michael Lisa,  Rongrong Ma, Ian Moult, Jean-Fran{\c c}ois Paquet, Ralf Rapp, Lijuan Ruan, Mike Sas, J{\"u}rgen Schukraft, Enrico Scomparin, Alba Soto-Ontoso, and Willam Zajc for valuable input during the writing of this article. 
We especially thank Hannah Bossi for assistance in various aspects of the preparation of figures for this manuscript.

We appreciate helpful comments on a draft version of the manuscript made by Yasuyuki Akiba, Frank Geurts, Peter Jacobs, Georgios Konstantinos Krintiras, Krishna Rajagopal, Lijuan Ruan,  Bj{\"o}rn Schenke, J{\"u}rgen Schukraft, Andre St{\aa}hl, Marco Van Leeuwen, and Urs Wiedemann.

We are indebted to the ALICE, ATLAS, CMS, PHENIX and STAR collaborations for their extensive experimental results.

We acknowledge support from the Office of Science of the U.S. Department of Energy, JH from grant DE-SC004168 and BM from grant DE-FG02-05ER41367. BM also acknowledges support by Yale University during Spring 2022 and Spring 2023.


\begin{thebibliography}{999}

\bibitem{Harris:1996zx}
J.~W.~Harris and B.~M\"uller,
``The Search for the quark - gluon plasma,''
Ann. Rev. Nucl. Part. Sci. \textbf{46}, 71 (1996)
[arXiv:hep-ph/9602235 [hep-ph]].

\bibitem{Aprahamian:2015qub}
A.~Aprahamian, A.~Robert, H.~Caines, G.~Cates, J.~A.~Cizewski, V.~Cirigliano, D.~J.~Dean, A.~Deshpande, R.~Ent and F.~Fahey, \textit{et al.}
``Reaching for the horizon: The 2015 Long Range Plan for Nuclear Science,''
[\url{https://science.osti.gov/-/media/np/nsac/pdf/2015LRP/2015_LRPNS_091815.pdf}] 

\bibitem{BRAHMS:2004adc}
I.~Arsene \textit{et al.} [BRAHMS],
``Quark gluon plasma and color glass condensate at RHIC? The Perspective from the BRAHMS experiment,''
Nucl. Phys. A \textbf{757}, 1-27 (2005)
[arXiv:nucl-ex/0410020 [nucl-ex]].

\bibitem{PHOBOS:2004zne}
B.~B.~Back \textit{et al.} [PHOBOS],
``The PHOBOS perspective on discoveries at RHIC,''
Nucl. Phys. A \textbf{757}, 28-101 (2005)
[arXiv:nucl-ex/0410022 [nucl-ex]].

\bibitem{STAR:2005gfr}
J.~Adams \textit{et al.} [STAR],
``Experimental and theoretical challenges in the search for the quark gluon plasma: The STAR Collaboration's critical assessment of the evidence from RHIC collisions,''
Nucl. Phys. A \textbf{757}, 102-183 (2005)
[arXiv:nucl-ex/0501009 [nucl-ex]].

\bibitem{PHENIX:2004vcz}
K.~Adcox \textit{et al.} [PHENIX],
``Formation of dense partonic matter in relativistic nucleus-nucleus collisions at RHIC: Experimental evaluation by the PHENIX collaboration,''
Nucl. Phys. A \textbf{757}, 184-283 (2005)
[arXiv:nucl-ex/0410003 [nucl-ex]].

\bibitem{ALICE:2022wpn}
ALICE Collaboration, ``The ALICE experiment--A journey through QCD,''
[ALICE], CERN-EP-2022-227
[arXiv:2211.04384 [nucl-ex]].

\bibitem{Karsch:1980mg}
F.~Karsch and H.~Satz,
``On the Thermodynamics of Confined Quarks,''
Phys. Rev. D \textbf{22}, 480 (1980).

\bibitem{Satz:1985vb}
H.~Satz,
``The Transition From Hadron Matter to Quark - Gluon Plasma,''
Ann. Rev. Nucl. Part. Sci. \textbf{35}, 245 (1985).

\bibitem{Borsanyi:2010cj}
S.~Borsanyi, G.~Endrodi, Z.~Fodor, A.~Jakovac, S.~D.~Katz, S.~Krieg, C.~Ratti and K.~K.~Szabo,
``The QCD equation of state with dynamical quarks,'' 
JHEP \textbf{11}, 077 (2010)
[arXiv:1007.2580 [hep-lat]].

\bibitem{HotQCD:2018pds}
A.~Bazavov \textit{et al.} [HotQCD],
``Chiral crossover in QCD at zero and non-zero chemical potentials,''
Phys. Lett. B \textbf{795}, 15-21 (2019)
[arXiv:1812.08235 [hep-lat]].

\bibitem{Bjorken:1982qr}
J.~D.~Bjorken, ``Highly Relativistic Nucleus-Nucleus Collisions: The Central Rapidity Region,'' Phys. Rev. D \textbf{27}, 140 (1983)

\bibitem{Busza:2018rrf}
W.~Busza, K.~Rajagopal and W.~van der Schee,
``Heavy Ion Collisions: The Big Picture, and the Big Questions,''
Ann. Rev. Nucl. Part. Sci. \textbf{68}, 339-376 (2018)
[arXiv:1802.04801 [hep-ph]].

\bibitem{Muller:2005en}
B.~M\"uller and K.~Rajagopal, ``From entropy and jet quenching to deconfinement?,''
Eur. Phys. J. C \textbf{43}, 15 (2005)
[arXiv:hep-ph/0502174 [hep-ph]].

\bibitem{PHENIX:2015tbb}
A.~Adare \textit{et al.} [PHENIX],
``Transverse energy production and charged-particle multiplicity at midrapidity in various systems from $\sqrt{s_{NN}}=7.7$ to 200 GeV,''
Phys. Rev. C \textbf{93}, 024901 (2016)
[arXiv:1509.06727 [nucl-ex]].

\bibitem{Andronic:2021dkw}
A.~Andronic, P.~Braun-Munzinger, K.~Redlich and J.~Stachel,
``Hadron yields in central nucleus-nucleus collisions, the statistical hadronization model and the QCD phase diagram,''
[arXiv:2101.05747 [nucl-th]].

\bibitem{ALICE:2016igk}
J.~Adam \textit{et al.} [ALICE],
``Measurement of transverse energy at midrapidity in Pb+Pb collisions at $\sqrt{s_{\rm NN}} = 2.76$ TeV,''
Phys. Rev. C \textbf{94}, 034903 (2016)
[arXiv:1603.04775 [nucl-ex]].

\bibitem{Song:2007ux}
H.~Song and U.~W.~Heinz,
``Causal viscous hydrodynamics in 2+1 dimensions for relativistic heavy-ion collisions,''
Phys. Rev. C \textbf{77}, 064901 (2008)
[arXiv:0712.3715 [nucl-th]].

\bibitem{STAR:2004moz}
J.~Adams \textit{et al.} [STAR],
``Measurements of transverse energy distributions in Au + Au collisions at 
$\sqrt{s_{\rm NN}} =$ 200 GeV,''
Phys. Rev. C \textbf{70}, 054907 (2004)
[arXiv:nucl-ex/0407003 [nucl-ex]].

\bibitem{PHENIX:2004vdg}
S.~S.~Adler \textit{et al.} [PHENIX],
``Systematic studies of the centrality and \snn\ dependence of the $dE(T)/d\eta$ and $d(N(ch)/d\eta$ in heavy ion collisions at mid-rapidity,''
Phys. Rev. C \textbf{71}, 034908 (2005)
[erratum: Phys. Rev. C \textbf{71}, 049901 (2005)]
[arXiv:nucl-ex/0409015 [nucl-ex]].

\bibitem{PHENIX:2013ehw}
S.~S.~Adler \textit{et al.} [PHENIX],
``Transverse-energy distributions at midrapidity in p+p , d+Au , and Au+Au collisions at \snn\ = $62.4-200$ GeV and implications for particle-production models,''
Phys. Rev. C \textbf{89}, 044905 (2014)
[arXiv:1312.6676 [nucl-ex]].

\bibitem{NA49:1994hfj}
T.~Alber \textit{et al.} [NA49],
``Transverse energy production in Pb-208 + Pb collisions at 158-GeV per nucleon,''
Phys. Rev. Lett. \textbf{75}, 3814 (1995)

\bibitem{WA98:2000mvt}
M.~M.~Aggarwal \textit{et al.} [WA98],
``Scaling of particle and transverse energy production in Pb-208 + Pb-208 collisions at 158-A-GeV,''
Eur. Phys. J. C \textbf{18}, 651 (2001)
[arXiv:nucl-ex/0008004 [nucl-ex]].

\bibitem{CMS:2012krf}
S.~Chatrchyan \textit{et al.} [CMS],
``Measurement of the pseudorapidity and centrality dependence of the transverse energy density in PbPb collisions at $\sqrt{s_{NN}}=2.76$ TeV,''
Phys. Rev. Lett. \textbf{109}, 152303 (2012)
[arXiv:1205.2488 [nucl-ex]].





\bibitem{Rafelski:1982ii}
J.~Rafelski,
``Formation and Observables of the Quark-Gluon Plasma,''
Phys. Rept. \textbf{88}, 331 (1982).

\bibitem{Rafelski:1982pu}
J.~Rafelski and B.~M\"uller,
``Strangeness Production in the Quark - Gluon Plasma,''
Phys. Rev. Lett. \textbf{48}, 1066 (1982)
[erratum: Phys. Rev. Lett. \textbf{56}, 2334 (1986)].

\bibitem{Koch:1986ud}
P.~Koch, B.~M\"uller and J.~Rafelski,
``Strangeness in Relativistic Heavy Ion Collisions,''
Phys. Rept. \textbf{142}, 167 (1986).

\bibitem{Petran:2013lja}
M.~Petr\'an, J.~Letessier, V.~Petr\'a\v{c}ek and J.~Rafelski,
``Hadron production and quark-gluon plasma hadronization in Pb-Pb collisions at $\sqrt{s_{NN}}=2.76$ TeV,''
Phys. Rev. C \textbf{88}, 034907 (2013)
[arXiv:1303.2098 [hep-ph]].

\bibitem{WA97:1999uwz}
E.~Andersen \textit{et al.} [WA97],
``Strangeness enhancement at mid-rapidity in Pb Pb collisions at 158-A-GeV/c,''
Phys. Lett. B \textbf{449}, 401 (1999)

\bibitem{NA57:2006aux}
F.~Antinori \textit{et al.} [NA57],
``Enhancement of hyperon production at central rapidity in 158-A-GeV/c Pb+Pb collisions,''
J. Phys. G \textbf{32}, 427 (2006)
[arXiv:nucl-ex/0601021 [nucl-ex]].

\bibitem{STAR:2007cqw}
B.~I.~Abelev \textit{et al.} [STAR],
``Enhanced strange baryon production in Au + Au collisions compared to p + p at 
$\sqrt{s} =$ 200 GeV,''
Phys. Rev. C \textbf{77}, 044908 (2008)
[arXiv:0705.2511 [nucl-ex]].

\bibitem{Becattini:2005xt}
F.~Becattini, J.~Manninen and M.~Gazdzicki,
``Energy and system size dependence of chemical freeze-out in relativistic nuclear collisions,''
Phys. Rev. C \textbf{73}, 044905 (2006)
[arXiv:hep-ph/0511092 [hep-ph]].

\bibitem{Becattini:2014hla}
F.~Becattini, E.~Grossi, M.~Bleicher, J.~Steinheimer and R.~Stock,
``Centrality dependence of hadronization and chemical freeze-out conditions in heavy ion collisions at $\sqrt s_{NN}$ = 2.76 TeV,''
Phys. Rev. C \textbf{90}, 054907 (2014)
[arXiv:1405.0710 [nucl-th]].

\bibitem{STAR:2017sal}
L.~Adamczyk \textit{et al.} [STAR],
``Bulk Properties of the Medium Produced in Relativistic Heavy-Ion Collisions from the Beam Energy Scan Program,''
Phys. Rev. C \textbf{96}, 044904 (2017)
[arXiv:1701.07065 [nucl-ex]].

\bibitem{Cleymans:1990mn}
J.~Cleymans, K.~Redlich and E.~Suhonen,
``Canonical description of strangeness conservation and particle production,''
Z. Phys. C \textbf{51}, 137 (1991)

\bibitem{Hamieh:2000tk}
S.~Hamieh, K.~Redlich and A.~Tounsi,
``Canonical description of strangeness enhancement from p-A to Pb Pb collisions,''
Phys. Lett. B \textbf{486}, 61 (2000)
[arXiv:hep-ph/0006024 [hep-ph]].

\bibitem{ALICE:2016fzo}
J.~Adam \textit{et al.} [ALICE],
``Enhanced production of multi-strange hadrons in high-multiplicity proton-proton collisions,''
Nature Phys. \textbf{13}, 535 (2017)
[arXiv:1606.07424 [nucl-ex]].

\bibitem{Cleymans:2020fsc}
J.~Cleymans, P.~M.~Lo, K.~Redlich and N.~Sharma,
``Multiplicity dependence of (multi)strange baryons in the canonical ensemble with phase shift corrections,''
Phys. Rev. C \textbf{103}, no.1, 014904 (2021)
[arXiv:2009.04844 [hep-ph]].

\bibitem{ALICE:2010khr}
K.~Aamodt \textit{et al.} [ALICE],
``Charged-particle multiplicity density at mid-rapidity in central Pb+Pb collisions at $\sqrt{s_{NN}} = 2.76$ TeV,''
Phys. Rev. Lett. \textbf{105}, 252301 (2010)
[arXiv:1011.3916 [nucl-ex]].



\bibitem{Matsui:1986dk}
T.~Matsui and H.~Satz,
``$J/\psi$ Suppression by Quark-Gluon Plasma Formation,''
Phys. Lett. B \textbf{178}, 416 (1986)

\bibitem{Digal:2001ue}
S.~Digal, P.~Petreczky and H.~Satz,
``Quarkonium feed down and sequential suppression,''
Phys. Rev. D \textbf{64}, 094015 (2001)
[arXiv:hep-ph/0106017 [hep-ph]].

\bibitem{Jakovac:2006sf}
A.~Jakovac, P.~Petreczky, K.~Petrov and A.~Velytsky,
``Quarkonium correlators and spectral functions at zero and finite temperature,''
Phys. Rev. D \textbf{75}, 014506 (2007)
[arXiv:hep-lat/0611017 [hep-lat]].

\bibitem{Petreczky:2021zmz}
P.~Petreczky, S.~Sharma and J.~H.~Weber,
``Bottomonium melting from screening correlators at high temperature,''
Phys. Rev. D \textbf{104}, 054511 (2021)
[arXiv:2107.11368 [hep-lat]].

\bibitem{Braaten:1996pv}
E.~Braaten, S.~Fleming and T.~C.~Yuan,
``Production of heavy quarkonium in high-energy colliders,''
Ann. Rev. Nucl. Part. Sci. \textbf{46} (1996), 197
[arXiv:hep-ph/9602374 [hep-ph]].

\bibitem{Thews:2000rj}
R.~L.~Thews, M.~Schroedter and J.~Rafelski,
``Enhanced $J/\psi$ production in deconfined quark matter,''
Phys. Rev. C \textbf{63}, 054905 (2001)
[arXiv:hep-ph/0007323 [hep-ph]].

\bibitem{Braun-Munzinger:2000csl}
P.~Braun-Munzinger and J.~Stachel,
``(Non)thermal aspects of charmonium production and a new look at J$/{\psi}$ suppression,''
Phys. Lett. B \textbf{490}, 196 (2000)
[arXiv:nucl-th/0007059 [nucl-th]].

\bibitem{NA38:1998udo}
M.~C.~Abreu \textit{et al.} [NA38],
``J$/{\psi}$, ${\psi}$' and Drell-Yan production in S+U interactions at 200-GeV per nucleon,''
Phys. Lett. B \textbf{449}, 128 (1999).

\bibitem{NA50:2000brc}
M.~C.~Abreu \textit{et al.} [NA50],
``Evidence for deconfinement of quarks and gluons from the J$/{\psi}$ suppression pattern measured in Pb + Pb collisions at the CERN SPS,''
Phys. Lett. B \textbf{477}, 28 (2000)

\bibitem{NA50:2004sgj}
B.~Alessandro \textit{et al.} [NA50],
``A New measurement of J/psi suppression in Pb+Pb collisions at 158-GeV per nucleon,''
Eur. Phys. J. C \textbf{39}, 335 (2005)
[arXiv:hep-ex/0412036 [hep-ex]].

\bibitem{NA60:2007ewx}
R.~Arnaldi \textit{et al.} [NA60],
``J/psi production in indium-indium collisions at 158-GeV/nucleon,''
Phys. Rev. Lett. \textbf{99}, 132302 (2007)

\bibitem{Kluberg:2009wc}
L.~Kluberg and H.~Satz,
``Color Deconfinement and Charmonium Production in Nuclear Collisions,''
[arXiv:0901.3831 [hep-ph]].

\bibitem{PHENIX:2008jgc}
A.~Adare \textit{et al.} [PHENIX],
``J/psi Production in $\sqrt{s_{NN}}=$ 200 GeV Cu+Cu Collisions,''
Phys. Rev. Lett. \textbf{101}, 122301 (2008)
[arXiv:0801.0220 [nucl-ex]].

\bibitem{PHENIX:2011img}
A.~Adare \textit{et al.} [PHENIX],
``$J/\psi$ suppression at forward rapidity in Au+Au collisions at $\sqrt{s_{NN}}=200$ GeV,''
Phys. Rev. C \textbf{84}, 054912 (2011)
[arXiv:1103.6269 [nucl-ex]].

\bibitem{ALICE:2019lga}
S.~Acharya \textit{et al.} [ALICE],
`Studies of J/$\psi$ production at forward rapidity in Pb+Pb collisions at $\sqrt{s_{\rm{NN}}}$ = 5.02 TeV,''
JHEP \textbf{02}, 041 (2020)
[arXiv:1909.03158 [nucl-ex]].

\bibitem{STAR:2019fge}
J.~Adam \textit{et al.} [STAR],
``Measurement of inclusive $J/\psi$ suppression in Au+Au collisions at $\sqrt{s_{NN}}$ = 200 GeV through the dimuon channel at STAR,''
Phys. Lett. B \textbf{797}, 134917 (2019)
[arXiv:1905.13669 [nucl-ex]].

\bibitem{ALICE:2019nrq}
S.~Acharya \textit{et al.} [ALICE],
``Centrality and transverse momentum dependence of inclusive J/\ensuremath{\psi} production at midrapidity in Pb\textendash{}Pb collisions at $\sqrt{s_{\rm{NN}}}$=5.02 TeV,''
Phys. Lett. B \textbf{805}, 135434 (2020)
[arXiv:1910.14404 [nucl-ex]].

\bibitem{ATLAS:2018hqe}
M.~Aaboud \textit{et al.} [ATLAS],
``Prompt and non-prompt $J/\psi $ and $\psi (2\mathrm {S})$ suppression at high transverse momentum in $5.02~\mathrm {TeV}$ Pb+Pb collisions with the ATLAS experiment,''
Eur. Phys. J. C \textbf{78}, 762 (2018)
[arXiv:1805.04077 [nucl-ex]].

\bibitem{CMS:2017uuv}
A.~M.~Sirunyan \textit{et al.} [CMS],
``Measurement of prompt and nonprompt charmonium suppression in $\text {PbPb}$ collisions at 5.02 $\,\text {Te}\text {V}$,''
Eur. Phys. J. C \textbf{78}, 509 (2018)
[arXiv:1712.08959 [nucl-ex]].

\bibitem{ALICE:2022_Psi-prime}
ALICE Collaboration
``$\psi$(2S) suppression in Pb–Pb collisions at the LHC,''
arXiv:2210.08893 [nucl-ex].

\bibitem{CMS:2018zza}
A.~M.~Sirunyan \textit{et al.} [CMS],
``Measurement of nuclear modification factors of $\Upsilon$(1S), $\Upsilon$(2S), and $\Upsilon$(3S) mesons in PbPb collisions at $\sqrt{s_{_\mathrm{NN}}} =$ 5.02 TeV,''
Phys. Lett. B \textbf{790}, 270 (2019)
[arXiv:1805.09215 [hep-ex]].

\bibitem{ATLAS:2022xso}
 [ATLAS],
``Production of $\varUpsilon(\textrm{nS})$ mesons in Pb+Pb and p+p collisions at 5.02 TeV,''
[arXiv:2205.03042 [nucl-ex]].

\bibitem{STAR_Upsilon_paper}
B.~Aboona \textit{et al.} [STAR],
``Observation of sequential $\Upsilon$ suppression in Au+Au collisions at $\sqrt{s_{_\mathrm{NN}}}$ = 200 GeV with the STAR experiment,'' Phys. Rev. Lett. \textbf{130}, 112301 (2023) [arXiv:2207.06568 [nucl-ex]].

\bibitem{PHENIX:2019brm}
U.~Acharya \textit{et al.} [PHENIX],
``Measurement of $J/\psi$ at forward and backward rapidity in $p+p$, $p+A$l, $p+A$u, and $^3$He$+$Au collisions at $\sqrt{s_{_{NN}}}=200~{\rm GeV}$,''
Phys. Rev. C \textbf{102}, 014902 (2020)
[arXiv:1910.14487 [hep-ex]].

\bibitem{STAR:2021zvb}
M.~Abdallah \textit{et al.} [STAR],
``Measurement of cold nuclear matter effects for inclusive J/\ensuremath{\psi} in p+Au collisions at sNN=200 GeV,''
Phys. Lett. B \textbf{825}, 136865 (2022)
[arXiv:2110.09666 [nucl-ex]].

\bibitem{Eskola:2021nhw}
K.~J.~Eskola, P.~Paakkinen, H.~Paukkunen and C.~A.~Salgado,
``EPPS21: a global QCD analysis of nuclear PDFs,''
Eur. Phys. J. C \textbf{82} (2022) 413
[arXiv:2112.12462 [hep-ph]].

\bibitem{PHENIX:2012czk}
A. Adare \textit{et al.} [PHENIX],
"{Transverse-Momentum Dependence of the $J/\psi$ Nuclear Modification in $d+$Au Collisions at $\sqrt{s_{NN}}=200$ GeV}",
Phys. Rev. C, \textbf{87}, 034904 (2013).

\bibitem{ALICE:2018mml}
S.~Acharya \textit{et al.} [ALICE],
``Inclusive J/$\psi$ production at forward and backward rapidity in p-Pb collisions at $\sqrt{s_{\rm NN}}$ = 8.16 TeV,''
JHEP \textbf{07}, 160 (2018)
[arXiv:1805.04381 [nucl-ex]].

\bibitem{ALICE:2020wwx}
S.~Acharya \textit{et al.} [ALICE],
``\Upsi\ production and nuclear modification at forward rapidity in Pb--Pb collisions at $\sqrt{s_{\rm NN}}$ = 5.02 TeV,''
Phys. Lett. B \textbf{822}, 136579 (2021)
[arXiv:2011.05758 [nucl-ex]].

\bibitem{ALICE:2019qie}
S.~Acharya \textit{et al.} [ALICE],
``$\Upsilon$ production in p--Pb collisions at $\sqrt{s_{NN}}$=8.16 TeV,''
Phys. Lett. B \textbf{806}, 135486 (2020)
[arXiv:1910.14405 [nucl-ex]].

\bibitem{Albacete:2017qng}
J.~L.~Albacete, F.~Arleo, G.~G.~Barnaf\"oldi, G.~B\'\i{}r\'o, D.~d'Enterria, B.~Duclou\'e, K.~J.~Eskola, E.~G.~Ferreiro, M.~Gyulassy and S.~M.~Harangoz\'o, \textit{et al.}
``Predictions for Cold Nuclear Matter Effects in $p+$Pb Collisions at $\sqrt{s_{_{NN}}} = 8.16$ TeV,''
Nucl. Phys. A \textbf{972}, 18-85 (2018)
[arXiv:1707.09973 [hep-ph]].

\bibitem{Ma:2017rsu}
Y.~Q.~Ma, R.~Venugopalan, K.~Watanabe and H.~F.~Zhang,
``$\psi(2S)$ versus $J/\psi$ suppression in proton-nucleus collisions from factorization violating soft color exchanges,''
Phys. Rev. C \textbf{97}, 014909 (2018)
[arXiv:1707.07266 [hep-ph]].

\bibitem{ATLAS:2017prf}
M.~Aaboud \textit{et al.} [ATLAS],
``Measurement of quarkonium production in proton\textendash{}lead and proton\textendash{}proton collisions at $5.02~\mathrm {TeV}$ with the ATLAS detector,''
Eur. Phys. J. C \textbf{78}, 171 (2018)
[arXiv:1709.03089 [nucl-ex]].

\bibitem{CMS:2017exb}
A.~M.~Sirunyan \textit{et al.} [CMS],
``Measurement of prompt and nonprompt $\mathrm{J}/{\psi }$ production in $\mathrm {p}\mathrm {p}$ and $\mathrm {p}\mathrm {Pb}$ collisions at $\sqrt{s_{\mathrm {NN}}} =5.02\,\text {TeV} $,''
Eur. Phys. J. C \textbf{77}, 269 (2017)
doi:10.1140/epjc/s10052-017-4828-3
[arXiv:1702.01462 [nucl-ex]].

\bibitem{CMS:2022wfi}
A.~Tumasyan \textit{et al.} [CMS],
``Nuclear modification of $\Upsilon$ states in pPb collisions at $\sqrt{s_\mathrm{NN}}$ = 5.02 TeV,''
Phys. Lett. B \textbf{835}, 137397 (2022)
doi:10.1016/j.physletb.2022.137397
[arXiv:2202.11807 [hep-ex]].

\bibitem{LHCb:2013gmv}
R.~Aaij \textit{et al.} [LHCb],
``Study of $J/\psi$ production and cold nuclear matter effects in $pPb$ collisions at $\sqrt{s_{NN}} = 5$ TeV,''
JHEP \textbf{02}, 072 (2014)
doi:10.1007/JHEP02(2014)072
[arXiv:1308.6729 [nucl-ex]].

\bibitem{LHCb:2014rku}
R.~Aaij \textit{et al.} [LHCb],
``Study of $\Upsilon$ production and cold nuclear matter effects in $p$Pb collisions at $\sqrt{s_{NN}}$=5 TeV,''
JHEP \textbf{07}, 094 (2014)
doi:10.1007/JHEP07(2014)094
[arXiv:1405.5152 [nucl-ex]].

\bibitem{NA60:2008dcb}
R.~Arnaldi \textit{et al.} [NA60],
``Evidence for the production of thermal-like muon pairs with masses above 1-GeV/c$^2$ in 158-A-GeV Indium-Indium Collisions,''
Eur. Phys. J. C \textbf{59}, 607 (2009)
[arXiv:0810.3204 [nucl-ex]].


\bibitem{STAR:2022Ye}
Zaochen~Ye [for the STAR Collaboration],
Talk at {\it Quark Matter} 2022.

\bibitem{Schnedermann:1993ws}
E.~Schnedermann, J.~Sollfrank and U.~W.~Heinz,
``Thermal phenomenology of hadrons from 200-A/GeV S+S collisions,''
Phys. Rev. C \textbf{48}, 2462 (1993)
[arXiv:nucl-th/9307020 [nucl-th]].


\bibitem{ALICE:2019hno}
S.~Acharya \textit{et al.} [ALICE], ``Production of charged pions, kaons, and (anti-)protons in Pb+Pb and inelastic p+p collisions at $\sqrt {s_{NN}}$ = 5.02 TeV,''
Phys. Rev. C \textbf{101}, 044907 (2020)
[arXiv:1910.07678 [nucl-ex]].

\bibitem{ALICE:2013mez}
B.~Abelev \textit{et al.} [ALICE],
``Centrality dependence of $\pi$, K, p production in Pb+Pb collisions at $\sqrt{s_{NN}}$ = 2.76 TeV,''
Phys. Rev. C \textbf{88}, 044910 (2013)
[arXiv:1303.0737 [hep-ex]].


\bibitem{Rybczynski:2012ee}
M.~Rybczynski and W.~Florkowski,
``Locally anisotropic momentum distributions of hadrons at freeze-out in relativistic heavy-ion collisions,''
J. Phys. G \textbf{40}, 025103 (2013)
[arXiv:1206.6587 [nucl-th]].

\bibitem{Mazeliauskas:2019ifr}
A.~Mazeliauskas and V.~Vislavicius,
``Temperature and fluid velocity on the freeze-out surface from $\pi$, $K$, $p$ spectra in pp, p-Pb and Pb+Pb collisions,''
Phys. Rev. C \textbf{101}, 014910 (2020)
[arXiv:1907.11059 [hep-ph]].

\bibitem{Khachatryan:2017dqo}
V.~Khachatryan [PHENIX],
``Low Momentum Direct Photons in Au+Au collisions at 39 GeV and 62.4  GeV measured by the PHENIX Experiment at RHIC,''
PoS \textbf{CPOD2017}, 079 (2018).

\bibitem{PHENIX:2014nkk}
A.~Adare \textit{et al.} [PHENIX],
``Centrality dependence of low-momentum direct-photon production in Au$+$Au collisions at $\sqrt{s_{_{NN}}}=200$ GeV,''
Phys. Rev. C \textbf{91}, 064904 (2015)
[arXiv:1405.3940 [nucl-ex]].

\bibitem{ALICE:2015xmh}
J.~Adam \textit{et al.} [ALICE],
``Direct photon production in Pb-Pb collisions at $\sqrt{s_{NN}} =$ 2.76 TeV,''
Phys. Lett. B \textbf{754}, 235 (2016)
[arXiv:1509.07324 [nucl-ex]].

\bibitem{vanHees:2014ida}
H.~van Hees, M.~He and R.~Rapp,
``Pseudo-critical enhancement of thermal photons in relativistic heavy-ion collisions?,''
Nucl. Phys. A \textbf{933}, 256 (2015)
[arXiv:1404.2846 [nucl-th]].

\bibitem{Paquet:2015lta}
J.~F.~Paquet, C.~Shen, G.~S.~Denicol, M.~Luzum, B.~Schenke, S.~Jeon and C.~Gale,
``Production of photons in relativistic heavy-ion collisions,''
Phys. Rev. C \textbf{93}, 044906 (2016)
[arXiv:1509.06738 [hep-ph]].

\bibitem{Paquet:2022wgu}
J.~F.~Paquet and S.~A.~Bass,
``Electromagnetic measurement of the temperature of quark-gluon plasma produced in central ultrarelativistic nuclear collisions,''
[arXiv:2205.12299 [nucl-th]].


\bibitem{Rajagopal:1993ah}
K.~Rajagopal and F.~Wilczek,
``Emergence of coherent long wavelength oscillations after a quench: Application to QCD,''
Nucl. Phys. B \textbf{404}, 577-589 (1993)
[arXiv:hep-ph/9303281 [hep-ph]].

\bibitem{Anselm:1991pi}
A.~A.~Anselm and M.~G.~Ryskin,
``Production of classical pion field in heavy ion high-energy collisions,''
Phys. Lett. B \textbf{266}, 482 (1991)
doi:10.1016/0370-2693(91)91073-5

\bibitem{Anselm:1996vm}
A.~A.~Anselm and M.~G.~Ryskin,
``The Production of a nonhomogeneous classical pion field and the distribution of the neutral and charged pions,''
Z. Phys. A \textbf{358}, 353 (1997)
[arXiv:hep-ph/9606319 [hep-ph]].

\bibitem{Mohanty:2005mv}
B.~Mohanty and J.~Serreau,
``Disoriented chiral condensate: theory and experiment,''
Phys. Rept. \textbf{414}, 263 (2005)
[arXiv:hep-ph/0504154 [hep-ph]].

\bibitem{STAR:2014xli}
L.~Adamczyk \textit{et al.} [STAR],
``Charged-to-neutral correlation at forward rapidity in Au+Au collisions at $\sqrt{s_{NN}}$=200 GeV,''
Phys. Rev. C \textbf{91}, 034905 (2015)
[arXiv:1408.5017 [nucl-ex]].

\bibitem{Gavin:2001uk}
S.~Gavin and J.~I.~Kapusta,
``Kaon and pion fluctuations from small disoriented chiral condensates,''
Phys. Rev. C \textbf{65}, 054910 (2002)
[arXiv:nucl-th/0112083 [nucl-th]].

\bibitem{Nayak:2019qzd}
R.~Nayak, S.~Dash, B.~Nandi and C.~Pruneau,
``Modeling of charged kaon and neutral kaon fluctuations as a signature for the production of a disoriented chiral condensate in $A-A$ collisions,''
Phys. Rev. C \textbf{101}, 054904 (2020)
[arXiv:1908.01130 [hep-ph]].

\bibitem{ALICE:2021fpb}
S.~Acharya \textit{et al.} [ALICE],
``Neutral to charged kaon yield fluctuations in Pb$-$Pb collisions at $\sqrt{s_{\rm NN}}$ = 2.76 TeV,''
Phys. Lett. B \textbf{832}, 137242 (2022)
[arXiv:2112.09482 [nucl-ex]].

\bibitem{Asakawa:2000wh}
M.~Asakawa, U.~W.~Heinz and B.~M\"uller,
``Fluctuation probes of quark deconfinement,''
Phys. Rev. Lett. \textbf{85}, 2072 (2000)
[arXiv:hep-ph/0003169 [hep-ph]].

\bibitem{Jeon:2000wg}
S.~Jeon and V.~Koch,
``Charged particle ratio fluctuation as a signal for QGP,''
Phys. Rev. Lett. \textbf{85}, 2076 (2000)
[arXiv:hep-ph/0003168 [hep-ph]].

\bibitem{Koch:2005vg}
V.~Koch, A.~Majumder and J.~Randrup,
``Baryon-strangeness correlations: A Diagnostic of strongly interacting matter,''
Phys. Rev. Lett. \textbf{95}, 182301 (2005)
[arXiv:nucl-th/0505052 [nucl-th]].

\bibitem{Shuryak:2000pd}
E.~V.~Shuryak and M.~A.~Stephanov,
``When can long range charge fluctuations serve as a QGP signal?,''
Phys. Rev. C \textbf{63}, 064903 (2001)
[arXiv:hep-ph/0010100 [hep-ph]].

\bibitem{Ratti:2018ksb}
C.~Ratti,
``Lattice QCD and heavy ion collisions: a review of recent progress,''
Rept. Prog. Phys. \textbf{81}, 084301 (2018)
[arXiv:1804.07810 [hep-lat]].

\bibitem{Ratti:2021ubw}
C.~Ratti and R.~Bellwied,
``The Deconfinement Transition of QCD: Theory Meets Experiment,''
Lect. Notes Phys. \textbf{981}, 1 (2021)
2021
(ISBN 978-3-030-67234-8, 978-3-030-67235-5).

\bibitem{Borsanyi:2020fev}
S.~Borsanyi, Z.~Fodor, J.~N.~Guenther, R.~Kara, S.~D.~Katz, P.~Parotto, A.~Pasztor, C.~Ratti and K.~K.~Szabo,
``QCD Crossover at Finite Chemical Potential from Lattice Simulations,''
Phys. Rev. Lett. \textbf{125}, 052001 (2020)
[arXiv:2002.02821 [hep-lat]].

\bibitem{Alba:2014eba}
P.~Alba, W.~Alberico, R.~Bellwied, M.~Bluhm, V.~Mantovani Sarti, M.~Nahrgang and C.~Ratti,
``Freeze-out conditions from net-proton and net-charge fluctuations at RHIC,''
Phys. Lett. B \textbf{738}, 305 (2014)
[arXiv:1403.4903 [hep-ph]].

\bibitem{Andronic:2005yp}
A.~Andronic, P.~Braun-Munzinger and J.~Stachel,
``Hadron production in central nucleus-nucleus collisions at chemical freeze-out,''
Nucl. Phys. A \textbf{772}, 167 (2006)
[arXiv:nucl-th/0511071 [nucl-th]].

\bibitem{Andronic:2018qqt}
A.~Andronic, P.~Braun-Munzinger, B.~Friman, P.~M.~Lo, K.~Redlich and J.~Stachel,
``The thermal proton yield anomaly in Pb-Pb collisions at the LHC and its resolution,''
Phys. Lett. B \textbf{792}, 304 (2019)
[arXiv:1808.03102 [hep-ph]].

\bibitem{Stephanov:1999zu}
M.~A.~Stephanov, K.~Rajagopal and E.~V.~Shuryak,
``Event-by-event fluctuations in heavy ion collisions and the QCD critical point,''
Phys. Rev. D \textbf{60}, 114028 (1999)
[arXiv:hep-ph/9903292 [hep-ph]].

\bibitem{Stephanov:2011pb}
M.~A.~Stephanov,
``On the sign of kurtosis near the QCD critical point,''
Phys. Rev. Lett. \textbf{107}, 052301 (2011)
[arXiv:1104.1627 [hep-ph]].

\bibitem{STAR:2013gus}
L.~Adamczyk \textit{et al.} [STAR],
``Energy Dependence of Moments of Net-proton Multiplicity Distributions at RHIC,''
Phys. Rev. Lett. \textbf{112}, 032302 (2014)
[arXiv:1309.5681 [nucl-ex]].

\bibitem{Liu:2022wme}
C.~Liu, P.~Adams, E.~Beebe, S.~Binello, I.~Blackler, M.~Blaskiewicz, K.~Brown, D.~Bruno, B.~Coe and K.~Drees, \textit{et al.}
``Summary of the 3-year Beam Energy Scan II operation at RHIC,''
JACoW \textbf{IPAC2022}, 1908-1911 (2022)

\bibitem{Pratt:2021xvg}
S.~Pratt and C.~Plumberg,
``Charge balance functions for heavy-ion collisions at energies available at the CERN Large Hadron Collider,''
Phys. Rev. C \textbf{104}, 014906 (2021)
[arXiv:2104.00628 [nucl-th]].

\bibitem{ALICE:2021hjb}
S.~Acharya \textit{et al.} [ALICE],
``General balance functions of identified charged hadron pairs of (\ensuremath{\pi},K,p) in Pb\textendash{}Pb collisions at sNN= 2.76 TeV,''
Phys. Lett. B \textbf{833}, 137338 (2022)
[arXiv:2110.06566 [nucl-ex]].

\bibitem{Aarts:2014nba}
G.~Aarts, C.~Allton, A.~Amato, P.~Giudice, S.~Hands and J.~I.~Skullerud,
``Electrical conductivity and charge diffusion in thermal QCD from the lattice,''
JHEP \textbf{02}, 186 (2015)
[arXiv:1412.6411 [hep-lat]].




\bibitem{Kapusta:1993hq}
J.~I.~Kapusta and E.~V.~Shuryak,
``Weinberg type sum rules at zero and finite temperature,''
Phys. Rev. D \textbf{49}, 4694 (1994)
[arXiv:hep-ph/9312245 [hep-ph]].

\bibitem{Holt:2012wr}
N.~P.~M.~Holt, P.~M.~Hohler and R.~Rapp,
``Quantitative Sum Rule Analysis of Low-Temperature Spectral Functions,''
Phys. Rev. D \textbf{87}, 076010 (2013)
[arXiv:1210.7210 [hep-ph]].

\bibitem{Bochkarev:1985ex}
A.~I.~Bochkarev and M.~E.~Shaposhnikov,
``Spectrum of the Hot Hadronic Matter and Finite Temperature QCD Sum Rules,''
Nucl. Phys. B \textbf{268}, 220-252 (1986).

\bibitem{Dominguez:1989bz}
C.~A.~Dominguez and M.~Loewe,
``Deconfinement and Chiral Symmetry Restoration at Finite Temperature,''
Phys. Lett. B \textbf{233}, 201-204 (1989).

\bibitem{NA60:2006ymb}
R.~Arnaldi \textit{et al.} [NA60],
``First measurement of the rho spectral function in high-energy nuclear collisions,''
Phys. Rev. Lett. \textbf{96}, 162302 (2006)
[arXiv:nucl-ex/0605007 [nucl-ex]].

\bibitem{NA60:2007lzy}
R.~Arnaldi \textit{et al.} [NA60],
``Evidence for radial flow of thermal dileptons in high-energy nuclear collisions,''
Phys. Rev. Lett. \textbf{100}, 022302 (2008)
[arXiv:0711.1816 [nucl-ex]].

\bibitem{NA60:2008ctj}
R.~Arnaldi \textit{et al.} [NA60],
``NA60 results on thermal dimuons,''
Eur. Phys. J. C \textbf{61}, 711 (2009)
[arXiv:0812.3053 [nucl-ex]].

\bibitem{NA60:2008iqj}
R.~Arnaldi \textit{et al.} [NA60],
``First results on angular distributions of thermal dileptons in nuclear collisions,''
Phys. Rev. Lett. \textbf{102}, 222301 (2009)
[arXiv:0812.3100 [nucl-ex]].

\bibitem{Rapp:1999ej}
R.~Rapp and J.~Wambach,
``Chiral symmetry restoration and dileptons in relativistic heavy ion collisions,''
Adv. Nucl. Phys. \textbf{25}, 1 (2000)
[arXiv:hep-ph/9909229 [hep-ph]].

\bibitem{Brown:2001nh}
G.~E.~Brown and M.~Rho,
``On the manifestation of chiral symmetry in nuclei and dense nuclear matter,''
Phys. Rept. \textbf{363}, 85 (2002)
[arXiv:hep-ph/0103102 [hep-ph]].

\bibitem{PHENIX:2015vek}
A.~Adare \textit{et al.} [PHENIX],
``Dielectron production in Au+Au collisions at $\sqrt{s_{NN}}$=200 GeV,''
Phys. Rev. C \textbf{93}, 014904 (2016)
[arXiv:1509.04667 [nucl-ex]].

\bibitem{STAR:2015tnn}
L.~Adamczyk \textit{et al.} [STAR],
``Measurements of Dielectron Production in Au$+$Au Collisions at $\sqrt{s_{\rm NN}}$ = 200 GeV from the STAR Experiment,''
Phys. Rev. C \textbf{92}, 024912 (2015)
[arXiv:1504.01317 [hep-ex]].

\bibitem{STAR:2023wta}
M.~I.~Abdulhamid \textit{et al.} [STAR],
``Measurements of dielectron production in Au+Au collisions at $\sqrt{s_{\rm NN}}$ = 27, 39, and 62.4 GeV from the STAR experiment,''
Phys. Rev. C \textbf{107}, L061901 (2023)
doi:10.1103/PhysRevC.107.L061901

\bibitem{ALICE:2022hvk}
S.~Acharya \textit{et al.} [ALICE],
``Dielectron production at midrapidity at low transverse momentum in peripheral and semi-peripheral Pb\textendash{}Pb collisions at $ {\sqrt{s}}_{\textrm{NN}} $ = 5.02 TeV,''
JHEP \textbf{06}, 024 (2023)
[arXiv:2204.11732 [nucl-ex]].

\bibitem{ALICE:2023jef}
S.~Acharya \textit{et al.} [ALICE],
``Dielectron production in central Pb$-$Pb collisions at $\sqrt{s_\mathrm{NN}}$ = 5.02 TeV,''
[arXiv:2308.16704 [nucl-ex]].




\bibitem{Gyulassy:1979yi}
M.~Gyulassy, S.~K.~Kauffmann and L.~W.~Wilson,
``Pion Interferometry of Nuclear Collisions. 1. Theory,''
Phys. Rev. C \textbf{20}, 2267 (1979).

\bibitem{Zajc:1992sz}
W.~A.~Zajc,
``A pedestrian's guide to interferometry,''
NATO Sci. Ser. B \textbf{303}, 435 (1993).

\bibitem{STAR:2020dav}
J.~Adam \textit{et al.} [STAR],
``Flow and interferometry results from Au+Au collisions at $\sqrt{s_{NN}} = 4.5$ GeV,''
Phys. Rev. C \textbf{103}, 034908 (2021)
[arXiv:2007.14005 [nucl-ex]].

\bibitem{Herrmann:1994rr}
M.~Herrmann and G.~F.~Bertsch,
``Source dimensions in ultrarelativistic heavy ion collisions,''
Phys. Rev. C \textbf{51}, 328 (1995)
[arXiv:hep-ph/9405373 [hep-ph]].

\bibitem{STAR:2005rpl}
J.~Adams \textit{et al.} [STAR],
``Proton - lambda correlations in central Au+Au collisions at $\sqrt{s_{\rm NN}}$ = 200 GeV,''
Phys. Rev. C \textbf{74}, 064906 (2006)
[arXiv:nucl-ex/0511003 [nucl-ex]].

\bibitem{ALICE:2021szj}
S.~Acharya \textit{et al.} [ALICE],
``Kaon\textendash{}proton strong interaction at low relative momentum via femtoscopy in Pb\textendash{}Pb collisions at the LHC,''
Phys. Lett. B \textbf{822}, 136708 (2021)
[arXiv:2105.05683 [nucl-ex]].

\bibitem{ALICE:2022yyh}
S.~Acharya \textit{et al.} [ALICE],
``Constraining the ${\overline{\textrm{K}}}{\textrm{N}}$ coupled channel dynamics using femtoscopic correlations at the LHC,''
Eur. Phys. J. C \textbf{83}, 340 (2023)
[arXiv:2205.15176 [nucl-ex]].

\bibitem{LHCb:2020sey}
R.~Aaij \textit{et al.} [LHCb],
``Observation of Multiplicity Dependent Prompt $\chi_{c1}(3872)$ and $\psi(2S)$ Production in $pp$ Collisions,''
Phys. Rev. Lett. \textbf{126}, 092001 (2021)
[arXiv:2009.06619 [hep-ex]].

\bibitem{CMS:2021znk}
A.~M.~Sirunyan \textit{et al.} [CMS],
``Evidence for X(3872) in Pb-Pb Collisions and Studies of its Prompt Production at $\sqrt {s_{NN}}$=5.02\,\,TeV,''
Phys. Rev. Lett. \textbf{128}, 032001 (2022)
[arXiv:2102.13048 [hep-ex]].

\bibitem{STAR:2011eej}
H.~Agakishiev \textit{et al.} [STAR],
``Observation of the antimatter helium-4 nucleus,''
Nature \textbf{473}, 353 (2011)
[erratum: Nature \textbf{475}, 412 (2011)]
[arXiv:1103.3312 [nucl-ex]].

\bibitem{STAR:2010gyg}
B.~I.~Abelev \textit{et al.} [STAR],
``Observation of an Antimatter Hypernucleus,''
Science \textbf{328}, 58-62 (2010)
[arXiv:1003.2030 [nucl-ex]].

\bibitem{ALICE:2015rey}
J.~Adam \textit{et al.} [ALICE],
``Precision measurement of the mass difference between light nuclei and anti-nuclei with the ALICE experiment at the LHC,''
Nature Phys. \textbf{11}, 811-814 (2015)
[arXiv:1508.03986 [nucl-ex]].

\bibitem{ALICE:2022veq}
S.~Acharya \textit{et al.} [ALICE],
``Light (anti)nuclei production in Pb-Pb collisions at $\sqrt{s_{\rm NN}}$ = 5.02 TeV,''
Phys. Rev. C \textbf{107}, 064904 (2023)
[arXiv:2211.14015 [nucl-ex]].

\bibitem{STAR:2021orx}
M.~Abdallah \textit{et al.} [STAR],
``Measurements of $H_\Lambda^3$ and $H_\Lambda^4$ Lifetimes and Yields in Au+Au Collisions in the High Baryon Density Region,''
Phys. Rev. Lett. \textbf{128}, 202301 (2022)
[arXiv:2110.09513 [nucl-ex]].


%
%

\bibitem{PHENIX:2001hpc}
K.~Adcox \textit{et al.} [PHENIX],
``Suppression of hadrons with large transverse momentum in central Au+Au collisions at \snn\ = 130 GeV,''
Phys. Rev. Lett. \textbf{88}, 022301 (2002)
[arXiv:nucl-ex/0109003 [nucl-ex]].

\bibitem{STAR:2002ggv}
C.~Adler \textit{et al.} [STAR],
``Centrality dependence of high $p_{T}$ hadron suppression in Au+Au collisions at \snn\ = 130 GeV,''
Phys. Rev. Lett. \textbf{89}, 202301 (2002)
[arXiv:nucl-ex/0206011 [nucl-ex]].

\bibitem{STAR:2002svs}
C.~Adler \textit{et al.} [STAR],
``Disappearance of back-to-back high $p_{T}$ hadron correlations in central Au+Au collisions at \snn\ = 200 GeV,''
Phys. Rev. Lett. \textbf{90}, 082302 (2003)
[arXiv:nucl-ex/0210033 [nucl-ex]].

\bibitem{STAR:2005ryu}
J.~Adams \textit{et al.} [STAR],
``Distributions of charged hadrons associated with high transverse momentum particles in pp and Au+Au collisions at \snn\ = 200 GeV,''
Phys. Rev. Lett. \textbf{95}, 152301 (2005)
[arXiv:nucl-ex/0501016 [nucl-ex]].

\bibitem{PHENIX:2006kkn}
A.~Adare \textit{et al.} [PHENIX],
``System Size and Energy Dependence of Jet-Induced Hadron Pair Correlation Shapes in Cu+Cu and Au+Au Collisions at \snn\ = 200 and 62.4 GeV,''
Phys. Rev. Lett. \textbf{98}, 232302 (2007)
[arXiv:nucl-ex/0611019 [nucl-ex]].

\bibitem{PHENIX:2008osq}
A.~Adare \textit{et al.} [PHENIX],
``Dihadron azimuthal correlations in Au+Au collisions at \snn\ = 200 GeV,''
Phys. Rev. C \textbf{78}, 014901 (2008)
[arXiv:0801.4545 [nucl-ex]].

\bibitem{STAR:2009ngv}
B.~I.~Abelev \textit{et al.} [STAR],
``Long range rapidity correlations and jet production in high energy nuclear collisions,''
Phys. Rev. C \textbf{80}, 064912 (2009)
[arXiv:0909.0191 [nucl-ex]].

\bibitem{CMS:2011cqy}
S.~Chatrchyan \textit{et al.} [CMS],
``Long-range and short-range dihadron angular correlations in central PbPb collisions at a nucleon-nucleon center of mass energy of 2.76 TeV,''
JHEP \textbf{07}, 076 (2011)
[arXiv:1105.2438 [nucl-ex]].

\bibitem{ALICE:2021nir}
S.~Acharya \textit{et al.} [ALICE],
``Long- and short-range correlations and their event-scale dependence in high-multiplicity pp collisions at \snn\ = 13 TeV,''
JHEP \textbf{05}, 290 (2021)
[arXiv:2101.03110 [nucl-ex]].

\bibitem{STAR:2013thw}
L.~Adamczyk \textit{et al.} [STAR],
``Jet-Hadron Correlations in \snn\ = 200 GeV $p+p$ and Central Au+Au Collisions,''
Phys. Rev. Lett. \textbf{112}, 122301 (2014)
[arXiv:1302.6184 [nucl-ex]].

\bibitem{ALICE:2019sqi}
S.~Acharya \textit{et al.} [ALICE],
``Jet-hadron correlations measured relative to the second order event plane in Pb-Pb collisions at $\sqrt{s_{\rm{NN}}}$ = 2.76 TeV,''
Phys. Rev. C \textbf{101}, 064901 (2020)
doi:10.1103/PhysRevC.101.064901
[arXiv:1910.14398 [nucl-ex]].

\bibitem{ALICE:2015mdb}
J.~Adam \textit{et al.} [ALICE],
``Measurement of jet quenching with semi-inclusive hadron-jet distributions in central Pb-Pb collisions at \snn\ = 2.76 TeV,''
JHEP \textbf{09}, 170 (2015)
[arXiv:1506.03984 [nucl-ex]].

\bibitem{STAR:2017hhs}
L.~Adamczyk \textit{et al.} [STAR],
``Measurements of jet quenching with semi-inclusive hadron+jet distributions in Au+Au collisions at \snn\ = 200 GeV,''
Phys. Rev. C \textbf{96}, 024905 (2017)
[arXiv:1702.01108 [nucl-ex]].

\bibitem{ATLAS:2010isq}
G.~Aad \textit{et al.} [ATLAS],
``Observation of a Centrality-Dependent Dijet Asymmetry in Lead-Lead Collisions at \snn\ = 2.77 TeV with the ATLAS Detector at the LHC,''
Phys. Rev. Lett. \textbf{105}, 252303 (2010)
[arXiv:1011.6182 [hep-ex]].

\bibitem{CMS:2012ulu}
S.~Chatrchyan \textit{et al.} [CMS],
``Jet momentum dependence of jet quenching in PbPb collisions at \snn\ = 2.76 TeV,''
Phys. Lett. B \textbf{712}, 176-197 (2012)
[arXiv:1202.5022 [nucl-ex]].

\bibitem{CMS:2011iwn}
S.~Chatrchyan \textit{et al.} [CMS],
``Observation and studies of jet quenching in PbPb collisions at nucleon-nucleon center-of-mass energy = 2.76 TeV,''
Phys. Rev. C \textbf{84}, 024906 (2011)
doi:10.1103/PhysRevC.84.024906
[arXiv:1102.1957 [nucl-ex]].

\bibitem{CMS:2015hkr}
V.~Khachatryan \textit{et al.} [CMS],
``Measurement of transverse momentum relative to dijet systems in PbPb and pp collisions at \snn\ = 2.76 TeV,''
JHEP \textbf{01}, 006 (2016)
[arXiv:1509.09029 [nucl-ex]].

\bibitem{ATLAS:2022zbu}
G.~Aad \textit{et al.} [ATLAS],
``Measurements of the suppression and correlations of dijets in Pb+Pb collisions at \snn\ = 5.02 TeV,''
Phys. Rev. C \textbf{107}, 054908 (2023)
[arXiv:2205.00682 [nucl-ex]].

\bibitem{Norman:2020grk}
J.~Norman [ALICE],
``Jet acoplanarity via hadron+jet measurements in Pb-Pb collisions at \snn\ = 5.02 TeV with ALICE,''
PoS \textbf{HardProbes2020}, 127 (2021)
[arXiv:2009.08261 [hep-ex]].

\bibitem{Anderson:2022nxb}
D.~Anderson [STAR],
``Measurement of Medium-induced Modification of Jet Yield and Acoplanarity Using Semi-inclusive $\gamma_{\mathrm{dir}}+$jet and $\pi^{0}+$jet Distributions in p+p and Central Au+Au Collisions at \snn\ = 200 GeV by STAR,''
Acta Phys. Polon. Supp. \textbf{16}, 55 (2023)
[arXiv:2212.09202 [nucl-ex]].

\bibitem{Appel:1985dq}
D.~A.~Appel,
``Jets as a Probe of Quark - Gluon Plasmas,''
Phys. Rev. D \textbf{33}, 717 (1986)

\bibitem{Blaizot:1986ma}
J.~P.~Blaizot and L.~D.~McLerran,
``Jets in Expanding Quark - Gluon Plasmas,''
Phys. Rev. D \textbf{34}, 2739 (1986)

\bibitem{Shibata:2022fyb}
M.~Shibata [PHENIX],
``Systematic study of energy loss in the quark-gluon plasma at RHIC-PHENIX,''
PoS \textbf{PANIC2021}, 249 (2022)

\bibitem{Beattie:2022ojg}
C.~Beattie, G.~Nijs, M.~Sas and W.~van der Schee,
``Hard probe path lengths and event-shape engineering of the quark-gluon plasma,''
Phys. Lett. B \textbf{836}, 137596 (2023)
[arXiv:2203.13265 [nucl-th]].

\bibitem{Beattie:2023mcz}
C.~Beattie,
``Pathlength-dependent jet quenching in the quark--gluon plasma at ALICE,''
CERN-THESIS-2023-030.



\bibitem{Gyulassy:1990ye}
M.~Gyulassy and M.~Plumer,
``Jet Quenching in Dense Matter,''
Phys. Lett. B \textbf{243} (1990), 432

\bibitem{Wang:1992qdg}
X.~N.~Wang and M.~Gyulassy,
``Gluon shadowing and jet quenching in A + A collisions at $\sqrt{s_{\rm NN}}$ = 200 GeV,''
Phys. Rev. Lett. \textbf{68} (1992), 1480

\bibitem{Baier:2000mf}
R.~Baier, D.~Schiff and B.~G.~Zakharov,
``Energy loss in perturbative QCD,''
Ann. Rev. Nucl. Part. Sci. \textbf{50} (2000), 37
[arXiv:hep-ph/0002198 [hep-ph]].

\bibitem{Renk:2014lza}
T.~Renk,
``A study of the constraining power of high $P_T$ observables in heavy-ion collisions,''
[arXiv:1408.6684 [hep-ph]].

\bibitem{BRAHMS:2003sns}
I.~Arsene \textit{et al.} [BRAHMS],
``Transverse momentum spectra in Au+Au and d+Au collisions at $\sqrt{s_{\rm NN}}$ = 200 GeV and the pseudorapidity dependence of high p(T) suppression,''
Phys. Rev. Lett. \textbf{91}, 072305 (2003)
[arXiv:nucl-ex/0307003 [nucl-ex]].

\bibitem{PHOBOS:2004juu}
B.~B.~Back \textit{et al.} [PHOBOS],
``Centrality dependence of charged hadron transverse momentum spectra in Au + Au collisions from $\sqrt{s_{\rm NN}}$ = 62.4 GeV to 200 GeV,''
Phys. Rev. Lett. \textbf{94}, 082304 (2005)
[arXiv:nucl-ex/0405003 [nucl-ex]].


\bibitem{ALICE:2018vuu}
S.~Acharya \textit{et al.} [ALICE],
``Transverse momentum spectra and nuclear modification factors of charged particles in pp, p-Pb and Pb+Pb collisions at the LHC,''
JHEP \textbf{11}, 013 (2018)
[arXiv:1802.09145 [nucl-ex]].

\bibitem{ALICE:2010yje}
K.~Aamodt \textit{et al.} [ALICE],
``Suppression of Charged Particle Production at Large Transverse Momentum in Central Pb+Pb Collisions at $\sqrt{s_{NN}} =$ 2.76 TeV,''
Phys. Lett. B \textbf{696}, 30 (2011)
[arXiv:1012.1004 [nucl-ex]].

\bibitem{CMS:2012aa}
S.~Chatrchyan \textit{et al.} [CMS],
``Study of high-pT charged particle suppression in PbPb compared to p+p collisions at $\sqrt{s_{NN}}=2.76$ TeV,''
Eur. Phys. J. C \textbf{72}, 1945 (2012)
[arXiv:1202.2554 [nucl-ex]].

\bibitem{ALICE:2012aqc}
B.~Abelev \textit{et al.} [ALICE],
``Centrality Dependence of Charged Particle Production at Large Transverse Momentum in Pb--Pb Collisions at $\sqrt{s_{\rm{NN}}} = 2.76$ TeV,''
Phys. Lett. B \textbf{720}, 52 (2013)
[arXiv:1208.2711 [hep-ex]].

\bibitem{ATLAS:2015qmb}
G.~Aad \textit{et al.} [ATLAS],
``Measurement of charged-particle spectra in Pb+Pb collisions at $\sqrt{{s}_\mathsf{{NN}}} = 2.76$ TeV with the ATLAS detector at the LHC,''
JHEP \textbf{09}, 050 (2015)
[arXiv:1504.04337 [hep-ex]].

\bibitem{PHENIX:2012oed}
A.~Adare \textit{et al.} [PHENIX],
``Evolution of $\pi^0$ suppression in Au+Au collisions from $\sqrt{s_{NN}} = 39$ to 200 GeV,''
Phys. Rev. Lett. \textbf{109} (2012), 152301
[erratum: Phys. Rev. Lett. \textbf{125} (2020) 049901]
[arXiv:1204.1526 [nucl-ex]].

\bibitem{Horvat:2013lza}
S.~Horvat [STAR],
``Charged Hadron Nuclear Modification Factors in the Beam Energy Scan from STAR,''
PoS \textbf{CPOD2013}, 002 (2013).

\bibitem{Cronin:1974zm}
J.~W.~Cronin \textit{et al.} [E100],
``Production of hadrons with large transverse momentum at 200, 300, and 400 GeV,''
Phys. Rev. D \textbf{11}, 3105-3123 (1975)

\bibitem{Kharzeev:2003wz}
D.~Kharzeev, Y.~V.~Kovchegov and K.~Tuchin,
``Cronin effect and high p(T) suppression in pA collisions,''
Phys. Rev. D \textbf{68}, 094013 (2003)
[arXiv:hep-ph/0307037 [hep-ph]].

\bibitem{Braun:1994bf}
M.~Braun and V.~Vechernin,
Nucl. Phys. B \textbf{427}, 614-640 (1994).

\bibitem{Fries:2003kq}
R.~J.~Fries, B.~M{\"u}ller, C.~Nonaka and S.~A.~Bass,
``Hadron production in heavy ion collisions: Fragmentation and recombination from a dense parton phase,''
Phys. Rev. C \textbf{68}, 044902 (2003)
[arXiv:nucl-th/0306027 [nucl-th]].

\bibitem{STAR:2006xud}
J.~Adams \textit{et al.} [STAR],
``Identified hadron spectra at large transverse momentum in p+p and d+Au collisions at s(NN)**(1/2) = 200-GeV,''
Phys. Lett. B \textbf{637}, 161-169 (2006)
doi:10.1016/j.physletb.2006.04.032
[arXiv:nucl-ex/0601033 [nucl-ex]].

\bibitem{PHENIX:2008saf}
A.~Adare \textit{et al.} [PHENIX],
``Suppression pattern of neutral pions at high transverse momentum in Au$+$Au collisions at $\sqrt{s_{NN}}=$ 200 GeV and constraints on medium transport coefficients,''
Phys. Rev. Lett. \textbf{101}, 232301 (2008)
[arXiv:0801.4020 [nucl-ex]].

\bibitem{ATLAS:2017rmz}
 [ATLAS],
``Measurement of nuclear modification factor $R_\mathrm{AA}$ in Pb+Pb collisions at $\sqrt{s_{NN}} = 5.02$ TeV with the ATLAS detector at the LHC,''
ATLAS-CONF-2017-012.

\bibitem{Sekihata:2018lwz}
D.~Sekihata [ALICE],
``Energy and system dependence of nuclear modification factors of inclusive charged particles and identified light hadrons measured in p-Pb, Xe-Xe and Pb+Pb collisions with ALICE,''
Nucl. Phys. A \textbf{982} (2019), 567
[arXiv:1807.11240 [hep-ex]].

\bibitem{CMS:2016xef}
V.~Khachatryan \textit{et al.} [CMS],
``Charged-particle nuclear modification factors in PbPb and pPb collisions at $ \sqrt{s_{\mathrm{N}\;\mathrm{N}}}=5.02 $ TeV,''
JHEP \textbf{04}, 039 (2017)
[arXiv:1611.01664 [nucl-ex]].

\bibitem{PHENIX:2006ujp}
S.~S.~Adler \textit{et al.} [PHENIX],
``Common suppression pattern of eta and pi0 mesons at high transverse momentum in Au+Au collisions at $\sqrt{s_{\rm NN}}$ = 200 GeV,''
Phys. Rev. Lett. \textbf{96} (2006), 202301
[arXiv:nucl-ex/0601037 [nucl-ex]].

\bibitem{ATLAS:2015rlt}
G.~Aad \textit{et al.} [ATLAS],
``Centrality, rapidity and transverse momentum dependence of isolated prompt photon production in lead-lead collisions at $\sqrt{s_{\mathrm{NN}}} = 2.76$ TeV measured with the ATLAS detector,''
Phys. Rev. C \textbf{93}, 034914 (2016)
doi:10.1103/PhysRevC.93.034914
[arXiv:1506.08552 [hep-ex]].

\bibitem{ATLAS:2019ibd}
G.~Aad \textit{et al.} [ATLAS],
``Measurement of $W^\pm $ boson production in Pb+Pb collisions at $\sqrt{s_{\mathrm{NN}}} = 5.02~\text {Te}\text {V}$ with the ATLAS detector,''
Eur. Phys. J. C \textbf{79}, 935 (2019)
doi:10.1140/epjc/s10052-019-7439-3
[arXiv:1907.10414 [nucl-ex]].


\bibitem{ATLAS:2019maq}
G.~Aad \textit{et al.} [ATLAS],
``$Z$ boson production in Pb+Pb collisions at $\sqrt{s_{\textrm{NN}}}$= 5.02 TeV measured by the ATLAS experiment,''
Phys. Lett. B \textbf{802}, 135262 (2020)
doi:10.1016/j.physletb.2020.135262
[arXiv:1910.13396 [nucl-ex]].

\bibitem{JET:2013cls}
K.~M.~Burke \textit{et al.} [JET],
``Extracting the jet transport coefficient from jet quenching in high-energy heavy-ion collisions,''
Phys. Rev. C \textbf{90}, 014909 (2014)
[arXiv:1312.5003 [nucl-th]].

\bibitem{JETSCAPE:2021ehl}
S.~Cao \textit{et al.} [JETSCAPE],
``Determining the jet transport coefficient $\hat{q}$ from inclusive hadron suppression measurements using Bayesian parameter estimation,''
Phys. Rev. C \textbf{104}, 024905 (2021)
[arXiv:2102.11337 [nucl-th]].



\bibitem{PHENIX:2006iih}
A.~Adare \textit{et al.} [PHENIX],
``Energy Loss and Flow of Heavy Quarks in Au+Au Collisions at $\sqrt{s_{\rm NN}}$ = 200 GeV,''
Phys. Rev. Lett. \textbf{98}, 172301 (2007)
[arXiv:nucl-ex/0611018 [nucl-ex]].

\bibitem{STAR:2006btx}
B.~I.~Abelev \textit{et al.} [STAR],
``Transverse momentum and centrality dependence of high-$p_T$ non-photonic electron suppression in Au+Au collisions at $\sqrt{s_{NN}} = 200$ GeV,''
Phys. Rev. Lett. \textbf{98}, 192301 (2007)
[erratum: Phys. Rev. Lett. \textbf{106}, 159902 (2011)]
[arXiv:nucl-ex/0607012 [nucl-ex]].

\bibitem{STAR:2014wif}
L.~Adamczyk \textit{et al.} [STAR],
``Observation of $D^0$ Meson Nuclear Modifications in Au+Au Collisions at $\sqrt{s_{NN}}=200$  GeV,''
Phys. Rev. Lett. \textbf{113}, 142301 (2014)
[erratum: Phys. Rev. Lett. \textbf{121}, 229901 (2018)]
[arXiv:1404.6185 [nucl-ex]].

\bibitem{Vanek:2020sbq}
J.~Vanek [STAR],
``Measurements of Open-Charm Hadrons in Au+Au Collisions at $\sqrt{s_{\mathrm {NN}}} = 200\,\text {GeV}$ by the STAR Experiment,''
Springer Proc. Phys. \textbf{250}, 115 (2020)

\bibitem{ALICE:2021rxa}
S.~Acharya \textit{et al.} [ALICE],
``Prompt D$^{0}$, D$^{+}$, and D$^{*+}$ production in Pb\textendash{}Pb collisions at $ \sqrt{s_{\mathrm{NN}}} $ = 5.02 TeV,''
JHEP \textbf{01}, 174 (2022)
[arXiv:2110.09420 [nucl-ex]].

\bibitem{PHENIX:2022wim}
U.~A.~Acharya \textit{et al.} [PHENIX],
``Charm- and Bottom-Quark Production in Au$+$Au Collisions at $\sqrt{s_{_{NN}}}$ = 200 GeV,''
[arXiv:2203.17058 [nucl-ex]].

\bibitem{STAR:2021uzu}
S.~Collaboration \textit{et al.} [STAR],
``Evidence of Mass Ordering of Charm and Bottom Quark Energy Loss in Au+Au Collisions at RHIC,''
[arXiv:2111.14615 [nucl-ex]].

\bibitem{ALICE:2019nuy}
S.~Acharya \textit{et al.} [ALICE],
``Measurement of electrons from semileptonic heavy-flavour hadron decays at midrapidity in p+p and Pb+Pb collisions at $\sqrt{s_{\rm{NN}}}$ = 5.02 TeV,''
Phys. Lett. B \textbf{804}, 135377 (2020)
[arXiv:1910.09110 [nucl-ex]].

\bibitem{ATLAS:2021xtw}
G.~Aad \textit{et al.} [ATLAS],
``Measurement of the nuclear modification factor for muons from charm and bottom hadrons in Pb+Pb collisions at 5.02 TeV with the ATLAS detector,''
Phys. Lett. B \textbf{829}, 137077 (2022)
[arXiv:2109.00411 [nucl-ex]].

\bibitem{ALICE:2020sjb}
S.~Acharya \textit{et al.} [ALICE],
``Production of muons from heavy-flavour hadron decays at high transverse momentum in Pb\textendash{}Pb collisions at $\sqrt{s_{\rm NN}}$ = 5.02 and 2.76 TeV,''
Phys. Lett. B \textbf{820}, 136558 (2021)
[arXiv:2011.05718 [nucl-ex]].

\bibitem{ALICE:2022xrg}
 [ALICE],
``Measurement of beauty-strange meson production in Pb$-$Pb collisions at $\sqrt{s_{\rm NN}} = 5.02$ TeV via non-prompt $\mathrm{D_s}^{+}$ mesons,''
[arXiv:2204.10386 [nucl-ex]].

\bibitem{CMS:2021mzx}
A.~Tumasyan \textit{et al.} [CMS],
``Observation of $B_s^0$ mesons and measurement of the $B_{s}^{0}/B^+$ yield ratio in PbPb collisions at Image 1 TeV,''
Phys. Lett. B \textbf{829}, 137062 (2022)
doi:10.1016/j.physletb.2022.137062
[arXiv:2109.01908 [hep-ex]].

\bibitem{CMS:2022sxl}
A.~Tumasyan \textit{et al.} [CMS],
``Observation of the $B_c^+$ Meson in Pb-Pb and pp Collisions at $\sqrt{s_{NN}}$=5.02\,\,TeV and Measurement of its Nuclear Modification Factor,''
Phys. Rev. Lett. \textbf{128}, 252301 (2022)
[arXiv:2201.02659 [hep-ex]].

\bibitem{Matteo Cacciari_2008}
M, Cacciari,  G.P. Salam and  G. Soyez,``The anti-kt jet clustering algorithm,''
Journal of High Energy Physics \textbf{04}, 063 (2008).

\bibitem{Ke:2020clc}
W.~Ke and X.~N.~Wang,
``QGP modification to single inclusive jets in a calibrated transport model,''
JHEP \textbf{05}, 041 (2021)
[arXiv:2010.13680 [hep-ph]].

\bibitem{JETSCAPE:2023hqn}
Y.~Tachibana \textit{et al.} [JETSCAPE],
``Hard Jet Substructure in a Multi-stage Approach,''
[arXiv:2301.02485 [hep-ph]].

\bibitem{ALICE:2019qyj}
S.~Acharya \textit{et al.} [ALICE],
``Measurements of inclusive jet spectra in p+p and central Pb+Pb collisions at $\sqrt{s_{\rm{NN}}}$ = 5.02 TeV,''
Phys. Rev. C \textbf{101}, 034911 (2020)
[arXiv:1909.09718 [nucl-ex]].

\bibitem{ATLAS:2018gwx}
M.~Aaboud \textit{et al.} [ATLAS],
``Measurement of the nuclear modification factor for inclusive jets in Pb+Pb collisions at $\sqrt{s_\mathrm{NN}}=5.02$ TeV with the ATLAS detector,''
Phys. Lett. B \textbf{790}, 108 (2019)
[arXiv:1805.05635 [nucl-ex]].

\bibitem{STAR:2020xiv}
J.~Adam \textit{et al.} [STAR],
``Measurement of inclusive charged-particle jet production in Au + Au collisions at $\sqrt{s_{NN}}=$200 GeV,''
Phys. Rev. C \textbf{102}, 054913 (2020)
[arXiv:2006.00582 [nucl-ex]].

\bibitem{CMS:2021vui}
A.~M.~Sirunyan \textit{et al.} [CMS],
``First measurement of large area jet transverse momentum spectra in heavy-ion collisions,''
JHEP \textbf{05}, 284 (2021)
[arXiv:2102.13080 [hep-ex]].

\bibitem{Bossi:2023nmu}
H.~Bossi,
``Novel Uses of Machine Learning for Differential Jet Quenching Measurements at the LHC,''
CERN-THESIS-2023-026.

\bibitem{CMS:2018jco}
A.~M.~Sirunyan \textit{et al.} [CMS],
``Jet Shapes of Isolated Photon-Tagged Jets in Pb+Pb and p+p Collisions at $\sqrt{s_\mathrm{NN}} =$ 5.02  TeV,''
Phys. Rev. Lett. \textbf{122}, 152001 (2019)
[arXiv:1809.08602 [hep-ex]].

\bibitem{ATLAS:2022cim}
 [ATLAS],
``Comparison of inclusive and photon-tagged jet suppression in 5.02 TeV Pb+Pb collisions with ATLAS,''
ATLAS-CONF-2022-019.

\bibitem{ATLAS:2023iad}
G.~Aad \textit{et al.} [ATLAS],
``Comparison of inclusive and photon-tagged jet suppression in 5.02 TeV Pb+Pb collisions with ATLAS,''
Phys. Lett. B \textbf{846}, 138154 (2023)
[arXiv:2303.10090 [nucl-ex]].

\bibitem{ATLAS:2018bvp}
M.~Aaboud \textit{et al.} [ATLAS],
``Measurement of jet fragmentation in Pb+Pb and p+p collisions at $\sqrt{s_{NN}} = 5.02$ TeV with the ATLAS detector,''
Phys. Rev. C \textbf{98}, 024908 (2018)
[arXiv:1805.05424 [nucl-ex]].

\bibitem{CMS:2013lhm}
S.~Chatrchyan \textit{et al.} [CMS],
``Modification of Jet Shapes in PbPb Collisions at $\sqrt {s_{NN}} = 2.76$ TeV,''
Phys. Lett. B \textbf{730}, 243 (2014)
[arXiv:1310.0878 [nucl-ex]].

\bibitem{CMS:2018zze}
A.~M.~Sirunyan \textit{et al.} [CMS],
``Jet properties in PbPb and p+p collisions at $\sqrt {s_{NN}} = 5.02 $ TeV,''
JHEP \textbf{05}, 006 (2018)
[arXiv:1803.00042 [nucl-ex]].

\bibitem{CMS:2022btc}
 [CMS],
``Search for medium effects using jets from bottom quarks in PbPb collisions at $\sqrt{s_\mathrm{NN}}$ = 5.02 TeV,''
[arXiv:2210.08547 [hep-ex]].

\bibitem{ATLAS:2022fgb}
 [ATLAS],
``Measurement of the nuclear modification factor of $b$-jets in 5.02 TeV Pb+Pb collisions with the ATLAS detector,''
[arXiv:2204.13530 [nucl-ex]].

\bibitem{Dasgupta:2013ihk}
M.~Dasgupta, A.~Fregoso, S.~Marzani and G.~P.~Salam,
``Towards an understanding of jet substructure,''
JHEP \textbf{09}, 029 (2013)
[arXiv:1307.0007 [hep-ph]].

\bibitem{Larkoski:2014wba}
A.~J.~Larkoski, S.~Marzani, G.~Soyez and J.~Thaler,
``Soft Drop,''
JHEP \textbf{05}, 146 (2014)
[arXiv:1402.2657 [hep-ph]].

\bibitem{Catani:1993hr}
S.~Catani, Y.~L.~Dokshitzer, M.~H.~Seymour and B.~R.~Webber,
``Longitudinally invariant $K_t$ clustering algorithms for hadron hadron collisions,''
Nucl. Phys. B \textbf{406}, 187 (1993)

\bibitem{ALICE:2021mqf}
S.~Acharya \textit{et al.} [ALICE],
``Measurement of the groomed jet radius and momentum splitting fraction in pp and Pb$-$Pb collisions at $\sqrt{s_{NN}} = 5.02$ TeV,''
Phys. Rev. Lett. \textbf{128}, 102001 (2022)
[arXiv:2107.12984 [nucl-ex]].

\bibitem{Dreyer:2018nbf}
F.~A.~Dreyer, G.~P.~Salam and G.~Soyez,
``The Lund Jet Plane,''
JHEP \textbf{12}, 064 (2018)
[arXiv:1807.04758 [hep-ph]].

\bibitem{ATLAS:2020bbn}
G.~Aad \textit{et al.} [ATLAS],
``Measurement of the Lund Jet Plane Using Charged Particles in 13 TeV Proton-Proton Collisions with the ATLAS Detector,''
Phys. Rev. Lett. \textbf{124}, 222002 (2020)
[arXiv:2004.03540 [hep-ex]].

\bibitem{ATLAS:2022vii}
 [ATLAS],
``Measurement of substructure-dependent jet suppression in Pb+Pb collisions at 5.02 TeV with the ATLAS detector,''
[arXiv:2211.11470 [nucl-ex]].

\bibitem{Gao:2019ojf}
A.~Gao, H.~T.~Li, I.~Moult and H.~X.~Zhu,
``Precision QCD Event Shapes at Hadron Colliders: The Transverse Energy-Energy Correlator in the Back-to-Back Limit,''
Phys. Rev. Lett. \textbf{123}, 062001 (2019)
[arXiv:1901.04497 [hep-ph]]

\bibitem{Dixon:2019uzg}
L.~J.~Dixon, I.~Moult and H.~X.~Zhu,
``Collinear limit of the energy-energy correlator,''
Phys. Rev. D \textbf{100}, 014009 (2019)
[arXiv:1905.01310 [hep-ph]].

\bibitem{Jaarsma:2022kdd}
M.~Jaarsma, Y.~Li, I.~Moult, W.~Waalewijn and H.~X.~Zhu,
``Renormalization group flows for track function moments,''
JHEP \textbf{06}, 139 (2022)
[arXiv:2201.05166 [hep-ph]].

\bibitem{Sveshnikov:1995vi}
N.~A.~Sveshnikov and F.~V.~Tkachov,
``Jets and quantum field theory,''
Phys. Lett. B \textbf{382}, 403 (1996)
[arXiv:hep-ph/9512370 [hep-ph]].

\bibitem{Chen:2022pdu}
H.~Chen, M.~Jaarsma, Y.~Li, I.~Moult, W.~J.~Waalewijn and H.~X.~Zhu,
``Multi-Collinear Splitting Kernels for Track Function Evolution,''
[arXiv:2210.10058 [hep-ph]].

\bibitem{Komiske:2022enw}
P.~T.~Komiske, I.~Moult, J.~Thaler and H.~X.~Zhu,
``Analyzing N-Point Energy Correlators inside Jets with CMS Open Data,''
Phys. Rev. Lett. \textbf{130}, 051901 (2023)
[arXiv:2201.07800 [hep-ph]].

\bibitem{Lee:2022ige}
K.~Lee, B.~Me\c{c}aj and I.~Moult,
``Conformal Colliders Meet the LHC,''
[arXiv:2205.03414 [hep-ph]].

\bibitem{Andres:2023xwr}
C.~Andres, F.~Dominguez, J.~Holguin, C.~Marquet and I.~Moult,
``A Coherent View of the Quark-Gluon Plasma from Energy Correlators,''
[arXiv:2303.03413 [hep-ph]].

\bibitem{Danielewicz:1984ww}
P.~Danielewicz and M.~Gyulassy,
``Dissipative Phenomena in Quark Gluon Plasmas,''
Phys. Rev. D \textbf{31}, 53 (1985).

\bibitem{Gustafsson:1984ka}
H.~A.~Gustafsson, H.~H.~Gutbrod, B.~Kolb, H.~Lohner, B.~Ludewigt, A.~M.~Poskanzer, T.~Renner, H.~Riedesel, H.~G.~Ritter and A.~Warwick, \textit{et al.}
``Collective Flow Observed in Relativistic Nuclear Collisions,''
Phys. Rev. Lett. \textbf{52}, 1590-1593 (1984)

\bibitem{Renfordt:1984et}
R.~E.~Renfordt, D.~Schall, R.~Bock, R.~Brockmann, J.~W.~Harris, A.~Sandoval, R.~Stock, H.~Strobele, D.~Bangert and W.~Rauch, \textit{et al.}
``Stopping Power and Collective Flow of Nuclear Matter in the Reaction Ar+Pb at 0.8-GeV/u,''
Phys. Rev. Lett. \textbf{53}, 763-766 (1984)

\bibitem{NA49:1997qey}
H.~Appelsh{\"a}user \textit{et al.} [NA49],
``Directed and elliptic flow in 158-GeV/nucleon Pb+Pb collisions,''
Phys. Rev. Lett. \textbf{80}, 4136-4140 (1998)
[arXiv:nucl-ex/9711001 [nucl-ex]].

\bibitem{Muller:2006ee}
B.~M{\"u}ller and J.~L.~Nagle,
``Results from the relativistic heavy ion collider,''
Ann. Rev. Nucl. Part. Sci. \textbf{56}, 93 (2006)
[arXiv:nucl-th/0602029 [nucl-th]].

\bibitem{Muller:2012zq}
B.~M{\"u}ller, J.~Schukraft and B.~Wyslouch,
``First Results from Pb+Pb collisions at the LHC,''
Ann. Rev. Nucl. Part. Sci. \textbf{62}, 361 (2012)
[arXiv:1202.3233 [hep-ex]].

\bibitem{ALICE:2018rtz}
S.~Acharya \textit{et al.} [ALICE],
``Energy dependence and fluctuations of anisotropic flow in Pb-Pb collisions at $ \sqrt{s_{\mathrm{NN}}}=5.02$ and 2.76 TeV,''
JHEP \textbf{07}, 103 (2018)
[arXiv:1804.02944 [nucl-ex]].

\bibitem{ATLAS:2018ezv}
M.~Aaboud \textit{et al.} [ATLAS],
``Measurement of the azimuthal anisotropy of charged particles produced in $\sqrt{s_{_\text {NN}}}$ = 5.02 TeV Pb+Pb collisions with the ATLAS detector,''
Eur. Phys. J. C \textbf{78}, 997 (2018)
[arXiv:1808.03951 [nucl-ex]].

\bibitem{Schenke:2020mbo}
B.~Schenke, C.~Shen and P.~Tribedy,
``Running the gamut of high energy nuclear collisions,''
Phys. Rev. C \textbf{102}, 044905 (2020)
[arXiv:2005.14682 [nucl-th]].

\bibitem{Kurkela:2015qoa}
A.~Kurkela and Y.~Zhu,
``Isotropization and hydrodynamization in weakly coupled heavy-ion collisions,''
Phys. Rev. Lett. \textbf{115}, no.18, 182301 (2015)
[arXiv:1506.06647 [hep-ph]].

\bibitem{Heller:2016gbp}
M.~P.~Heller,
``Holography, Hydrodynamization and Heavy-Ion Collisions,''
Acta Phys. Polon. B \textbf{47}, 2581 (2016)
[arXiv:1610.02023 [hep-th]].

\bibitem{Kumar:2011de}
L.~Kumar [STAR],
``Results from the STAR Beam Energy Scan Program,''
Nucl. Phys. A \textbf{862}, 125 (2011)
[arXiv:1101.4310 [nucl-ex]].

\bibitem{Fries:2003vb}
R.~J.~Fries, B.~M{\"u}ller, C.~Nonaka and S.~A.~Bass,
``Hadronization in heavy ion collisions: Recombination and fragmentation of partons,''
Phys. Rev. Lett. \textbf{90}, 202303 (2003)
[arXiv:nucl-th/0301087 [nucl-th]].

\bibitem{STAR:2022tfp}
M.~Abdallah \textit{et al.} [STAR],
``Azimuthal anisotropy measurement of (multi)strange hadrons in Au+Au collisions at $\sqrt{s_{NN}} =  54.4 GeV$,''
Phys. Rev. C \textbf{107}, 024912 (2023)
[arXiv:2205.11073 [nucl-ex]].

\bibitem{ALICE:2018yph}
S.~Acharya \textit{et al.} [ALICE],
``Anisotropic flow of identified particles in Pb-Pb collisions at $ {\sqrt{s}}_{\mathrm{NN}}=5.02 $ TeV,''
JHEP \textbf{09}, 006 (2018)
[arXiv:1805.04390 [nucl-ex]].

\bibitem{Staig:2010pn}
P.~Staig and E.~Shuryak,
``The Fate of the Initial State Fluctuations in Heavy Ion Collisions. II The Fluctuations and Sounds,''
Phys. Rev. C \textbf{84}, 034908 (2011)
[arXiv:1008.3139 [nucl-th]].

\bibitem{CMS:2017xnj}
A.~M.~Sirunyan \textit{et al.} [CMS],
``Pseudorapidity and transverse momentum dependence of flow harmonics in pPb and PbPb collisions,''
Phys. Rev. C \textbf{98}, 044902 (2018)
[arXiv:1710.07864 [nucl-ex]].

\bibitem{ALICE:2020sup}
S.~Acharya \textit{et al.} [ALICE],
``Higher harmonic non-linear flow modes of charged hadrons in Pb-Pb collisions at $\sqrt{s_{\rm{NN}}}$ = 5.02 TeV,''
JHEP \textbf{05}, 085 (2020)
[arXiv:2002.00633 [nucl-ex]].

\bibitem{Bernhard:2019bmu}
J.~E.~Bernhard, J.~S.~Moreland and S.~A.~Bass,
``Bayesian estimation of the specific shear and bulk viscosity of quark\textendash{}gluon plasma,''
Nature Phys. \textbf{15}, 1113 (2019)

\bibitem{Gyulassy:2004zy}
M.~Gyulassy and L.~McLerran,
``New forms of QCD matter discovered at RHIC,''
Nucl. Phys. A \textbf{750}, 30 (2005)
[arXiv:nucl-th/0405013 [nucl-th]].

\bibitem{Shuryak:2004kh}
E.~Shuryak,
``A strongly coupled quark-gluon plasma,''
J. Phys. G \textbf{30}, S1221-S1224 (2004).

\bibitem{Kovtun:2004de}
P.~Kovtun, D.~T.~Son and A.~O.~Starinets,
``Viscosity in strongly interacting quantum field theories from black hole physics,''
Phys. Rev. Lett. \textbf{94}, 111601 (2005)
[arXiv:hep-th/0405231 [hep-th]].

\bibitem{Baier:1996kr}
R.~Baier, Y.~L.~Dokshitzer, A.~H.~Mueller, S.~Peigne and D.~Schiff,
``Radiative energy loss of high-energy quarks and gluons in a finite volume quark - gluon plasma,''
Nucl. Phys. B \textbf{483}, 291-320 (1997)
[arXiv:hep-ph/9607355 [hep-ph]].

\bibitem{Baier:1996sk}
R.~Baier, Y.~L.~Dokshitzer, A.~H.~Mueller, S.~Peigne and D.~Schiff,
``Radiative energy loss and p(T) broadening of high-energy partons in nuclei,''
Nucl. Phys. B \textbf{484}, 265-282 (1997)
[arXiv:hep-ph/9608322 [hep-ph]].

\bibitem{ALICE:2015efi}
J.~Adam \textit{et al.} [ALICE],
``Azimuthal anisotropy of charged jet production in $\sqrt{s_{\rm NN}}$ = 2.76 TeV Pb-Pb collisions,''
Phys. Lett. B \textbf{753}, 511-525 (2016)
[arXiv:1509.07334 [nucl-ex]].

\bibitem{CMS:2012tqw}
S.~Chatrchyan \textit{et al.} [CMS],
``Azimuthal anisotropy of charged particles at high transverse momenta in PbPb collisions at $\sqrt{s_{NN}}=2.76$ TeV,''
Phys. Rev. Lett. \textbf{109}, 022301 (2012)
[arXiv:1204.1850 [nucl-ex]].

\bibitem{ATLAS:2021ktw}
G.~Aad \textit{et al.} [ATLAS],
``Measurements of azimuthal anisotropies of jet production in Pb+Pb collisions at $\sqrt{s_{NN}} =$ 5.02 TeV with the ATLAS detector,''
Phys. Rev. C \textbf{105}, 064903 (2022)
[arXiv:2111.06606 [nucl-ex]].

\bibitem{CMS:2023lgq}
CMS Collaboration,
``Observation of enhanced long-range elliptic anisotropies inside high-multiplicity jets in pp collisions at the LHC,''
CMS-PAS-HIN-21-013.

\bibitem{Gardim:2019xjs}
F.~G.~Gardim, G.~Giacalone, M.~Luzum and J.~Y.~Ollitrault,
``Thermodynamics of hot strong-interaction matter from ultrarelativistic nuclear collisions,''
Nature Phys. \textbf{16}, 615-619 (2020)
[arXiv:1908.09728 [nucl-th]].

\bibitem{Gardim:2019brr}
F.~G.~Gardim, G.~Giacalone and J.~Y.~Ollitrault,
``The mean transverse momentum of ultracentral heavy-ion collisions: A new probe of hydrodynamics,''
Phys. Lett. B \textbf{809}, 135749 (2020)
[arXiv:1909.11609 [nucl-th]].

\bibitem{CMS:2023byu}
CMS Collaboration,
``Extracting the speed of sound in the strongly interacting matter created in relativistic nuclear collisions,''
CMS-PAS-HIN-23-003.

\bibitem{Liang:2004ph}
Z.~T.~Liang and X.~N.~Wang,
``Globally polarized quark-gluon plasma in non-central A+A collisions,''
Phys. Rev. Lett. \textbf{94}, 102301 (2005)
[erratum: Phys. Rev. Lett. \textbf{96}, 039901 (2006)]
[arXiv:nucl-th/0410079 [nucl-th]].

\bibitem{STAR:2017ckg}
L.~Adamczyk \textit{et al.} [STAR],
``Global $\Lambda$ hyperon polarization in nuclear collisions: evidence for the most vortical fluid,''
Nature \textbf{548}, 62-65 (2017)
[arXiv:1701.06657 [nucl-ex]].

\bibitem{Karpenko:2016jyx}
I.~Karpenko and F.~Becattini,
``Study of $\Lambda $ polarization in relativistic nuclear collisions at $\sqrt{s_\mathrm {NN}}=7.7$ \textendash{}200 GeV,''
Eur. Phys. J. C \textbf{77}, 213 (2017)
[arXiv:1610.04717 [nucl-th]].

\bibitem{Florkowski:2017ruc}
W.~Florkowski, B.~Friman, A.~Jaiswal and E.~Speranza,
``Relativistic fluid dynamics with spin,''
Phys. Rev. C \textbf{97}, 041901 (2018)
[arXiv:1705.00587 [nucl-th]].

\bibitem{STAR:2022fan}
M.~S.~Abdallah \textit{et al.} [STAR],
``Pattern of global spin alignment of \ensuremath{\phi} and K$^{*0}$ mesons in heavy-ion collisions,''
Nature \textbf{614}, 244-248 (2023)
[arXiv:2204.02302 [hep-ph]].

\bibitem{Kundu:2021lra}
S.~Kundu [ALICE],
``Spin alignment measurements of vector mesons with ALICE at the LHC,''
Nucl. Phys. A \textbf{1005}, 121912 (2021)

\bibitem{Chapline:1973kkq}
G.~F.~Chapline, M.~H.~Johnson, E.~Teller and M.~S.~Weiss,
``Highly excited nuclear matter,''
Phys. Rev. D \textbf{8}, 4302-4308 (1973)

\bibitem{Harris:1984up}
J.~W.~Harris, R.~Stock, R.~Bock, R.~Brockmann, A.~Sandoval, H.~Strobele, G.~Odyniec, H.~G.~Pugh, L.~S.~Schroeder and R.~E.~Renfordt, \textit{et al.}
``Pion Production as a Probe of the Nuclear Matter Equation of State,''
Phys. Lett. B \textbf{153}, 377-381 (1985)

\bibitem{Ratti:2022qgf}
C.~Ratti,
``Equation of state for QCD from lattice simulations,''
Prog. Part. Nucl. Phys. \textbf{129}, 104007 (2023)

\bibitem{Moreland:2015dvc}
J.~S.~Moreland and R.~A.~Soltz,
``Hydrodynamic simulations of relativistic heavy-ion collisions with different lattice quantum chromodynamics calculations of the equation of state,''
Phys. Rev. C \textbf{93}, 044913 (2016)
[arXiv:1512.02189 [nucl-th]].

\bibitem{Annala:2019puf}
E.~Annala, T.~Gorda, A.~Kurkela, J.~N\"attil\"a and A.~Vuorinen,
``Evidence for quark-matter cores in massive neutron stars,''
Nature Phys. \textbf{16}, 907-910 (2020)
[arXiv:1903.09121 [astro-ph.HE]].

\bibitem{LIGOScientific:2017ync}
B.~P.~Abbott \textit{et al.} 
``Multi-messenger Observations of a Binary Neutron Star Merger,''
Astrophys. J. Lett. \textbf{848}, L12 (2017)
[arXiv:1710.05833 [astro-ph.HE]].

\bibitem{LIGOScientific:2017zic}
B.~P.~Abbott \textit{et al.} 
``Gravitational Waves and Gamma-rays from a Binary Neutron Star Merger: GW170817 and GRB 170817A,''
Astrophys. J. Lett. \textbf{848}, L13 (2017)

\bibitem{Oechslin:2004yj}
R.~Oechslin, K.~Uryu, G.~S.~Poghosyan and F.~K.~Thielemann,
``The Influence of quark matter at high densities on binary neutron star mergers,''
Mon. Not. Roy. Astron. Soc. \textbf{349}, 1469 (2004)
[arXiv:astro-ph/0401083 [astro-ph]].

\bibitem{Huth:2021bsp}
S.~Huth, P.~T.~H.~Pang, I.~Tews, T.~Dietrich, A.~L.~F\`evre, A.~Schwenk, W.~Trautmann, K.~Agarwal, M.~Bulla and M.~W.~Coughlin, \textit{et al.}
``Constraining Neutron-Star Matter with Microscopic and Macroscopic Collisions,''
Nature \textbf{606}, 276-280 (2022)
[arXiv:2107.06229 [nucl-th]].

\bibitem{Most:2018eaw}
E.~R.~Most, L.~J.~Papenfort, V.~Dexheimer, M.~Hanauske, S.~Schramm, H.~St\"ocker and L.~Rezzolla,
``Signatures of quark-hadron phase transitions in general-relativistic neutron-star mergers,''
Phys. Rev. Lett. \textbf{122}, 061101 (2019)
[arXiv:1807.03684 [astro-ph.HE]].

\bibitem{Bauswein:2018bma}
A.~Bauswein, N.~U.~F.~Bastian, D.~B.~Blaschke, K.~Chatziioannou, J.~A.~Clark, T.~Fischer and M.~Oertel,
``Identifying a first-order phase transition in neutron star mergers through gravitational waves,''
Phys. Rev. Lett. \textbf{122}, 061102 (2019)
[arXiv:1809.01116 [astro-ph.HE]].

\bibitem{Tan:2020ics}
H.~Tan, J.~Noronha-Hostler and N.~Yunes,
``Neutron Star Equation of State in light of GW190814,''
Phys. Rev. Lett. \textbf{125}, 261104 (2020)
[arXiv:2006.16296 [astro-ph.HE]].

\cite{Most:2022wgo}
\bibitem{Most:2022wgo}
E.~R.~Most, A.~Motornenko, J.~Steinheimer, V.~Dexheimer, M.~Hanauske, L.~Rezzolla and H.~Stoecker,
``Probing neutron-star matter in the lab: Similarities and differences between binary mergers and heavy-ion collisions,''
Phys. Rev. D \textbf{107}, 043034 (2023)
[arXiv:2201.13150 [nucl-th]].

\bibitem{CBM:2016kpk}
T.~Ablyazimov \textit{et al.} [CBM],
``Challenges in QCD matter physics --The scientific programme of the Compressed Baryonic Matter experiment at FAIR,''
Eur. Phys. J. A \textbf{53}, 60 (2017)
[arXiv:1607.01487 [nucl-ex]].

\bibitem{Fermi:1950jd}
E.~Fermi,
``High-energy nuclear events,''
Prog. Theor. Phys. \textbf{5}, 570-583 (1950)

\bibitem{Belenkij:1955pgn}
S.~Z.~Belenkij and L.~D.~Landau,
``Hydrodynamic theory of multiple production of particles,''
Usp. Fiz. Nauk \textbf{56}, 309 (1955)

\bibitem{Hagedorn:1965st}
R.~Hagedorn,
``Statistical thermodynamics of strong interactions at high-energies,''
Nuovo Cim. Suppl. \textbf{3}, 147-186 (1965).

\bibitem{E735:1994hbl}
N.~T.~Porile \textit{et al.} [E735],
``Recent results from E735: Search for quark - gluon plasma in p anti-p collisions at 0.3-TeV to 1.8-TeV,''
Nucl. Phys. A \textbf{566}, 431C-434C (1994)

\bibitem{CMS:2010ifv}
V.~Khachatryan \textit{et al.} [CMS],
``Observation of Long-Range Near-Side Angular Correlations in Proton-Proton Collisions at the LHC,''
JHEP \textbf{09}, 091 (2010)
[arXiv:1009.4122 [hep-ex]].

\bibitem{ATLAS:2015hzw}
G.~Aad \textit{et al.} [ATLAS],
``Observation of Long-Range Elliptic Azimuthal Anisotropies in $\sqrt{s}=$13 and 2.76 TeV $pp$ Collisions with the ATLAS Detector,''
Phys. Rev. Lett. \textbf{116}, 172301 (2016)
[arXiv:1509.04776 [hep-ex]].

\bibitem{CMS:2015fgy}
V.~Khachatryan \textit{et al.} [CMS],
``Measurement of long-range near-side two-particle angular correlations in pp collisions at $\sqrt s =$13 TeV,''
Phys. Rev. Lett. \textbf{116}, 172302 (2016)
[arXiv:1510.03068 [nucl-ex]].


\bibitem{ALICE:2012eyl}
B.~Abelev \textit{et al.} [ALICE],
``Long-range angular correlations on the near and away side in $p$-Pb collisions at $\sqrt{s_{NN}}=5.02$ TeV,''
Phys. Lett. B \textbf{719}, 29-41 (2013)
[arXiv:1212.2001 [nucl-ex]].

\bibitem{ATLAS:2012cix}
G.~Aad \textit{et al.} [ATLAS],
``Observation of Associated Near-Side and Away-Side Long-Range Correlations in $\sqrt{s_{NN}}$=5.02  TeV Proton-Lead Collisions with the ATLAS Detector,''
Phys. Rev. Lett. \textbf{110}, 182302 (2013)
[arXiv:1212.5198 [hep-ex]].

\bibitem{CMS:2012qk}
S.~Chatrchyan \textit{et al.} [CMS],
``Observation of Long-Range Near-Side Angular Correlations in Proton-Lead Collisions at the LHC,''
Phys. Lett. B \textbf{718}, 795-814 (2013)
[arXiv:1210.5482 [nucl-ex]].

\bibitem{CMS:2015yux}
V.~Khachatryan \textit{et al.} [CMS],
``Evidence for Collective Multiparticle Correlations in p-Pb Collisions,''
Phys. Rev. Lett. \textbf{115}, 012301 (2015)
[arXiv:1502.05382 [nucl-ex]].

\bibitem{PHENIX:2018lia}
C.~Aidala \textit{et al.} [PHENIX],
``Creation of quark\textendash{}gluon plasma droplets with three distinct geometries,''
Nature Phys. \textbf{15}, 214-220 (2019)
[arXiv:1805.02973 [nucl-ex]].

\bibitem{Nagle:2018nvi}
J.~L.~Nagle and W.~A.~Zajc,
``Small System Collectivity in Relativistic Hadronic and Nuclear Collisions,''
Ann. Rev. Nucl. Part. Sci. \textbf{68}, 211-235 (2018)
[arXiv:1801.03477 [nucl-ex]].

\bibitem{Weller:2017tsr}
R.~D.~Weller and P.~Romatschke,
``One fluid to rule them all: viscous hydrodynamic description of event-by-event central p+p, p+Pb and Pb+Pb collisions at $\sqrt {s_{NN}} = 5.02$ TeV,''
Phys. Lett. B \textbf{774}, 351-356 (2017)
[arXiv:1701.07145 [nucl-th]].

\bibitem{ALICE:2017svf}
S.~Acharya \textit{et al.} [ALICE],
``Constraints on jet quenching in p-Pb collisions at $\sqrt {s_{NN}}$ = 5.02 TeV measured by the event-activity dependence of semi-inclusive hadron-jet distributions,''
Phys. Lett. B \textbf{783}, 95-113 (2018)
[arXiv:1712.05603 [nucl-ex]].

\bibitem{Bierlich:2017vhg}
C.~Bierlich, G.~Gustafson and L.~L\"onnblad,
``Collectivity without plasma in hadronic collisions,''
Phys. Lett. B \textbf{779}, 58-63 (2018)
[arXiv:1710.09725 [hep-ph]].

\bibitem{Davidson:2016mhd}
P.~A.~Davidson,
``Introduction to Magnetohydrodynamics,''
(Cambridge University Press, Cambridge, 2017).

\bibitem{Hernandez:2017mch}
J.~Hernandez and P.~Kovtun,
``Relativistic magnetohydrodynamics,''
JHEP \textbf{05}, 001 (2017)
[arXiv:1703.08757 [hep-th]].

\bibitem{Shi:2013jha}
S.~Shi [STAR],
``Recent elliptic flow results from beam energy scan at STAR,''
J. Phys. Conf. Ser. \textbf{422}, 012002 (2013)

\bibitem{Parfenov:2020fuo}
P.~Parfenov [STAR],
``Elliptic ($v_2$) and triangular ($v_3$) anisotropic flow of identified hadrons from the STAR Beam Energy Scan program,''
J. Phys. Conf. Ser. \textbf{1690}, 012128 (2020)
[arXiv:2012.06759 [hep-ex]].

\bibitem{ALICE:2022wwr}
 [ALICE],
``Letter of intent for ALICE 3: A next-generation heavy-ion experiment at the LHC,''
[arXiv:2211.02491 [physics.ins-det]].

\bibitem{NA60:2022sze}
C.~Ahdida \textit{et al.} [NA60+],
``Letter of Intent: the NA60+ experiment,''
[arXiv:2212.14452 [nucl-ex]].

\bibitem{PHENIX:2015siv}
A.~Adare \textit{et al.} [PHENIX],
``An Upgrade Proposal from the PHENIX Collaboration,''
[arXiv:1501.06197 [nucl-ex]].

\bibitem{Mantysaari:2022ffw}
H.~M\"antysaari, B.~Schenke, C.~Shen and W.~Zhao,
``Bayesian inference of the fluctuating proton shape,''
Phys. Lett. B \textbf{833}, 137348 (2022)
[arXiv:2202.01998 [hep-ph]].

\bibitem{STAR:2021wwq}
M.~Abdallah \textit{et al.} [STAR],
``Probing the Gluonic Structure of the Deuteron with $J/\psi$ Photoproduction in d+Au Ultraperipheral Collisions,''
Phys. Rev. Lett. \textbf{128}, 122303 (2022)
[arXiv:2109.07625 [nucl-ex]].

\bibitem{ALICE:2019tqa}
S.~Acharya \textit{et al.} [ALICE],
``Coherent J/$\psi$ photoproduction at forward rapidity in ultra-peripheral Pb-Pb collisions at $\sqrt{s_{\rm{NN}}}=5.02$ TeV,''
Phys. Lett. B \textbf{798}, 134926 (2019)
[arXiv:1904.06272 [nucl-ex]].

\bibitem{Accardi:2012qut}
A.~Accardi, J.~L.~Albacete, M.~Anselmino, N.~Armesto, E.~C.~Aschenauer, A.~Bacchetta, D.~Boer, W.~K.~Brooks, T.~Burton and N.~B.~Chang, \textit{et al.}
``Electron Ion Collider: The Next QCD Frontier: Understanding the glue that binds us all,''
Eur. Phys. J. A \textbf{52}, 268 (2016)
[arXiv:1212.1701 [nucl-ex]].


\end{thebibliography}
\end{document}